\documentclass[preprint,10pt]{aastex}

\usepackage{graphicx}
\usepackage{epsfig}
\usepackage{multirow}

\def\ha{\relax \ifmmode {\rm H}\alpha\else H$\alpha$\fi}
\def\pa{\relax \ifmmode {\rm Pa}\alpha\else Pa$\alpha$\fi}
\def\arcsec{\hbox{$^{\prime\prime}$}}
\def\nii{\relax \ifmmode {\rm N\,{\sc ii}}\else N\,{\sc ii}\fi}
\def\hii{\relax \ifmmode {\rm H\,{\sc ii}}\else H\,{\sc ii}\fi}
\def\hi{\relax \ifmmode {\rm H\,{\sc i}}\else H\,{\sc i}\fi}
\def\deg{\hbox{$^{\circ}$}}

\shorttitle{How much mass is contained in thick disks?}
\shortauthors{Comer\'on et al.}

\begin{document}


\title{Thick disks of edge-on galaxies seen through the Spitzer Survey of Stellar Structure in Galaxies (S$^4$G): Lair of missing baryons?}


\author{
S\'ebastien Comer\'on,\altaffilmark{1,2}
Bruce G.~Elmegreen,\altaffilmark{3}
Johan H.~Knapen,\altaffilmark{4,5}
Heikki Salo,\altaffilmark{6}
Eija Laurikainen,\altaffilmark{2,6}
Jarkko Laine,\altaffilmark{6}
E.~Athanassoula,\altaffilmark{7}
Albert Bosma,\altaffilmark{7}
Kartik Sheth,\altaffilmark{8,9,10}
Michael W.~Regan,\altaffilmark{11}
Joannah L.~Hinz,\altaffilmark{12}
Armando Gil de Paz,\altaffilmark{13}
Kar\'in Men\'endez-Delmestre,\altaffilmark{14}
Trisha Mizusawa,\altaffilmark{8}
Juan-Carlos Mu\~noz-Mateos,\altaffilmark{8}
Mark Seibert,\altaffilmark{14}
Taehyun Kim,\altaffilmark{8,15}
Debra M.~Elmegreen,\altaffilmark{16}
Dimitri A.~Gadotti,\altaffilmark{17}
Luis C.~Ho,\altaffilmark{14}
Benne W.~Holwerda,\altaffilmark{18}
Jani Lappalainen,\altaffilmark{6}
Eva Schinnerer\altaffilmark{19}
and Ramin Skibba\altaffilmark{12}}

\altaffiltext{1}{Korea Astronomy and Space Science Institute, 61-1 Hwaam-dong, Yuseong-gu, Daejeon 305-348, Republic of Korea}
\altaffiltext{2}{Finnish Centre of Astronomy with ESO (FINCA), University of Turku, V\"ais\"al\"antie 20, FI-21500, Piikki\"o, Finland}
\altaffiltext{3}{IBM T.~J.~Watson Research Center, 1101 Kitchawan Road, Yorktown Heights, NY 10598, USA}
\altaffiltext{4}{Instituto de Astrof\'isica de Canarias, E-38200 La Laguna, Spain}
\altaffiltext{5}{Departamento de Astrof\'isica, Universidad de La Laguna, E-38205 La Laguna, Tenerife, Spain}
\altaffiltext{6}{Astronomy Division, Department of Physical Sciences, P.~O.~Box 3000, FIN-90014 University of Oulu, Finland}
\altaffiltext{7}{Laboratoire d'Astrophysique de Marseille (LAM), UMR6110, Universit\'e de Provence/CNRS, Technop\^ole de Marseille \'Etoile, 38 rue Fr\'ed\'eric Joliot Curie, 13388 Marseille C\'edex 20, France}
\altaffiltext{8}{National Radio Astronomy Observatory / NAASC, 520, Edgemont Road, Charlottesville, VA 22903, USA}
\altaffiltext{9}{Spitzer Space Center, California Institute of Technology, Pasadena, CA 91125, USA}
\altaffiltext{10}{California Institute of Technology, 1200 East California Boulevard, Pasadena, CA 91125, USA}
\altaffiltext{11}{Space Telescope Science Institute, 3700 San Martin Drive, Baltimore, MD 21218, USA}
\altaffiltext{12}{Steward Observatory, University of Arizona, 933 North Cherry Avenue, Tucson, AZ 85721, USA}
\altaffiltext{13}{Departamento de Astrof\'isica, Universidad Complutense de Madrid, 28040, Madrid Spain}
\altaffiltext{14}{The Observatories of the Carnegie Institution of Washington, 813 Santa Barbara Street, Pasadena, CA 91101, USA}
\altaffiltext{15}{Astronomy Program, Department of Physics and Astronomy, Seoul National University, Seoul, 151-742, Republic of Korea}
\altaffiltext{16}{Vassar College, Department of Physics and Astronomy, Poughkeepsie, NY 12604}
\altaffiltext{17}{European Southern Observatory, Casilla 19001, Santiago 19, Chile}
\altaffiltext{18}{Astrophysics, Cosmology and Gravity Center (ACGC), Astronomy Department, University of Cape Town, Private Bag X3, 7700 Rondebosch, Republic of South Africa}
\altaffiltext{19}{Max-Planck-Institut f\"ur Astronomie, K\"onigstuhl 17, 69117 Heidelberg, Germany}


\begin{abstract}

Most, if not all, disk galaxies have a thin (classical) disk and a thick disk. In most models thick disks are thought to be a necessary consequence of the disk formation and/or evolution of the galaxy. We present the results of a study of the thick disk properties in a sample of carefully selected edge-on galaxies with types ranging from $T=3$ to $T=8$. We fitted  one-dimensional luminosity profiles with physically motivated functions -- the solutions of two stellar and one gaseous isothermal coupled disks in equilibrium -- which are likely to yield more accurate results than other functions used in previous studies. The images used for the fits come from the Spitzer Survey of Stellar Structure in Galaxies (S$^4$G). We found that thick disks are on average more massive than previously reported, mostly due to the selected fitting function. Typically, the thin and the thick disk have similar masses. We also found that thick disks do not flare significantly within the observed range in galactocentric radii and that the ratio of thick to thin disk scaleheights is higher for galaxies of earlier types.

Our results tend to favor an {\it in situ} origin for most of the stars in the thick disk. In addition the thick disk may contain a significant amount of stars coming from satellites accreted after the initial build-up of the galaxy and an extra fraction of stars coming from the secular heating of the thin disk by its own overdensities.

Assigning thick disk light to the thin disk component may lead to an underestimate of the overall stellar mass in galaxies, because of different mass to light ratios in the two disk components. On the basis of our new results, we estimate that disk stellar masses are between 10\% and 50\% higher than previously thought and we suggest that thick disks are a reservoir of ``local missing baryons''.

\end{abstract}




\section{Introduction}

Thick disks in lenticular and spiral galaxies are defined as disk-like components with a scaleheight larger than that of the thin or `canonical' disk. They are typically detected as exponential excesses of light a few thin disk scaleheights above the galaxy mid-plane in edge-on galaxies. They have been first detected by Tsikoudi (1979) and recognized as a distinct galaxy structural component by Burstein (1979). A thick disk component was soon found in the Milky Way (Gilmore \& Reid 1983) and it was afterwards proved to be made of old and metal-poor stars when compared to the thin disk stellar population (Reid \& Majewski 1993; Chiba \& Beers 2000). Later studies found thick disks to be nearly ubiquitous (Yoachim \& Dalcanton 2006; Comer\'on et al.~2011a) and to be systematically older than their thin counterparts (Yoachim \& Dalcanton 2008a).

The origin of thick disks is still a matter of debate, with four formation mechanisms being discussed (see the introduction of Yoachim \& Dalcanton 2008a and references therein). The first two possibilities consider two different mechanisms to dynamically heat an originally thin disk, meaning that its vertical stellar velocity dispersion is increased. In the first possibility this dynamical heating could be a consequence of internal evolution due to gravitational interaction with thin disk overdensities like giant molecular clouds or spiral arms (Villumsen 1985; H\"anninen \& Flynn 2002; Sch\"onrich \& Binney 2009; Bournaud et al.~2009). In the second the heating is considered to be due to galaxy-galaxy or dark matter subhalo-galaxy interactions and mergers (Quinn et al.~1993;  Robin et al.~1996; Walker et al.~1996; Vel\'azquez \& White 1999; Chen et al.~2001; Benson et al.~2004; Hayashi \& Chiba 2006; Kazantzidis et al.~2008; Read et al.~2008; Villalobos \& Helmi 2008; Qu et al.~2011). The third possibility is that the thick disk is a consequence of {\it in situ} star formation (Brook et al.~2004; Robertson et al.~2006; Elmegreen \& Elmegreen 2006; Brook et al.~2007; Richard et al.~2010) or star formation with a high initial velocity dispersion in very massive star clusters (Kroupa 2002). Possibilities one and three are related if thick disk star formation occurs in gas that has a high turbulent velocity dispersion from its own clump stirring. The fourth possibility is that the thick disk is formed by accretion of stars from disrupted small satellite galaxies after the initial build-up of the galaxy (Statler 1988; Gilmore et al.~2002; Abadi et al.~2003; Navarro et al.~2004; Martin et al.~2004; Read et al.~2008).

Another topic of debate is that of ``missing'' or ``lost''  baryons (Persic \& Salucci 1992). The fraction of baryons detected at high redshift (see, e.g., WMAP predictions by Spergel et al.~2003), does apparently not match the present-day fraction of baryons. Part of the missing baryons are expected to be found in  hot gas in the warm-hot intergalactic medium (WHIM; see, e.g., Fukugita et al.~1998, simulations from Cen \& Ostriker and WHIM emission detection/constraining by Hickox \& Markevitch 2007 and Zappacosta et al.~2010). Additionally, dark matter haloes in which galaxies lie are also found to be very baryon deficient compared to the cosmic baryon fraction (McGaugh 2008). This apparent lack of baryons inside the galaxy potential well is what has been called the ``local missing baryon'' problem. Part of these baryons have probably escaped due to winds caused by supernova feedback, but the local baryon budget is far from being accounted for (McGaugh 2008 and references therein).

\begin{figure}[t]
\begin{center}
\begin{tabular}{c}
\includegraphics[width=0.45\textwidth]{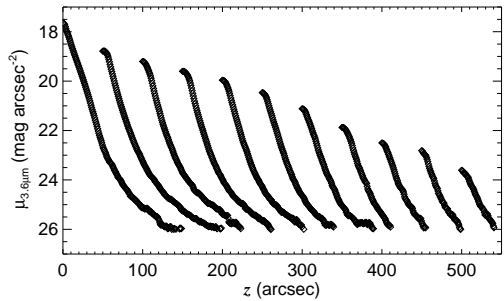}
\end{tabular}
\caption{\label{simpleNGC4565} Reproduction of Figure 12 in van der Kruit \& Searle (1981) using S$^4$G data. The plot displays luminosity profiles of NGC~4565, going from a galactocentric distance $R=0$ outward using a radial bin width of 46\arcsec. The data of the four quadrants have been averaged in order to produce the plot.}
\end{center}
\end{figure}

In this Paper we make use of {\it Spitzer} Survey of Stellar Structure in Galaxies (S$^{4}$G; Sheth et al.~2010) mid-IR imaging of edge-on disk galaxies to assess their stellar vertical mass distribution. The main observational goal of S$^{4}$G is obtaining IRAC (Fazio et al.~2004) $3.6\mu {\rm m}$ and $4.5\mu {\rm m}$-band images for 2331 galaxies with a radial velocity $V_{\rm radio}<3000\,{\rm km\,s^{-1}}$. These passbands are not strongly affected by dust (Draine \& Lee 1984; Sheth et al.~2010) and are not as sensitive to star-forming regions as optical wavelengths, making them especially well suited for tracing the underlying stellar population of galaxies. In addition, having a typical surface brightness sensitivity of $\mu_{\rm AB}\sim27\,{\rm mag\,arcsec}^{-2}$ in the $3.6\mu{\rm m}$ band the S$^{4}$G images have an unprecedented depth for such a large number of galaxies. This high sensitivity allows us to reproduce luminosity profiles down to $\mu_{\rm AB}\sim26\,{\rm mag\,arcsec}^{-2}$ at $3.6\mu{\rm m}$ (Figure~\ref{simpleNGC4565}) with a much better spatial resolution than presented in previous papers (e.g., Burstein 1979; van der Kruit \& Searle 1981). In order to study these edge-on galaxies, we made fits of vertical luminosity profiles in the $3.6\mu {\rm m}$ band. The fitted function results from coupling two luminous (stellar) isothermal disks in equilibrium. We also considered a third component made of cold gas in some cases. We tested the effect that the inclusion of a dark matter halo would have in our fits and found it to cause biases smaller than the error introduced by choosing the mass-to-light ratios of the thin and thick disk. In addition, we tested the effect of fitting the luminosity profiles of non perfectly edge-on galaxies and found it to introduce a small bias which goes opposite to that introduced by a dark matter halo.

This Paper is structured as follows. We present the selected sample in Section~2. In Section~3 we describe the luminosity profile fitting we adopted and describe its potential caveats. The results of our fits are presented in Section~4 and are discussed in Section~5. The conclusions and a brief summary of the Paper are presented in Section~6.

\section{Sample selection}

The sample studied in this Paper was selected from the 817-galaxy S$^{4}$G subsample for which reduced data were available at the time (December 2010; Salo et al.~2011). Ellipticity profiles were constructed during the S$^4$G data processing from the $3.6\mu{\rm m}$ images by running {\sc iraf}'s {\sc ellipse} with the galaxy center position fixed. We selected all those disk galaxies with types $-3 \leq T < 8$ whose maximum disk ellipticity is $\epsilon>0.8$. Galaxies of morphological type $8\leq T<9$ (the so-called Magellanic galaxies) were rejected because their structure is generally ill-defined. The resulting 61 selected galaxies were visually inspected in order to detect any sign of spiral structure or (pseudo)rings, which would indicate a disk which is not edge-on. The images used for this rejection process were those from S$^{4}$G as well as some from the Sloan Digital Sky Survey Data Release 7 (Abazajian et al.~2009) and the {\it Hubble Space Telescope} Legacy Archive (hosted at http://hla.stsci.edu). As a result, fourteen galaxies were removed from the sample. We removed a final galaxy from the sample because of the presence of a very bright star close to the galactic disk affecting it at most galactocentric radii.

The final sample contains 46 galaxies with types $3.0 \leq T \leq 7.5$ at a median distance of $\widetilde{D}=24.7\,{\rm Mpc}$ (using the NASA/IPAC Extragalactic Database mean value of redshift-independent distances except for IC~1553, NGC~5470 and PGC~012349 for which we estimated the distance using HyperLEDA's heliocentric radial velocity from radio measurement and using a Hubble constant $H_{0}=70\,{\rm km\,s^{-1}}$). The sample lacks early-type galaxies for two reasons. First, a large fraction of the S$^4$G galaxies are late-types (see Figure 1 of Sheth et al.~2010). Second, the ellipticity criterion used for selecting the sample biases against galaxies with large bulges or stellar haloes.

As a comparison, one other recent study on thick disks in late-type galaxies with a statistically significant sample is that from Yoachim \& Dalcanton (2006), who study 34 edge-on galaxies of types $5 \leq T \leq 9$ with a median distance of $\widetilde{D}=74.3\,{\rm Mpc}$.

\begin{figure}[t]
\begin{center}
\begin{tabular}{c}
\includegraphics[width=0.45\textwidth]{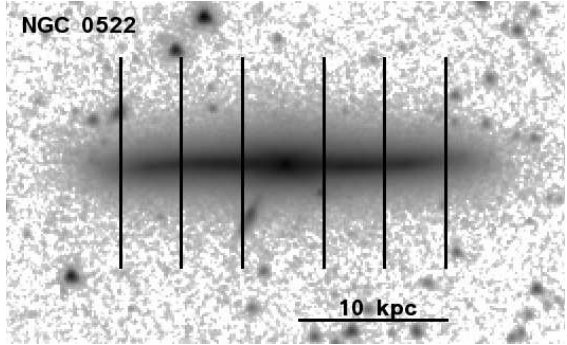}
\end{tabular}
\caption{\label{ngc0522} $3.6\mu{\rm m}$-band S$^4$G image of NGC~0522, one of the galaxies in our sample. NGC~0522 is similar to the Milky Way both in morphological type and in size. The vertical lines indicate the limits of the bins for which luminosity profiles have been produced, from left to right, at galactocentric distances of $-0.8\,r_{25}<R<-0.5\,r_{25}$, $-0.5\,r_{25}<R<-0.2\,r_{25}$, $-0.2\,r_{25}<R<0.2\,r_{25}$, $0.2\,r_{25}<R<0.5\,r_{25}$ and $0.5\,r_{25}<R<0.8\,r_{25}$. In order to avoid the influence of the bulge we have ignored the central ($-0.2\,r_{25}<R<0.2\,r_{25}$) bin throughout the Paper, in all galaxies.}
\end{center}
\end{figure}

An S$^4$G $3.6\mu{\rm m}$-band image of one of the selected galaxies, NGC~0522, is shown in Figure~\ref{ngc0522}.

\section{Fitting procedure}

\subsection{Choice of the fitting function}

Traditionally, single-component edge-on disks have been fitted by functions proportional to ${\rm sech}^2(z/z_0)$, where $z_0$ is a scaleheight. This function was derived on theoretical grounds by Spitzer (1942) and van der Kruit \& Searle (1981) by solving the equations of equilibrium of a single isothermal sheet. In this context isothermal means that the stars in the disk are assumed to have a constant vertical dispersion velocity, $\langle(v_{\rm z})^{2}\rangle^{\frac{1}{2}}$, at all $z$. This function can be approximated as an exponential with scaleheight $z_{\rm exp}=z_0/2$ at large vertical distance, which justifies the exponential fits used by several authors (Morrison et al.~1997; Comer\'on et al.~2011a). Later on, due to discrepancies between the real luminosity profiles and the ${\rm sech}^2(z/z_0)$ function at low $z$, van der Kruit (1988) proposed the rather {\it ad hoc} function ${\rm sech}^{2/n}(nz/2z_0)$, as a generalization of his previously proposed fit. Recently Banerjee \& Jog (2007) have suggested that these discrepancies are due to the gravitational interaction of the thin disk with the gas disk.

The functions described previously are reasonable approximations for the behavior of a single disk, and superpositions of them have been used to describe edge-on galaxies with both a thin and a thick disk component (Shaw \& Gilmore 1989; Shaw \& Gilmore 1990; de Grijs \& van der Kruit 1996;  Morrison et al.~1997; N\"aslund \& J\"ors\"ater 1997; Wu et al.~2002; Pohlen et al.~2004a; Yoachim \& Dalcanton 2006; Comer\'on et al.~2011a). However, Morrison et al.~(1997) make a warning about the limitations of one-dimensional fits arising from degeneracies. While these superpositions are handy because they are analytic, they are not as physically justified as in the case of a single-component disk, as they ignore the gravitational interaction between the different disk components. This is why, for this Paper, we choose to integrate the equations of equilibrium for a set of gravitationally coupled isothermal stellar and gas disks.

\subsection{Creation of synthetic luminosity profiles}

We assumed that disks are relaxed structures whose stars or gas clumps behave like particles of an isothermal fluid in equilibrium. The assumption of equilibrium was recently proven to be reasonable for the Milky Way (S\'anchez-Salcedo et al.~2011). We assume each disk (stellar thin disk, stellar thick disk and gas disk) has its own internal velocity dispersion. Under these conditions the equations which the disks follow are:

\begin{equation}
\frac{{\rm d}^{2}\rho_{\rm i}}{{\rm d}z^{2}}=\frac{\rho_{\rm i}}{\langle(v_{\rm z})^{2}_{\rm i}\rangle}\left(-4\pi G(\rho_{\rm t}+\rho_{\rm T}+\rho_{\rm g})+\frac{{\rm d}K_{\rm DM}}{{\rm d}z}\right)+\frac{1}{\rho_{\rm i}}\left(\frac{{\rm d}\rho_{\rm i}}{{\rm d}z}\right)^{2},
\end{equation}
\noindent where $t$ refers to the properties of the thin disk, $T$ to those of the thick disk, $g$ to those of the gas disk, the subindex $i$ can be either $t$, $T$ or $g$, $\rho$ stands for the mass density, and $\langle(v_{\rm z})^{2}\rangle^{\frac{1}{2}}$ for the vertical velocity dispersion of the particles -- stars or gas clumps -- as described by Narayan \& Jog (2002). $\frac{{\rm d}K_{\rm DM}}{{\rm d}z}$ is a term which describes dark matter effects. From now on we will define the vertical velocity dispersion as $\sigma_{\rm i}\equiv\langle(v_{\rm z})^{2}_{\rm i}\rangle^{\frac{1}{2}}$. To solve this set of three second-order differential equations we used the Newmark-$\beta$ method with $\beta=0.25$ and $\gamma=0.5$, where $\beta$ and $\gamma$ are internal parameters of the algorithm which have been set to make it unconditionally stable (Newmark 1959).

We set the mid-plane to be at $z=0$ and then we integrated the equations from $z=0$ to a large $z=z_{\rm f}$, where $z_{\rm f}$ causes that $\rho(z=0)\gg\rho(z=z_{\rm f})$, $\rho$ being the sum of the densities of all the disks. For practical purposes $z_{\rm f}$ can be considered as infinity. In order to solve these three coupled second order equations six boundary conditions are required. The selected ones were $\rho_{\rm i}(z=0)$ ($\rho_{\rm i0}$ thereafter) and $\frac{{\rm d}\rho_{\rm i}}{{\rm d}z}|_{z=0}$. The first set of derivative boundary conditions is naturally $\frac{{\rm d}\rho_{\rm i}}{{\rm d}z}|_{z=0}=0$ as the mid-plane is by definition the location of the maximum density in the disk.

We solved our differential equations for a grid of models with different central density ratios (ratios of initial conditions; $\rho_{\rm T0}/\rho_{\rm t0}$) and different vertical velocity dispersion ratios, $\sigma_{\rm T}/\sigma_{\rm t}$. We made integrations for 150 values of $\rho_{\rm T0}/\rho_{\rm t0}$, equally spaced from 0.015 to 2.25, and for 150 values of $\sigma_T/\sigma_t$, equally spaced from 1.1 to 16.0. The limiting values of the grids were selected {\it a posteriori}, after testing grids of different sizes and verifying that the selected one was covering the whole parameter space of the studied sample. 

Two cases were studied, namely one with $\int_{z=0}^{z=\infty}\rho_{\rm g}\,{\rm d}z=0$ and one other with $\int_{z=0}^{z=\infty}\rho_{\rm g}\,{\rm d}z=0.2\int_{z=0}^{z=\infty}\rho_{\rm t}\,{\rm d}z$. The first case represents a disk with no gas and the second one represents a disk with an average abundance of gas (the gas mass fraction at Solar radius is 0.17 according to Banerjee \& Jog 2007). We refer to these cases as `without-gas' and `with-gas', respectively. The gas velocity dispersion was fixed to be $\sigma_{\rm g} = (1/3) \sigma_{\rm t}$, in rough agreement with the values found for the disks of the Milky Way, of $\sigma_{\rm g} = 8\,{\rm km\,s}^{-1}$ for the local \hi\ gas (Spitzer 1978), $\sigma_{\rm g} = 5\,{\rm km\,s}^{-1}$ for the local \hii\ gas (Stark 1984; Clemens 1985), and a thin disk stellar velocity dispersion $\sigma_{\rm t} = 18\,{\rm km\,s}^{-1}$ (Lewis \& Freeman 1989; Narayan \& Jog 2002; Banerjee \& Jog 2007).

We set the dark matter halo term in Equation~1 to be $\frac{{\rm d}K_{\rm DM}}{{\rm d}z}=0$. This will be further justified in Section~3.6.2.

\begin{figure}[t]
\begin{center}
\begin{tabular}{c}
\includegraphics[width=0.45\textwidth]{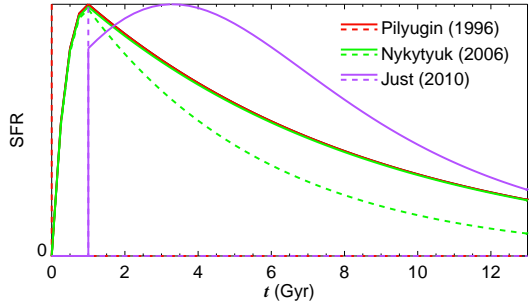}
\end{tabular}
\caption{\label{sfr}  SFHs inferred for the Solar neighborhood, from Pilyugin \& Edmunds (1996; red), Nykytyuk \& Mishenina (2006; green) and Just \& Jahrei\ss\ (2010; purple). The horizontal axis denotes time with the origin 13\,Gyr ago. The vertical axis shows the star formation rate with an arbitrary scaling. Solid curves denote the thin disk SFH, and dashed curves denote the thick disk one.}
\end{center}
\end{figure}

We transformed the resulting synthetic stellar mass density profiles into synthetic luminosity profiles by assigning a different mass-to-light ratio ($\Upsilon$) for the thin and the thick disk stellar components. In order to calculate $\Upsilon$ we used the Bruzual \& Charlot (2003) spectral synthesis models with the set of stellar evolutionary tracks from the `Padova 1994' library (Alongi et al.~1993; Bressan et al.~1993; Fagotto et al.~1994a; Fagotto et al.~1994b) with a Salpeter Initial Mass Function (IMF; Salpeter 1955). Once selected a given $\Upsilon_{\rm T}/\Upsilon_{\rm t}$, we made the synthetic luminosity profiles and scaled them to have a central luminosity equal to one and to $\frac{\rho(z=200)}{\rho(z=0)}=0.1$, where the units of $z$ are arbitrary. We considered three different star formation histories (SFHs) calculated for the solar neighborhood. These three SFHs are shown in Figure~\ref{sfr} and are described in the next subsection.

In summary, the equations need six boundary conditions and a scaling factor for the absolute value of the intensity. The intensity scaling was done by setting the mid-plane intensity equal to 1 in both the models and the observations. Then the six boundary conditions consisted of three differentials at $z=0$, i.e., $\frac{{\rm d}\rho_{\rm i}}{{\rm d}z}|_{z=0}=0$ for each component, two normalizations, which are the thick-to-thin disk density ratio and velocity dispersion ratio, and one other normalization for the vertical pixel scale, which was taken to be $\frac{\rho(z=200)}{\rho(z=0)}=0.1$.

\subsection{Adopted star formation histories}

\subsubsection{Pilyugin \& Edmunds star formation history}
Pilyugin \& Edmunds (1996) used the following function to fit the stellar population of the Solar neighborhood, which we assumed to be representative of a typical thin disk population:

\begin{equation}
\psi(t)\propto\left\{
\begin{array}{l l}
\alpha\,(t/\tau_{\rm top})\,{\rm exp}(-t/\tau_{\rm top}) & {\rm for}\,\,t\leq \tau_{\rm top}\\
{\rm exp}(-t/\tau_{\rm SFR}) & {\rm for}\,\,t> \tau_{\rm top}\\
\end{array}
\right.
\end{equation}
\noindent where $\alpha={\rm exp}(1-\tau_{\rm top}/\tau_{\rm SFR})$ is a normalization constant and where $\tau_{\rm top}$ and $\tau_{\rm SFR}$ are free parameters which define the timescale of a first burst of star formation and the timescale of the subsequent star formation decay. We used the parameters obtained from their best fit, namely $\tau_{\rm top}=1\,{\rm Gyr}$ and $\tau_{\rm SFR}=8\,{\rm Gyr}$ for a 13\,Gyr old thin disk. Approximating the results from their chemical evolution model (Figure~2 of Pilyugin \& Edmunds 1996) we set thin disk stars formed at time $0\,{\rm Gyr}<t<2.75\,{\rm Gyr}$ to have a metallicity of $Z=0.004$, those formed at $2.75\,{\rm Gyr}<t<7\,{\rm Gyr}$ to have $Z=0.008$ and those formed at $7\,{\rm Gyr}<t<13\,{\rm Gyr}$ to have $Z=0.02$. As the authors of this study make no statement about the thick disk population, we assumed it to be formed quickly at the beginning of the galaxy history -- in practice as a result of a single burst of star formation 13\,Gyr ago ($t=0$) and with a low metallicity ($Z=0.0004$). As a result we found that $\Upsilon_{\rm T}/\Upsilon_{\rm t}=2.4$. The calculated $\Upsilon$ accounts only for the stellar mass (stars and stellar remnants) and ignores the gas expelled from stars in the form of stellar winds, supernovae and planetary nebulae. This is not a problem since part of the gas was assumed to be expelled from the galaxy by the supernovae winds and the remaining gas was taken into account in the modeled gas component.  We used the same IMF as Pilyugin \& Edmunds (1996), namely that described by Salpeter (1955).

As the star formation history is probably different for every galaxy, we have calculated how $\Upsilon_{\rm T}/\Upsilon_{\rm t}$ would vary with $\tau_{\rm SFR}$. We have found that $\Upsilon_{\rm T}/\Upsilon_{\rm t}=2.0$ for $\tau_{\rm SFR}=5\,{\rm Gyr}$ and  $\Upsilon_{\rm T}/\Upsilon_{\rm t}=2.5$ for $\tau_{\rm SFR}=10\,{\rm Gyr}$.

\subsubsection{Nykytyuk \& Mishenina star formation history}

Nykytyuk \& Mishenina (2006) used the same SFH -- $\psi(t)$ --, IMF, and Galaxy age as Pilyugin \& Edmunds (1996) for the thin disk. For the thick disk they used as similar function with $\tau_{\rm SFR}=5\,{\rm Gyr}$. Approximating the results from their chemical evolution model (Figures~5 and 8 of Nykytyuk \& Mishenina 2006) we set thin disk stars formed at time $0\,{\rm Gyr}<t<2.75\,{\rm Gyr}$ to have a metallicity of $Z=0.004$, those formed at $2.75\,{\rm Gyr}<t<7\,{\rm Gyr}$ to have $Z=0.008$ and those formed at $7\,{\rm Gyr}<t<13\,{\rm Gyr}$ to have $Z=0.02$. For the thick disk we set the stars formed at time $0\,{\rm Gyr}<t<0.75\,{\rm Gyr}$ to have a metallicity of $Z=0.0004$, those formed at $0.75\,{\rm Gyr}<t<1.25\,{\rm Gyr}$ to have $Z=0.004$ and those formed at $1.25\,{\rm Gyr}<t<13\,{\rm Gyr}$ to have $Z=0.008$. For this SFH we calculated $\Upsilon_{\rm T}/\Upsilon_{\rm t}=1.2$.

\subsubsection {Just \& Jahrei\ss\ star formation history}

In the model of Just \& Jahrei\ss\ (2010), the Milky Way is assumed to have an age of 12\,Gyr and a SFH as follows:

\begin{equation}
\psi(t)\propto\frac{(t+t_0)t^3_{\rm n}}{(t^2+t^2_1)^2}
\end{equation}

where $t_0=5.6\,{\rm Gyr}$, $t_1=8.2\,{\rm Gyr}$ and $t_n=9.9\,{\rm Gyr}$. Approximating the results from their chemical evolution model (Figure~16 of Just \& Jahrei\ss\ 2010) we set thin disk stars formed at time $0\,{\rm Gyr}<t<0.75\,{\rm Gyr}$ to have a metallicity of $Z=0.004$, those formed at $0.75\,{\rm Gyr}<t<7.5\,{\rm Gyr}$ to have $Z=0.008$ and those formed at $7.5\,{\rm Gyr}<t<12\,{\rm Gyr}$ to have $Z=0.02$. Just \& Jahrei\ss\ (2010) adopted the thick disk to have an age of 12\,Gyr and here we have set it to have a metallicity $Z=0.0004$. For this SFH we calculated $\Upsilon_{\rm T}/\Upsilon_{\rm t}=2.4$. This value is only an approximation as Just \& Jahrei\ss\ used a Scalo (1986) IMF when fitting the stellar population of the solar neighborhood and using a Salpeter (1955) IMF as we have done introduces some internal incoherence in the calculations.

\subsubsection{Which SFH should we use?}

We can summarize the three previous subsections by stating that there are basically two kinds of SFH (see Figure~2). The first class of models implies that the stars in the thick disk were created nearly instantaneously early in the galaxy history like in our implementation of a thick disk in the Pilyugin \& Edmunds (1996) model and in Just \& Jahrei\ss\ (2010). The two models yield a very similar result, namely that $\Upsilon_{\rm T}/\Upsilon_{\rm t}=2.4$. The second class of models implies that the thick disk is made of stars created over a longer time, like what is proposed by Nykytyuk \& Mishenina (2006) yielding $\Upsilon_{\rm T}/\Upsilon_{\rm t}=1.2$. Choosing one of these two kinds of SFHs would imply some assumptions on the formation mechanism of the thick disk. Moreover, different galaxies are likely to have different star formation histories which will yield a different $\Upsilon_{\rm T}/\Upsilon_{\rm t}$.

A rough estimate of $\Upsilon_{\rm T}/\Upsilon_{\rm t}$ can be made from the data collected in Yoachim \& Dalcanton (2006). The authors used their $B-R$ measurements for the thin and the thick disks in order to estimate $\left(\Upsilon_{\rm T}/\Upsilon_{\rm t}\right)_{R}$ using some recipes included in Bell \& de Jong (2001), a paper which also includes recipes to convert $B-R$ to $\left(\Upsilon_{\rm T}/\Upsilon_{\rm t}\right)_{K}$. As the $K$-band is relatively close to the $3.6\mu{\rm m}$-band, $\left(\Upsilon_{\rm T}/\Upsilon_{\rm t}\right)_{K}$ can be used as an approximation for the $\left(\Upsilon_{\rm T}/\Upsilon_{\rm t}\right)$ in this paper. For the galaxies in Yoachim \& Dalcanton (2006), $\overline{\left(\Upsilon_{\rm T}/\Upsilon_{\rm t}\right)_{K}}=1.6$ with a dispersion of 0.3. Using their data we found that there is a certain trend of $\left(\Upsilon_{\rm T}/\Upsilon_{\rm t}\right)_{K}$ with the maximum circular velocity of the galaxy, $v_{\rm c}$, with $\left(\Upsilon_{\rm T}/\Upsilon_{\rm t}\right)_{K}=2.1\pm0.2-(0.005\pm0.002)v_{\rm c}$. The linear correlation coefficient is $r=0.5$. This relationship means that more massive galaxies tend to have disks with a lower $\left(\Upsilon_{\rm T}/\Upsilon_{\rm t}\right)_{K}$.

In order to avoid a $\Upsilon_{\rm T}/\Upsilon_{\rm t}$ choice, and also due to the large scatter in the $\Upsilon_{\rm T}/\Upsilon_{\rm t}$ derived from Yoachim \& Dalcanton (2006), we decided to study two cases, namely $\Upsilon_{\rm T}/\Upsilon_{\rm t}=1.2$ and $\Upsilon_{\rm T}/\Upsilon_{\rm t}=2.4$, which are the results of different fitted Milky Way SFHs. The $\Upsilon_{\rm T}/\Upsilon_{\rm t}$ of most galaxies is probably found between these values, as discussed in the previous paragraph. As we also studied two gas disk concentrations, we produced four grids of 150 by 150 luminosity profile models. The $\Upsilon_{\rm T}/\Upsilon_{\rm t}$ ratio affects the thick and thin disk stellar mass ratios by nearly a constant factor. In what follows, for many figures, we only show the $\Upsilon_{\rm T}/\Upsilon_{\rm t}=1.2$ results and then mention the appropriate factors for scaling these results to the $\Upsilon_{\rm T}/\Upsilon_{\rm t}=2.4$ fits.

\subsubsection{Limitations and caveats of the selected fitting function}

We assumed the thin and the thick disk to be made of isothermal sheets with two distinct and unique vertical velocity dispersions. Although this is true in first approximation, a small but significant increase of the vertical velocity dispersion has been found for Milky Way thick disk stars at large vertical distance above the mid-plane (Moni Bidin et al.~2010).

The line of sight integration has not been taken into account when producing the synthetic luminosity profiles. This is a valid approximation as long as the radial scalelengths of all the disks are similar and the scale heights have no radial dependence. The first assumption allows the ratios of mid-plane densities of the different disks to be constant. In the sample studied by Yoachim \& Dalcanton (2006) the thick to thin disk scalelength ratio is on average $\left<h_{\rm T}/h_{\rm t}\right>\sim1.2$, which justifies our assumption. The second assumption prevents the integration of different scale heights on a line of sight that covers a range of radii. Our results here are consistent with this assumption, as we see no significant change in the thick disk scaleheight with radius.

The third caveat is that we neglected the effect of the bulge. This can be justified because our sample is made of late-type galaxies and because the fits were made at $R>0.2\,r_{25}$, where the bulge is relatively unimportant.

We also ignored the effects of dark matter and not perfectly edge-on galaxy geometries, which will be justified in Sections~3.6.2 and 3.6.3.

\subsection{Compilation of observed luminosity profiles}

Using S$^{4}$G $3.6\mu{\rm m}$-band images, we produced four luminosity profiles for each galaxy: at each side of the bulge along the long disk axis for projected galactocentric distances $0.2\,r_{25}<|R|<0.5\,r_{25}$ and $0.5\,r_{25}<|R|<0.8\,r_{25}$, as shown in Figure~\ref{ngc0522} ($r_{25}$ from HyperLEDA; Paturel et al.~2003). The profiles were prepared by adding the counts above and below the mid-plane in order to get one unique profile for each bin and averaging for each $z$ over the non-masked pixels. We used manually refined masks coming from the S$^{4}$G Pipeline~2 (Sheth et al.~2010).

\subsection{Comparison between observed and synthetic luminosity profiles}

\begin{figure}[!t]
\begin{center}
\begin{tabular}{c}
\includegraphics[width=0.45\textwidth]{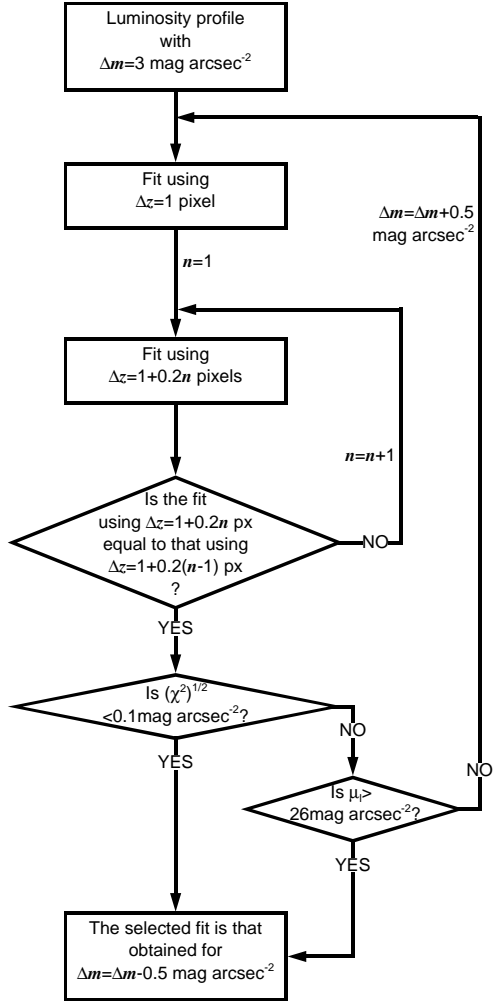}
\end{tabular}
\caption{\label{diagram} Flow diagram showing the procedure used to compare the real and the synthetic luminosity profiles, as described in Section~3.5.}
\end{center}
\end{figure}

The observed luminosity profiles were scaled in luminosity and in height above the mid-plane, $z$, using the same procedure as for the synthetic profiles. We ignored the mid-plane pixel in our fits to avoid the region of highest extinction. Note that at $3.6\mu{\rm m}$, and especially at $R>0.2\,r_{25}$ where we study vertical profiles, there is very little dust extinction. However, to allow for a small amount of extinction, we scaled the luminosity profile outside of the mid-plane by various factors, from 0.80 to 1.00 in steps of 0.01. These 20 different scalings for the luminosity profiles introduce a third dimension to the parameters of the fit. The synthetic profiles were then convolved with a Gaussian kernel with a Full Width at Half Maximum (FWHM) of 2.2\arcsec, which is the FWHM of a synthetic $3.6\mu{\rm m}$-band S$^{4}$G Point Spread Function (PSF), made by stacking several hundreds non-saturated stars (S$^4$G ``super-PSF''; Sheth et al.~2010). Each pixel in S$^{4}$G has a size of 0.75\arcsec.

The code which compared the observed and synthetic luminosity profiles was made of two nested loops which are illustrated in the flow diagram shown in Figure~\ref{diagram}.

The external loop compared the brightest section of the observed luminosity profiles, with different assumed dust extinctions, with a Gaussian convolved synthetic profile. The brightest section was defined to have a given dynamical range, $\Delta m$, with an initial $\Delta m=3\,{\rm mag\,arcsec}^{-2}$. We term the faintest fitted magnitude as $\mu_{\rm l}$. The best fitted profile -- that with the lowest $sqrt{\chi^2}$ in magnitudes -- was considered to be that with the `correct' central dust extinction. If the best fit had $\sqrt{\chi^2}<0.1\,{\rm mag}$ and if $\mu_{\rm l}<26\,{\rm mag\,arcsec}^{-2}$, we made an extra iteration of the loop with $\Delta m=\Delta m+0.5\,{\rm mag\,arcsec}^{-2}$. If one  of these conditions was not fulfilled, we considered as the best fit that obtained for $\Delta m=\Delta m-0.5\,{\rm mag\,arcsec}^{-2}$. The limit $\mu_{\rm l}=26\,{\rm mag\,arcsec}^{-2}$ was selected in order to avoid any noise effect at low surface brightness regions.

The inner loop was in charge of detecting the thickness of an eventual mid-plane dust layer. The synthetic and the real luminosity profiles were compared for $z>1\,{\rm pixel}$ in order to avoid comparing regions which are presumably affected by dust. We selected as a best fit the synthetic profile for which the $\sqrt{\chi^2}$ was minimum. Then we compared again the real profiles with the synthetic grid, but for $z>1.2\,{\rm pixels}$ (we made linear interpolation between pixels in order to use fractional pixels). If the best fitting synthetic profile was the same as when comparing for $z>1\,{\rm pixel}$, we took this fit as a good result; if not, we considered that at $z=1\,{\rm pixel}$ the dust effect is still important and thus repeated the procedure comparing the best fits for $z>1.2\,{\rm pixels}$ and for $z>1.4\,{\rm pixels}$. We continued this procedure with steps of $\Delta z=0.2\,{\rm pixels}$ until the best fit for $z=z_{\rm a}$ was the same as for $z=z_{\rm a}+\Delta z$.

The statistical uncertainties in the fits are negligible compared to those introduced by considering a given $\Upsilon_{\rm T}/\Upsilon_{\rm t}$.

\begin{figure*}[t]
\begin{center}
\begin{tabular}{c c}
\includegraphics[width=0.45\textwidth]{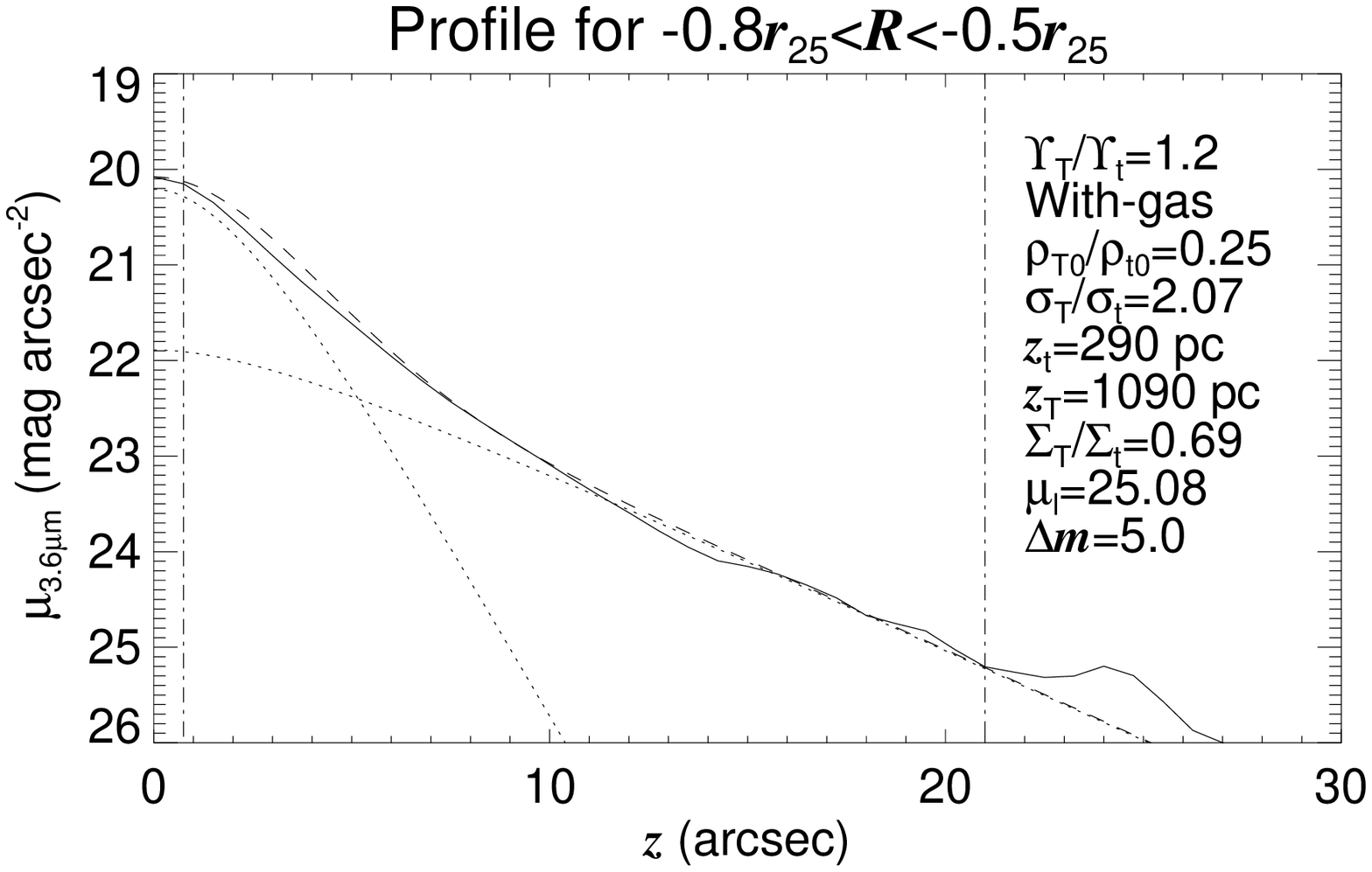}&
\includegraphics[width=0.45\textwidth]{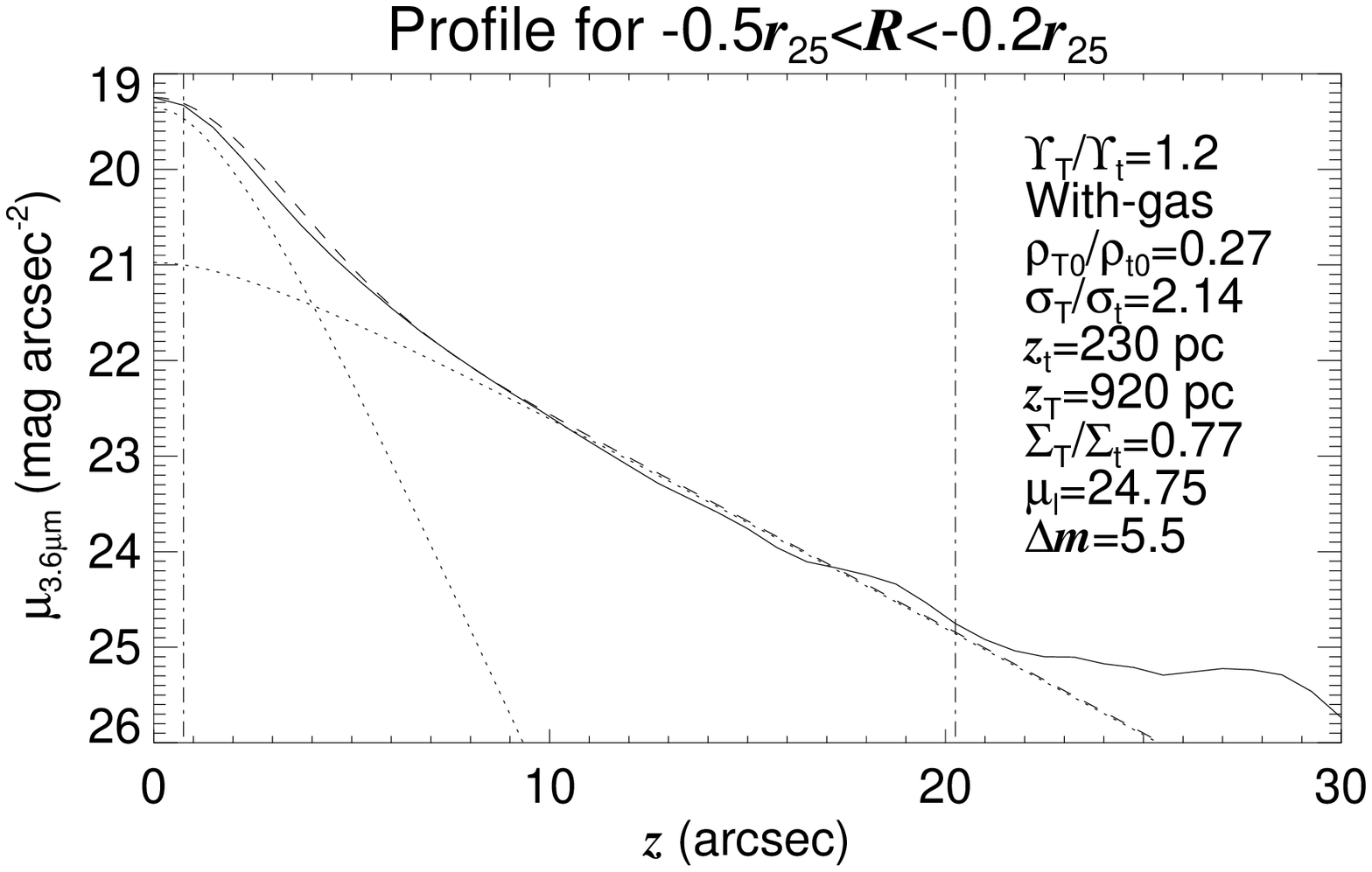}\\
\includegraphics[width=0.45\textwidth]{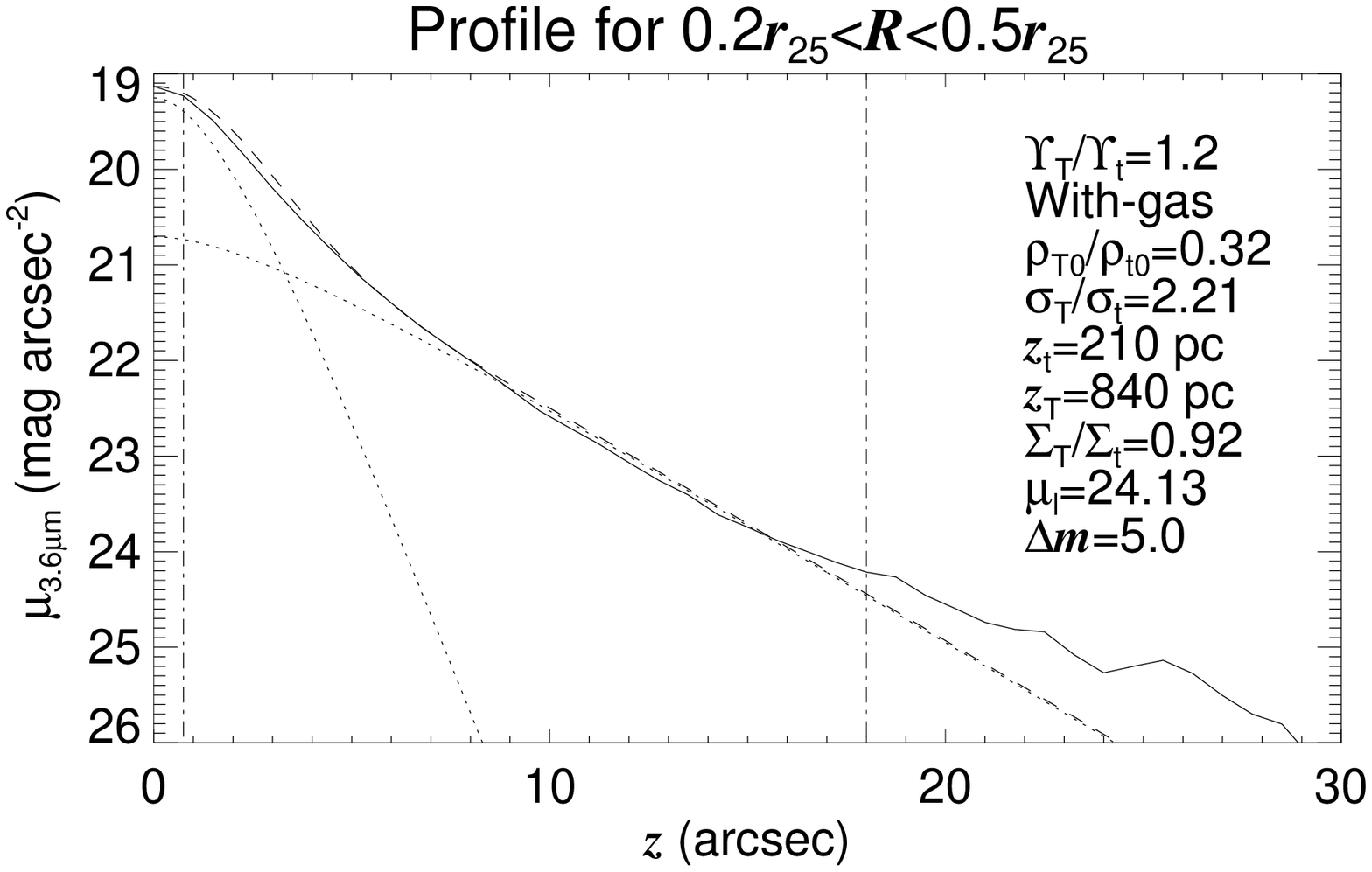}&
\includegraphics[width=0.45\textwidth]{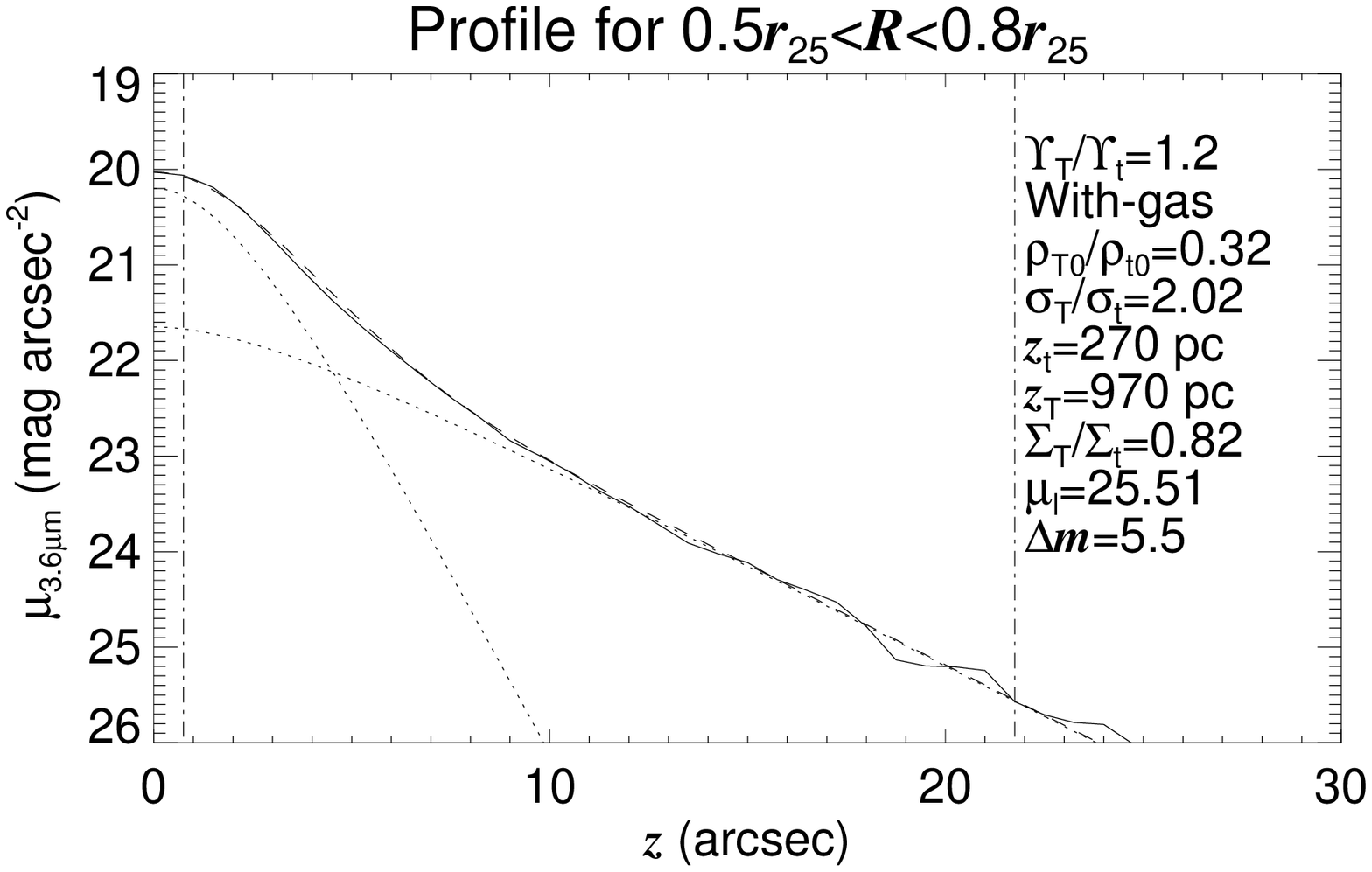}
\end{tabular}
\caption{\label{example} Fits to the surface brightness profiles of NGC~0522 for the four fitted bins for the case $\Upsilon_T/\Upsilon_t=1.2$ and with-gas. The solid curve represents the observed luminosity profile, and the dashed curve the best fit. The dotted curves indicate the contributions of the thin and the thick disk. The dash-dotted vertical lines indicate the limits of the range in vertical distance above the mid-plane used for the fit.}
\end{center}
\end{figure*}

An example of the fitted profiles is presented in Figure~\ref{example}.

\subsection{Reliability of the fits}

\subsubsection{Degeneracies}

\begin{figure*}[t]
\begin{center}
\begin{tabular}{c c}
\includegraphics[width=0.45\textwidth]{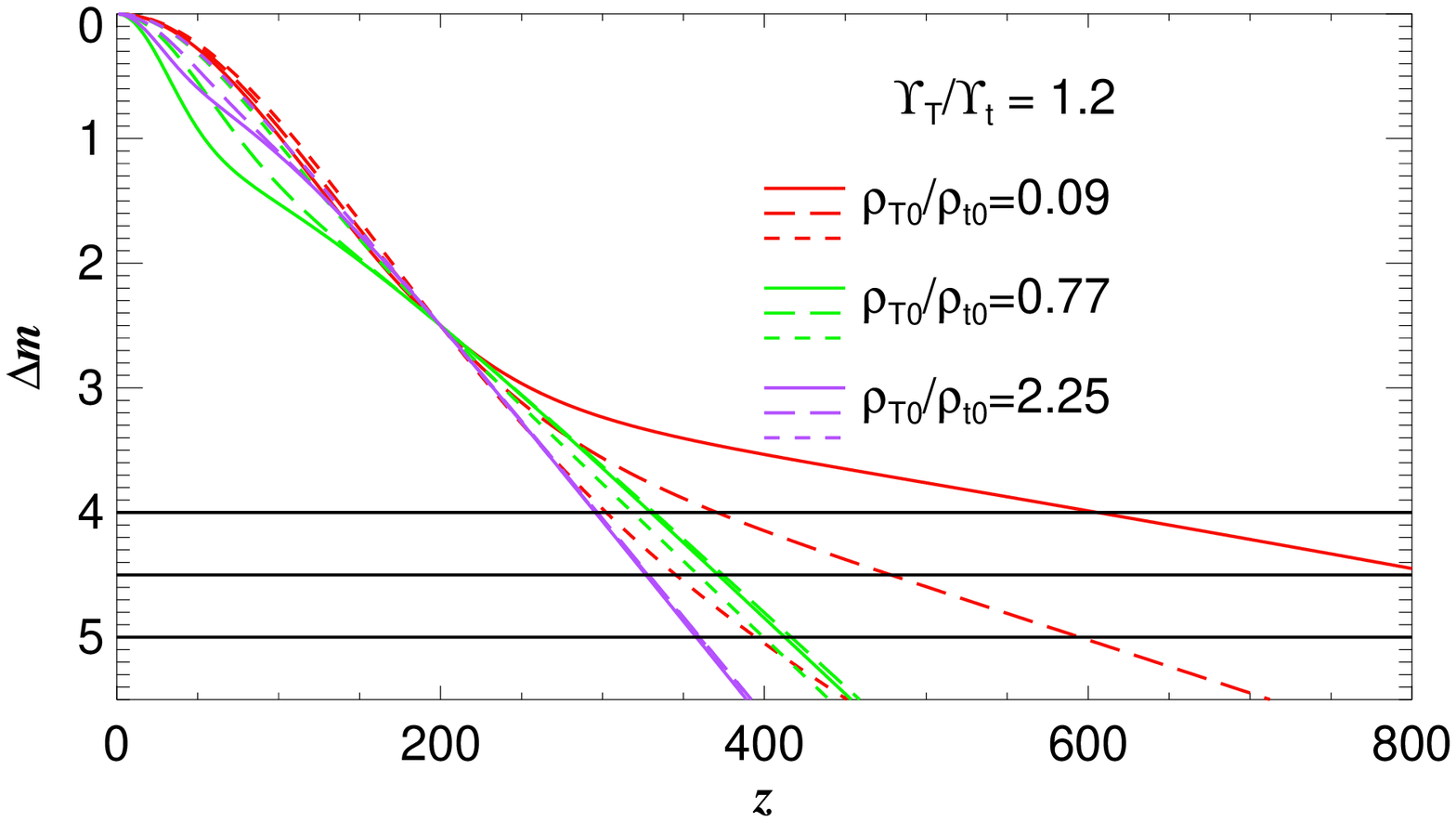}&
\includegraphics[width=0.45\textwidth]{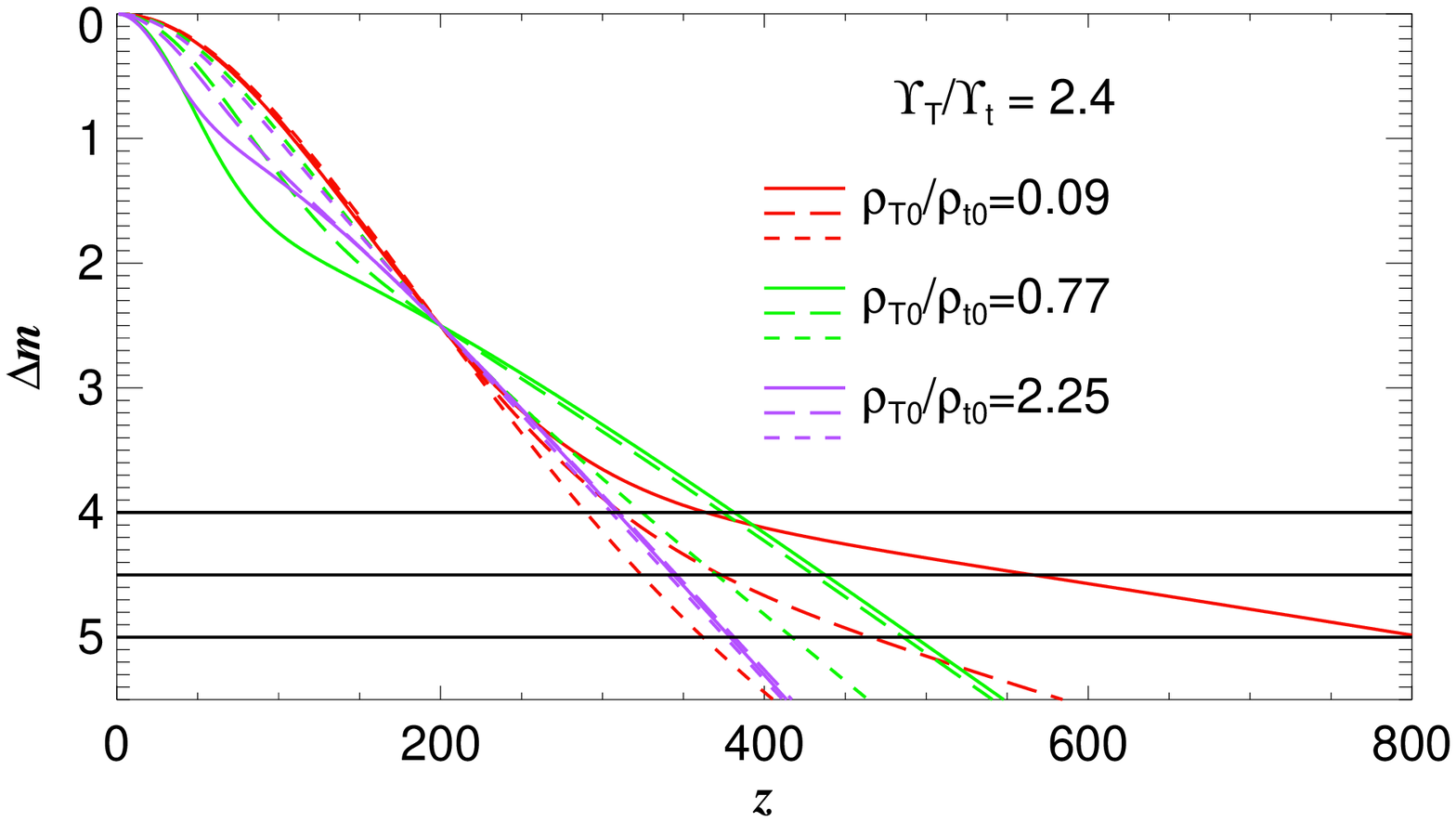}\\
\end{tabular}
\caption{\label{prof} Plot of several of the synthetic surface brightness profiles in the without-gas case. The panel on left shows profiles computed with $\Upsilon_{\rm T}/\Upsilon_{\rm t}=1.2$ and those at the right with $\Upsilon_{\rm T}/\Upsilon_{\rm t}=2.4$. The horizontal axis shows the vertical distance above the mid-plane in the arbitrary units we have used for our scaling. The vertical axis shows the decrement of surface brightness in magnitudes relative to the mid-plane. Red, green and violet curves indicate profiles with $\rho_{\rm T0}/\rho_{\rm t0}$ equal to 0.09, 0.77 and 2.25, respectively. Short-dashed, long-dashed and solid curves denote profiles with $\sigma_{\rm T}/\sigma_{\rm t}$ equal to 1.76, 2.47 and 3.62, respectively. The horizontal black lines indicate, from top to bottom, $\Delta m$ equal to 4.0, 4.5 and 5.0\,magnitudes.}
\end{center}
\end{figure*}

\begin{figure}[!t]
\begin{center}
\begin{tabular}{c c}
\includegraphics[width=0.45\textwidth]{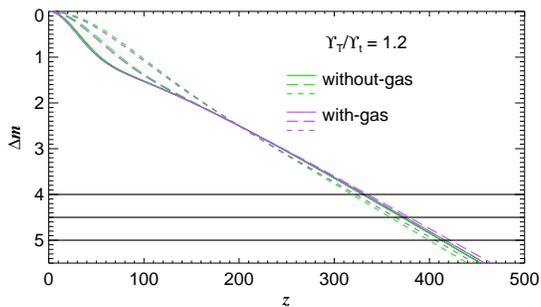}&
\end{tabular}
\caption{\label{profgas} Plot comparing the synthetic surfaced brightness profiles in the without-gas case with those in the with-gas case. The profiles have $\Upsilon_{\rm T}/\Upsilon_{\rm t}=1.2$ and $\rho_{\rm T0}/\rho_{\rm t0}=0.77$, like the green profiles in the left panel in Figure~\ref{prof}. The green curves denote the without-gas case and the violet ones the with-gas case.}
\end{center}
\end{figure}

\begin{figure}[!t]
\begin{center}
\begin{tabular}{c}
\includegraphics[width=0.45\textwidth]{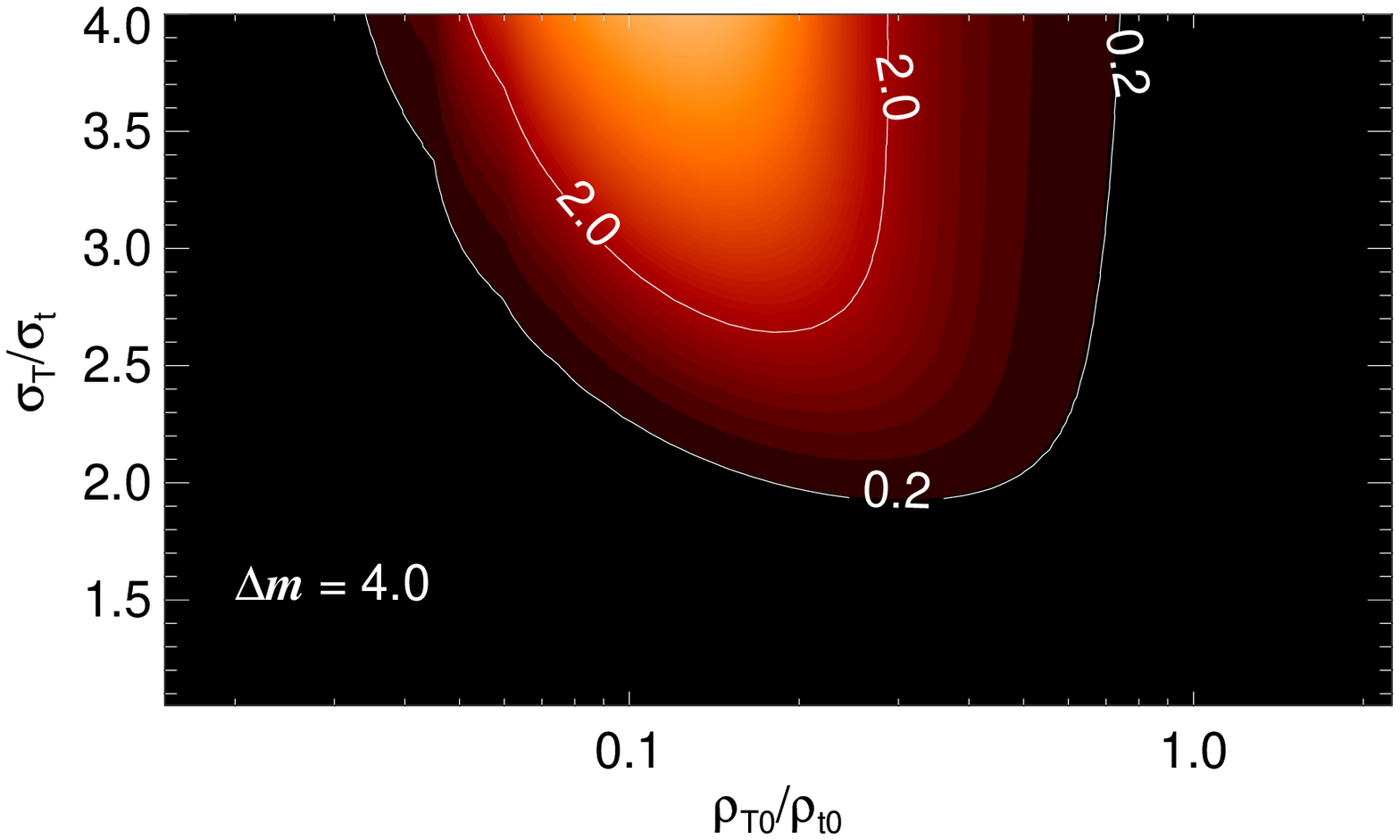}\\
\includegraphics[width=0.45\textwidth]{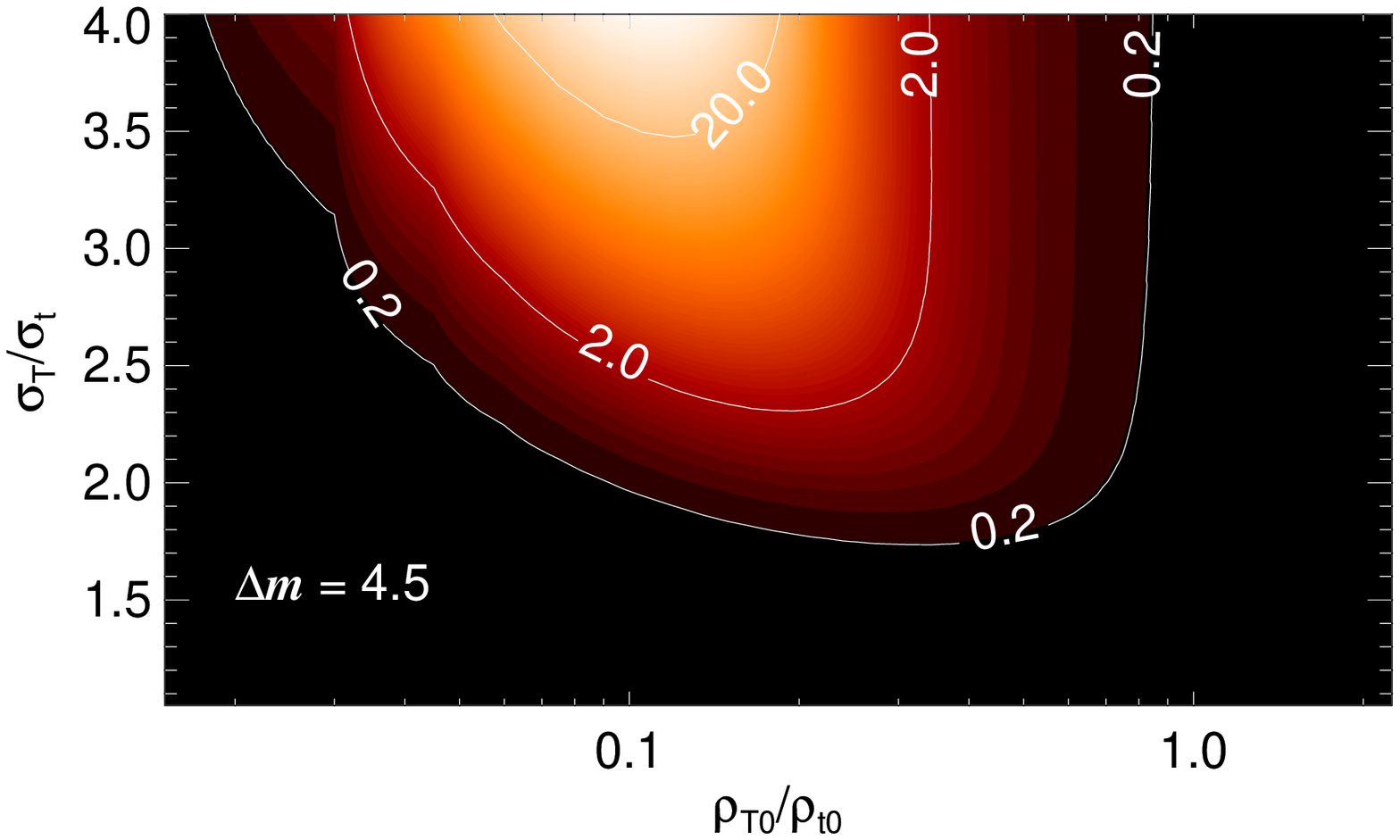}\\
\includegraphics[width=0.45\textwidth]{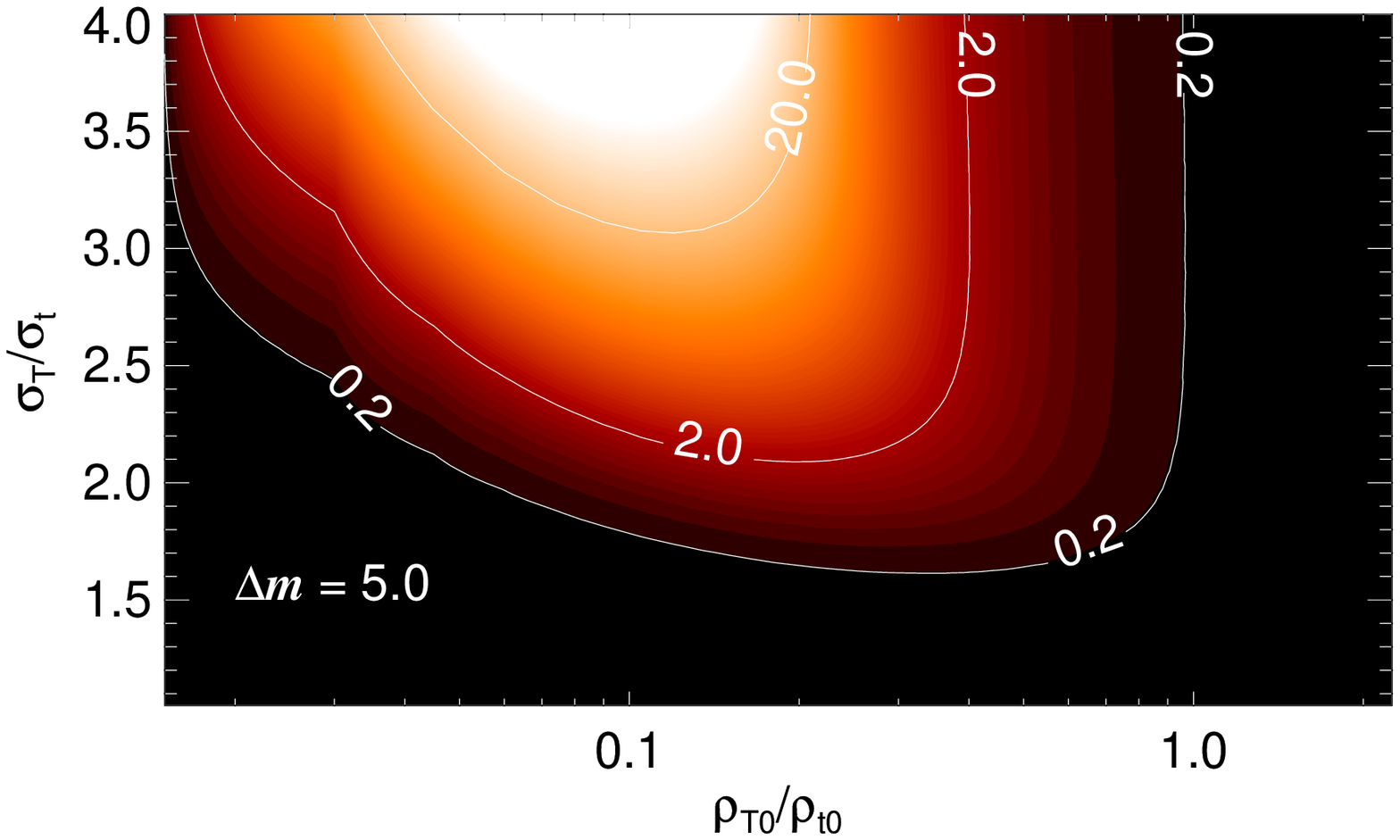}
\end{tabular}
\caption{\label{single_disk} Square root of the mean square difference between a single-disk profile and our grid of synthetic two-disk profiles. The vertical axis displays the ratio of the mean vertical dispersion velocity between the thin and the thick disk, $\sigma_{\rm T}/\sigma_{\rm t}$, and the horizontal axis displays the ratio between the thin and thick disk mid-plane density, $\rho_{\rm T0}/\rho_{\rm t0}$. The selected $\Delta m$ is given in the bottom-left corner of each panel. All plots are for $\Upsilon_{\rm T}/\Upsilon_{\rm t}=1.2$ and with-gas. For $\Upsilon_{\rm T}/\Upsilon_{\rm t}=2.4$, the distribution is very similar, but shifted by a factor two to the right. The with-gas and without-gas distributions are nearly identical.}
\end{center}
\end{figure}

We tested our fits in order to detect possible degeneracies and to test the conditions in which our fitting method can be applied.

Figure~\ref{prof} shows the synthetic profiles obtained for a selection of $\rho_{\rm T0}/\rho_{\rm t0}$ and $\sigma_{\rm T}/\sigma_{\rm t}$. For $\rho_{\rm T0}/\rho_{\rm t0}=0.77$, a central density typical of what can be found in our fits, the profiles with a different ratio of velocity dispersions are easily distinguishable because the inflection point which indicates the transition from the thin to thick disk-dominated area can be detected even when the fit is produced over a low $\Delta m$ (where $\Delta m$ is the range of magnitudes over which the fit is produced). The profiles are much more degenerate for $\rho_{\rm T0}/\rho_{\rm t0}=2.25$, the case with a higher central thick disk density in our grid, as the luminosity is thick disk-dominated at every $z$, hiding the inflection point. At the other side of the studied range of central density ratios, the profiles made for thin disk-dominated galaxies -- $\rho_{\rm T0}/\rho_{\rm t0}=0.09$ -- show a clear inflection point, but only if $\Delta m$ is large enough. In the case of $\Upsilon_{\rm T}/\Upsilon_{\rm t}=2.4$, a galaxy with $\rho_{\rm T0}/\rho_{\rm t0}=0.09$ shows a profile very similar to that of a galaxy with $\rho_{\rm T0}/\rho_{\rm t0}=2.25$ unless $\Delta m$ becomes large enough to unveil the inflection point.

Figure~\ref{profgas} shows how adding a gas disk in the model of a galaxy affects the synthetic luminosity profiles for a model with an intermediate $\rho_{\rm T0}/\rho_{\rm t0}$. The effect is small, and causes the model thin disk scaleheight to become smaller.

To quantify the possibility of a two-disk galaxy to be identified as a single-disk one, we estimated the mean square difference between our sets of synthetic luminosity profiles and an isothermal single-disk profile in equilibrium for different total magnitude ranges in the fit, $\Delta m$ (Figure~\ref{single_disk}). We show only the case $\Upsilon_{\rm T}/\Upsilon_{\rm t}=1.2$ with-gas; for $\Upsilon_{\rm T}/\Upsilon_{\rm t}=2.4$, the distribution is very similar and shifted a factor two to the right. The with-gas and without-gas distributions are nearly the same. We have considered a profile to be compatible with a single-disk luminosity distribution when $\sqrt{\chi^2}<0.2\,{\rm mag\,arcsec}^{-2}$. As indicated by Figure~\ref{prof}, we have found that for high values of $\rho_{\rm T0}/\rho_{\rm t0}$ the profiles are degenerate and a two-disk profile cannot be distinguished from a single-disk profile. When $\Delta m$ is low, this degeneracy also affects profiles with a low $\rho_{\rm T0}/\rho_{\rm t0}$. In order to avoid degeneracies, we have decided to include in our study only galaxies for which the luminosity profiles have been successful over a dynamical range of $\Delta m \ge 4.5\,{\rm mag\,arcsec}^{-2}$. This restriction, combined with that of fitting until the faintest level of $\mu_{\rm l}=26\,{\rm mag\,arcsec}^{-2}$, ensures that only profiles with a central brightness, $\mu(z=0)<21.5\,{\rm mag\,arcsec}^{-2}$, are considered. Two galaxies, even though they have $\mu(z=0)<21.5\,{\rm mag\,arcsec}^{-2}$, are excluded from the study, because the fit could only be done over $\Delta m<4.5\,{\rm mag\,arcsec}^{-2}$ due to their profile not being compatible with that of two coupled isothermal stellar disks in equilibrium.

\subsubsection{Effect of ignoring a dark matter halo}

We ignored the effect that a dark matter halo would have in our fits. Including them would be problematic due to our ignorance of the properties of the dark matter haloes, with possibilities ranging from maximum disks (for which no significant amount of dark matter amount is needed to explain the shape of the rotation curve within the optical radius) to clearly submaximal disks, going through studies which indicate that disks with large circular velocities are maximum and those with lower circular velocities are submaximal (see discussion in Bosma 2004). Examples of studies which have found at least disks in high-luminosity galaxies to be maximum are those by Salucci \& Persic (1999) and Palunas \& Williams (2000). Moni Bidin et al.~(2010) have found no significant dynamical effect of dark matter in our Galaxy at Solar radius, implying that its disk is likely to be maximal. On the other hand, studies over statistically significant samples which find indications of an opposite result include those by Courteau \& Rix (1999) and Pizagno et al.~(2005).

We produced a ``worst-case'' test on how ignoring the presence of a dark matter halo affected our fits. As the main results of this paper relates to the ratio of column mass densities between the thin and the thick disk, $\Sigma_{\rm T}/\Sigma_{\rm t}$, our tests were directed to know how this parameter varies with the inclusion of a dark matter halo. To do so we have made several tests on NGC~5470. We choose this galaxy because it is the faintest one in our sample for which it has been possible to do fits with $\Delta m\ge 4.5$, and it thus is the studied galaxy which is more likely to have a submaximal disk. We have used the Narayan \& Jog (2002) formalism, with a pseudo-isothermal dark matter halo (van Albada et al.~1985) term as follows:
\begin{equation}
\frac{{\rm d}K_{\rm DM}}{{\rm d}z}=-\frac{v^2_{\rm max}}{r^2}\left(1-\frac{2z^2}{r^2}-\frac{z^2R^2_{\rm c}}{r^2\left(R^2_{\rm c}+r^2\right)}+\frac{R_{\rm c}}{r}\left(\frac{3z^2}{r^2}-1\right){\rm arctan}\left(\frac{r}{R_{\rm c}}\right)\right),
\end{equation}
\noindent where $v_{\rm max}$ is the extrapolated rotation speed at infinity, $R_{\rm c}$ is the core radius of the dark matter halo and $r=\sqrt{\rho^2+z^2}$ is the 3D radius or distance to the galaxy center.

\begin{table*}[t]
\begin{center}
\caption{\label{testdm} Column density ratios ($\Sigma_{\rm T}/\Sigma_{\rm t}$) obtained from fitting NGC~5470 luminosity profiles including the effect of a dark matter halo.}
\begin{tabular}{c c c| c c |c c}
\hline
\hline
$R_{\rm c}$&$v_{\rm max}$&$R_{\rm D}$& \multicolumn{2}{c|}{$\Sigma_{\rm T}/\Sigma_{\rm t}$} & \multicolumn{2}{c}{$\Sigma_{\rm T}/\Sigma_{\rm t}$}\\
(pc)&(km\,s$^{-1}$)&(pc)& \multicolumn{2}{c|}{($-0.8\,r_{25}<R<-0.5\,r_{25}$)}&\multicolumn{2}{c}{($0.5\,r_{25}<R<0.8\,r_{25}$)}\\
& & & $\Upsilon_{\rm T}/\Upsilon_{\rm t}=1.2$ & $\Upsilon_{\rm T}/\Upsilon_{\rm t}=2.4$ &$\Upsilon_{\rm T}/\Upsilon_{\rm t}=1.2$ &$\Upsilon_{\rm T}/\Upsilon_{\rm t}=2.4$\\
\hline
\nodata & 0(a) & \nodata & 0.92 & 1.73 & 1.03 & 1.95\\
\nodata & 0(b) & \nodata & 0.89 & 1.77 & 1.01 & 2.01\\
1930 & 120 & 500 & 0.69 & 1.45 & 0.84 & 1.72\\
1930 & 120 & 610 & 0.70 & 1.45 & 0.84 & 1.72\\
1930 & 120 & 720 & 0.73 & 1.46 & 0.84 & 1.72\\
1930 & 120 & 900 & 0.74 & 1.56 & 0.89 & 1.79\\
1930 & 120 & 1080& 0.79 & 1.62 & 0.91 & 1.85\\
3520 & 60  & 500 & 0.85 & 1.67 & 0.97 & 1.94\\
3520 & 60  & 610 & 0.83 & 1.89 & 0.95 & 1.98\\
3520 & 60  & 720 & 0.85 & 1.17 & 0.70 & 2.02\\
3520 & 60  & 900 & 0.92 & 2.03 & 0.77 & 2.09\\
3520 & 60  & 1080& 0.98 & 1.96 & 0.82 & 1.70\\
3520 & 120 & 500 & 0.63 & 1.52 & 0.75 & 1.58\\
3520 & 120 & 610 & 0.65 & 1.30 & 0.76 & 2.08\\
3520 & 120 & 720 & 0.66 & 1.34 & 0.80 & 1.60\\
3520 & 120 & 900 & 0.70 & 1.46 & 0.85 & 2.17\\
3520 & 120 & 1080& 0.76 & 1.52 & 0.87 & 1.78\\
16860& 120 & 720 & 0.83 & 1.72 & 0.99 & 1.99\\
16860& 230 & 720 & 0.86 & 1.22 & 0.98 & 2.00\\
\hline
\end{tabular}
\end{center}
Note.~-- The fit made with $v_{\rm max}=0$(a) corresponds to that described in Section~3.5. The fit made with $v_{\rm max}=0$(b) corresponds to a fit made with no vertical pixel scaling, such as in the case of fits with dark matter, implying an extra fitting parameter ($\sigma_{\rm t}$ and $\sigma_{\rm T}$ instead of $\sigma_{\rm t}/\sigma_{\rm T}$).
\end{table*}

As the inclusion of a dark matter halo implies the need for a line-of-sight integration, we needed to know the scalelength of the disk. NGC~5470 has a disk truncation (downbending) at $R\sim40\arcsec$, which corresponds to $R\sim3$\,kpc. The bins with $0.5\,r_{25}<|R|<0.8\,r_{25}$ -- those which are more likely to be affected by the dark matter halo -- are located in the outer, steeper disk. Using van der Kruit \& Searle (1981) approximations for measuring the scalelength of edge-on disks when $R$ is much larger than the scalelength, we found that the scalelength of the outer disk is $h\sim700$\,pc. We also found the galaxy to have no significant emission in $3.6\mu$m for $R>6.2$\,kpc, a distance we have set as the end of the stellar disks. In addition, we assumed that the mid-plane stellar mass density of NGC~5470 at $\rho=0.65r_{25}$ is $\rho_{\rm t0}+\rho_{\rm T0}=0.1\,M_{\bigodot}\,{\rm pc^{-3}}$, which is a value on the order of the mass density in the Solar neighborhood. In order to reduce the computation time we tested without-gas cases only. The inclusion of a dark matter halo forces us to use the thin and the thick disk vertical velocity dispersions as free parameters ($\sigma_{\rm t}$ and $\sigma_{\rm T}$) instead of their ratio, as we have been doing until now.

As we aimed to use realistic dark matter halo parameters, we searched for galaxies similar to NGC~5470 in de Blok et al.~(2008), where the authors deduce halo properties from \hi\ rotation curves. We found that NGC~7793 has a luminosity and a rotation speed similar to that of NGC~5470, so we used the values of $R_{\rm c}$ corresponding to this galaxy found in their Table~3 ($R_{\rm c}=1.93$\,kpc and $R_{\rm c}=3.52$\,kpc) for testing the effect of dark matter. We considered cases with a maximum circular velocity $v_{\rm c}=v_{\rm max}\sim120\,{\rm km\,s^{-1}}$ and $\frac{1}{2}v_{\rm c}=v_{\rm max}\sim60\,{\rm km\,s^{-1}}$. In addition we made tests with the halo of a much larger galaxy, NGC~0925, which has $R_{\rm c}=16.86$\,pc. We also considered several disk scalelengths around the value $h\sim700$\,pc, as we have found our results to be dependent on this parameter. The results are presented in Table~\ref{testdm}.

We found that, in general for a submaximal disk, by neglecting the halo we introduce a relatively small bias, namely we overestimate the ratio of $\Sigma_{\rm T}/\Sigma_{\rm t}$ by $10-30\%$ However, for some peculiar haloes, the effect may be larger, up to $\sim40-50\%$. Considering that NGC~5470 is the faintest galaxy in the sample and that brighter galaxies are probably less affected by dark matter (closer to maximum) we conclude that ignoring the effect of a dark matter halo introduces an bias smaller than that caused by choosing a given $\Upsilon_{\rm T}/\Upsilon_{\rm t}$ (which has an effect on the order of 50\% for the range of ``reasonable'' $\Upsilon_{\rm T}/\Upsilon_{\rm t}$ we have been studying), and that it is thus justified to ignore its influence.

\subsubsection{Effect of the line-of-sight integration in not perfectly edge-on galaxies}

Even though we have been quite restrictive at selecting galaxies to be included in our sample, some galaxies may be far from an ideal edge-on geometry. Thus, we tested the effect of line-of-sight integration in galaxies that are not perfectly edge-on.

The test has been done for a ``typical'' galaxy in our sample, namely a galaxy with a radius $r_{25}=14.3$\,kpc, a truncation radius somewhat larger than $r_{\rm 25}$, $r_{\rm trunc}=20$\,kpc, a thin and thick disk scalelength $h=2$\,kpc and a thick disk scaleheight $z_{\rm T}=700$\,pc. In addition we set the distance to the test galaxy to be $D=25$\,Mpc. All the galaxy properties are scalable, and the distance to the galaxy has only been included to account for the effects derived from the PSF. We assumed $\Upsilon_{\rm T}/\Upsilon_{\rm t}=1.2$ and selected the galaxy to have $\rho_{\rm T0}/\rho_{\rm t0}=0.50$ and $\sigma_{T}/\sigma_{t}=2.19$ which are also typical values obtained in the fits to the galaxies in our sample (see Section~4).

\begin{figure*}[!t]
\begin{center}
\begin{tabular}{c c}
\includegraphics[width=0.45\textwidth]{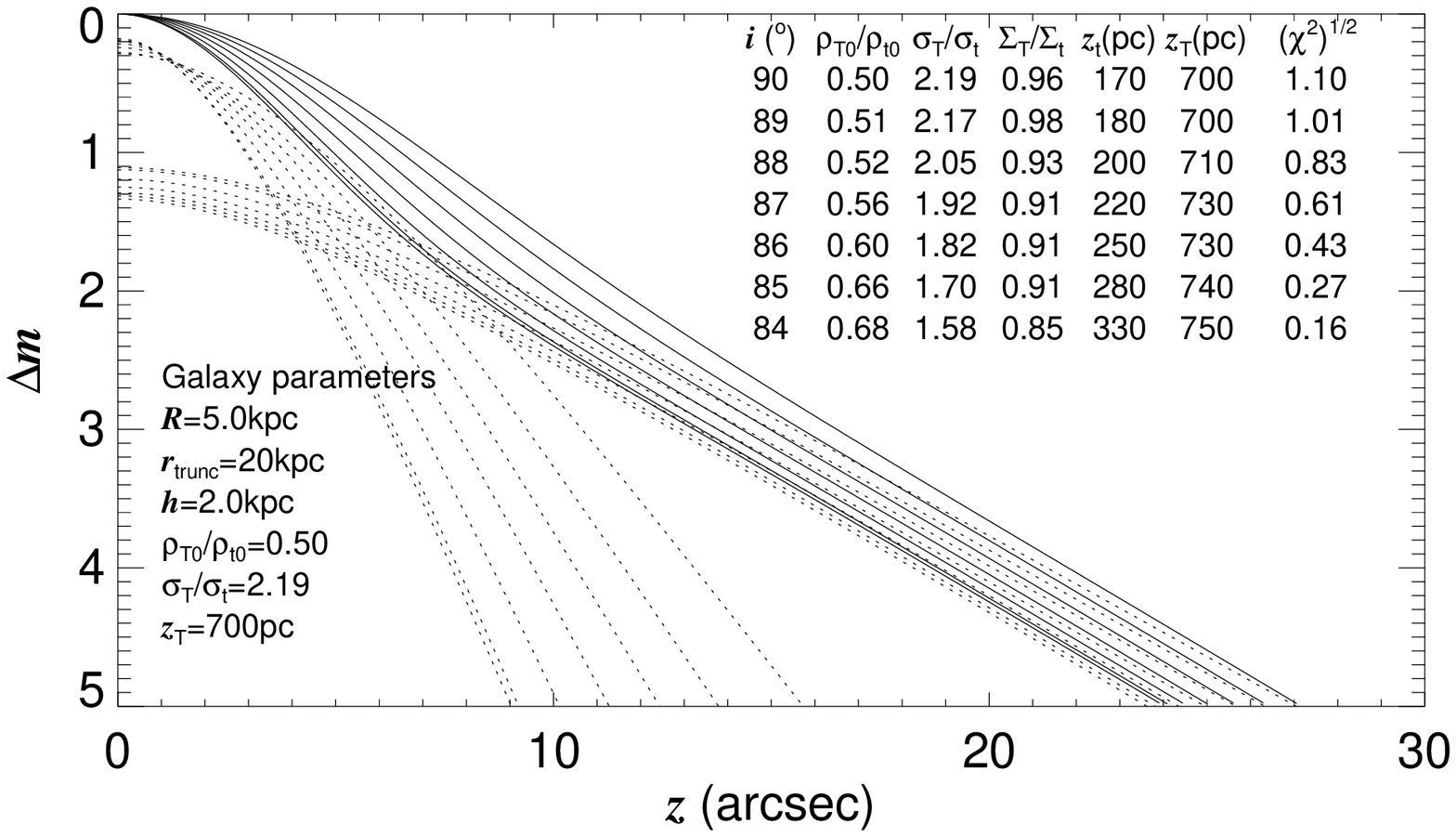}&
\includegraphics[width=0.45\textwidth]{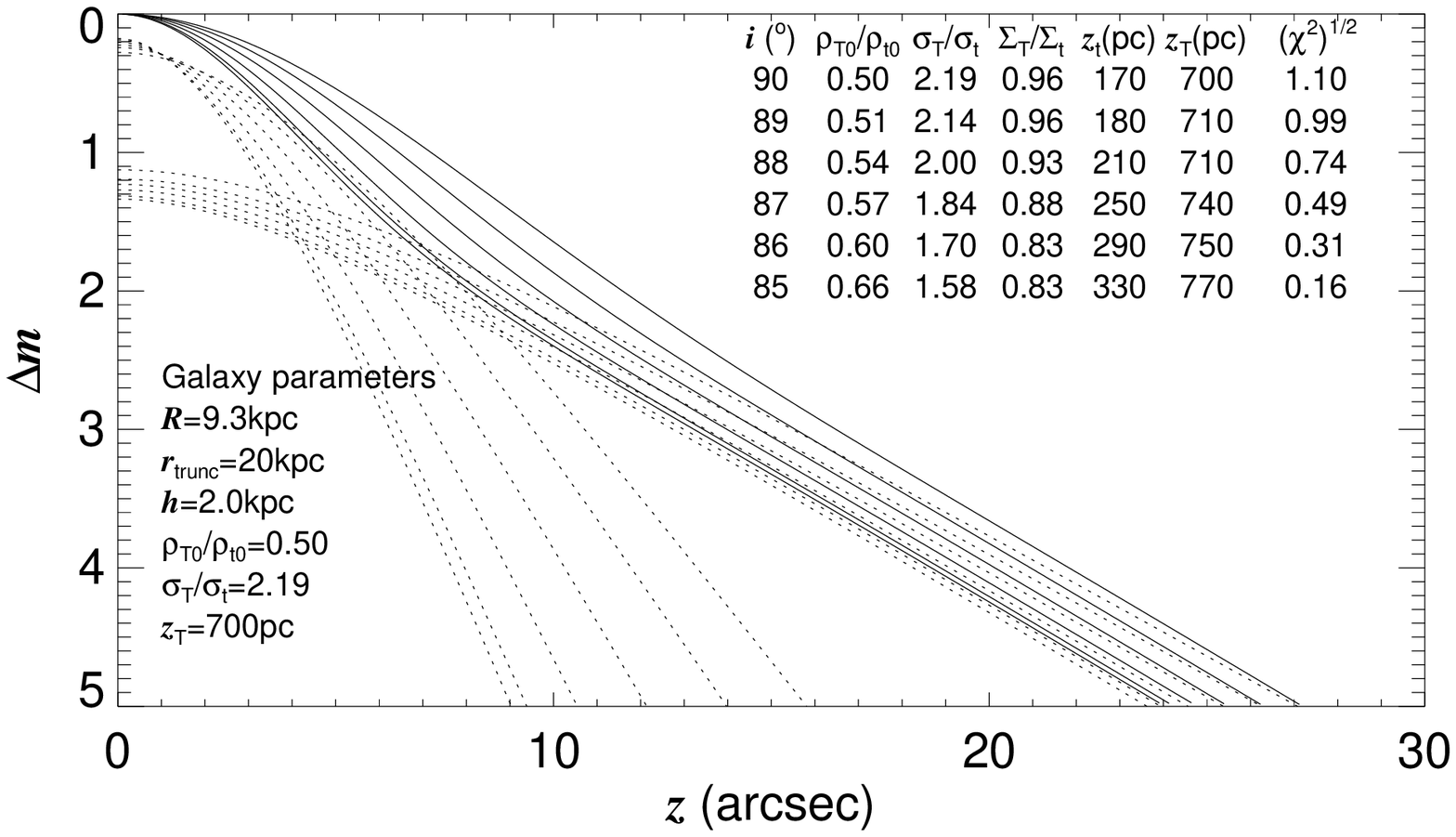}
\end{tabular}
\caption{\label{inclination} Plots representing the effect of fitting the luminosity profiles of a galaxy (with the parameters appearing in the bottom-left corner) which is not perfectly edge-on. From bottom to top, the solid lines represent how the luminosity profile would appear for decreasing inclination angles starting at $i=90\deg$ and with a step of $\Delta i=1\deg$. The dotted lines represent the fitted thin and thick disk components. The table in the top-right corner shows parameters derived from the fits for different inclination angles. $R$ stands for the galactocentric distance at which we are making the fit and $(\chi^2)^{1/2}$ is the square root of the mean square difference between a single-disk profile and the fitted synthetic luminosity profile in units of ${\rm mag\,arcsec^{-2}}$. The values of $R$ we selected correspond to the middle of the $0.2\,r_{25}<|R|<0.5\,r_{25}$ bin (left panel) and the middle of the $0.5\,r_{25}<|R|<0.8\,r_{25}$ bin (right panel). The luminosity profiles have been plotted for decreasing inclination until ${\sqrt \chi^2}=0.2\,{\rm mag\,arcsec^{-2}}$, which is value for which we considered a two stellar disk structure to be indistinguishable from a single-disk structure.}
\end{center}
\end{figure*}

The result of the test is shown in Figure~\ref{inclination} in which luminosity profiles have been fitted for different inclination angles, $i$, at the galactocentric radii, $R$, corresponding to the middle of the $0.2\,r_{25}<|R|<0.5\,r_{25}$ and the $0.5\,r_{25}<|R|<0.8\,r_{25}$ bins and for $\Delta m=5.0\,{\rm mag\,arcsec^{-2}}$. The effect of lowering the inclination is to increase the fitted $\rho_{\rm T0}/\rho_{\rm t0}$ and to reduce the fitted $\sigma_{\rm T}/\sigma_{\rm t}$, which translates into the fitted thin and the thick disks being less differentiated until to a certain $i$ for which the sum of a thin and a thick disk is indistinguishable from a single-disk profile (the square root of the mean square difference between a single-disk profile and the fitted synthetic luminosity profile is $\sqrt{\chi^2}<0.2\,{\rm mag\,arcsec^{-2}}$). The change in the fitted $\rho_{\rm T0}/\rho_{\rm t0}$ and $\sigma_{\rm T}/\sigma_{\rm t}$ happens in such a way that $\Sigma_{\rm T}/\Sigma_{\rm t}$ remains roughly constant. The small changes in $\Sigma_{\rm T}/\Sigma_{\rm t}$ go in the direction of underestimating it (an effect on the order of $\sim10\%$), thus partly counteracting the effect of a dark matter halo in the case of a submaximal disk. The fitted thick disk scaleheight ($z_{\rm T}$) remains roughly constant when decreasing $i$, but the fitted thin disk scaleheight ($z_{\rm t}$) increases significantly.

We have repeated the test for a variety of reasonable galaxy parameters. In general two disks can be distinguished down to $80\deg<i<85\deg$. The fitted $\Sigma_{\rm T}/\Sigma_{\rm t}$ is in general in agreement with that of an edge-on geometry down to at least $i=86\deg$, although it tends to be underestimated by a factor up to $\sim20\%$. For lower $i$ values $\Sigma_{\rm T}/\Sigma_{\rm t}$ can be heavily underestimated, especially when measured at high $R$ for galaxies with low $r_{\rm trunc}$ (e.g. $r_{\rm trunc}=r_{25}$) and for all $R$ in those galaxies with low $\rho_{\rm T0}/\rho_{\rm t0}$ and/or low $\sigma_{\rm T}/\sigma_{\rm t}$. 

\subsubsection{Testing the selected $\Upsilon_{\rm T}/\Upsilon_{\rm t}$}

A way to test whether the two $\Upsilon_{\rm T}/\Upsilon_{\rm t}$ chosen are two good limiting cases is to calculate the $\Upsilon_{\rm T}/\Upsilon_{\rm t}$ at a wavelength other than $3.6\mu{\rm m}$ and compare the results with those obtained using S$^4$G $3.6\mu{\rm m}$ images. If the adopted SFH, and thus the $\Upsilon_{\rm T}/\Upsilon_{\rm t}$, was accurate, then $\Sigma_{\rm T}/\Sigma_{\rm t}$ should be similar in both cases.

We produced the test using SDSS $r$-band imaging for NGC~5470. We choose this galaxy because it has a thin and symmetric dust lane, thus facilitating the analysis. Using the SFHS in Section~3.3 and the Bruzual \& Charlot (2003) spectral synthesis models we calculated that when $\Upsilon_{\rm T}/\Upsilon_{\rm t}=1.2$ in $3.6\mu{\rm m}$, $\left(\Upsilon_{\rm T}/\Upsilon_{\rm t}\right)_{r}=1.1$ and that when $\Upsilon_{\rm T}/\Upsilon_{\rm t}=2.4$, $\left(\Upsilon_{\rm T}/\Upsilon_{\rm t}\right)_{r}=1.3$. As our fitting code was designed to deal with low extinction levels and the $r$-band image of NGC~5470 is strongly affected by a mid-plane dust lane, we manually set the code to ignore the pixels for which $z<4\arcsec$. The fact that the dust has a large effect for more than half of the range in $z$ where the thin disk luminosity dominates illustrates the importance of infrared imaging for this kind of study. We have been able to produce satisfactory fits for galactocentric bins except for that at $0.5\,r_{25}<R<0.8\,r_{25}$. We found that on average when the $3.6\mu{\rm m}$ $\Upsilon_{\rm T}/\Upsilon_{\rm t}=1.2$, $\left(\Sigma_{\rm T}/\Sigma_{\rm t}\right)_{r}=1.42\left(\Sigma_{\rm T}/\Sigma_{\rm t}\right)$ and that when $\Upsilon_{\rm T}/\Upsilon_{\rm t}=2.4$, $\left(\Sigma_{\rm T}/\Sigma_{\rm t}\right)_{r}=0.78\left(\Sigma_{\rm T}/\Sigma_{\rm t}\right)$, with a scatter compatible with the fitting uncertainties. This proves that for NGC~5470 the actual $\Upsilon_{\rm T}/\Upsilon_{\rm t}$ (that which makes $\left(\Sigma_{\rm T}/\Sigma_{\rm t}\right)_{r}=\left(\Sigma_{\rm T}/\Sigma_{\rm t}\right)$) is found somewhere between the two limiting values used in this Paper. 

\subsubsection{The case of NGC~4565}

We found an argument which, although maybe circumstantial, seems to prove that using the solution of equations of equilibrium at fitting the luminosity profiles gives a more accurate result than a sum of somewhat {\it ad hoc} analytical functions. One of the most studied edge-on galaxies, NGC~4565, has been decomposed several times. Jensen \& Thuan (1982) and N\"aslund \& J\"ors\"ater (1997) needed three disks to fit it (or two disks plus a ``corona'' or halo component), but we only needed two to fit the luminosity profiles of this galaxy down to $26\,{\rm mag\,arcsec^{-2}}$, which is similar to the depth of previous studies. The simpler fit certainly advocates for the use of non-analytic physically based functions as done in this study. Alternatively, the need for a third disk may be introduced by some peculiar dust geometry which we avoided by studying the galaxy in the infrared.

\section{Results}

Of the 46 galaxies selected to be fitted, 14 have a mid-plane brightness too low to be fitted down to $\Delta m = 4.5\,{\rm mag\,arcsec}^{-2}$ at any of the vertical bins that we have studied. In addition, two more galaxies, although bright enough, could not be fitted down to $\Delta m = 4.5\,{\rm mag\,arcsec}^{-2}$ because their profiles differed significantly from any of our synthetic models. Thus, we obtained good fits for only 30 galaxies in our sample. The parameters of the fits with $\Delta m \ge 4.5\,{\rm mag\,arcsec}^{-2}$ are presented in Tables~\ref{table1} and \ref{table2}.

Most of the fits are compatible with a double stellar disk. However, some of the fits for ESO~548-063, IC~5052, NGC~1827, PGC~013646 and UGC~10297 are compatible with a single-disk structure (see discussion in Section~3.6.1). All these fits have $\Delta m = 4.5\,{\rm mag\,arcsec}^{-2}$, so they have been calculated over a relatively small dynamic range, which may explain the non-detection of a thick disk. Alternatively, these galaxies may not be inclined enough ($i\lessapprox85$) for allowing the detection of a two stellar disk structure. The results of fits compatible with a single disk have not been included in the plots of this Section.

Two galaxies, ESO~079-003 and NGC~4013, have some bins with lower fitted magnitudes of $\mu_{\rm l}<24\,{\rm mag\,arcsec}^{-2}$. An inspection to their luminosity profiles and the images of these galaxies shows the presence of a third extended component, maybe a third disk or an especially bright stellar halo. NGC~3628 also appears to have an extra extended component but faint enough to allow fitting the luminosity profiles down to levels fainter than $\mu_{\rm l}<24\,{\rm mag\,arcsec}^{-2}$. NGC~4013 is a peculiar galaxy in the sense it is known to have a very prominent \hi\ warp (Bottema et al.~1987) and a `giant tidal stream' (Mart\'inez-Delgado et al.~2009) and it has been studied in detail in a follow-up letter (Comer\'on et al.~2011b). For the other galaxies, our set of synthetic fitting functions appears to describe reasonably well all the disk components.

Although for the same luminosity profile fits with different $\Upsilon_{\rm T}/\Upsilon_{\rm t}$ and gas content usually have the same $\Delta m$, in a few cases fits including a gas disk seem to fit slightly better the luminosity profiles. For $\Upsilon_{\rm T}/\Upsilon_{\rm t}=1.2$, in the without-gas case the luminosity profiles were fitted over $\langle\Delta m\rangle=5.12\,{\rm mag\,arcsec}^{-2}$ on average, while when including a gas disk they were fitted over $\langle\Delta m\rangle=5.19\,{\rm mag\,arcsec}^{-2}$. In the case of $\Upsilon_{\rm T}/\Upsilon_{\rm t}=2.4$ the average dynamical range of the fits is $\langle\Delta m\rangle=5.06\,{\rm mag\,arcsec}^{-2}$ and $\langle\Delta m\rangle=5.16\,{\rm mag\,arcsec}^{-2}$, respectively. These calculations have been performed over the luminosity profiles for which the fit was successful for the two values $\Upsilon_{\rm T}/\Upsilon_{\rm t}$ and the two gas disk fractions used. As discussed by Banerjee \& Jog (2007), this is probably due to the fact that the inclusion of an invisible gas disk helps to fit better the inner parts of the stellar luminosity profile under the assumption of equilibrium.

We have found that for most cases the dust extinction in the mid-plane is small or negligible, as can be seen in columns 19 to 22 of Table~\ref{table1}.

\begin{figure*}[!t]
\begin{center}
\begin{tabular}{c c}
\includegraphics[width=0.45\textwidth]{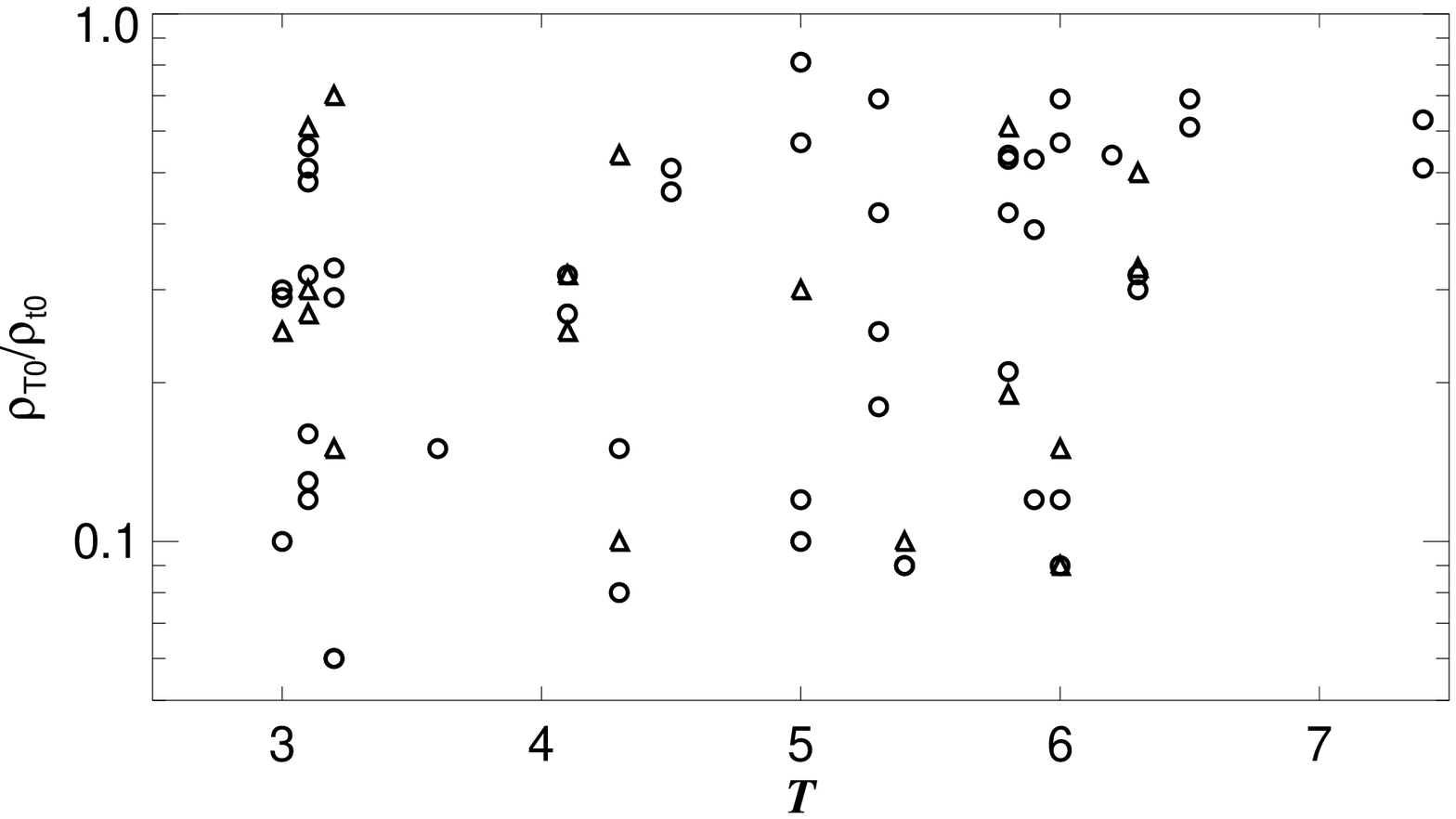}&
\includegraphics[width=0.45\textwidth]{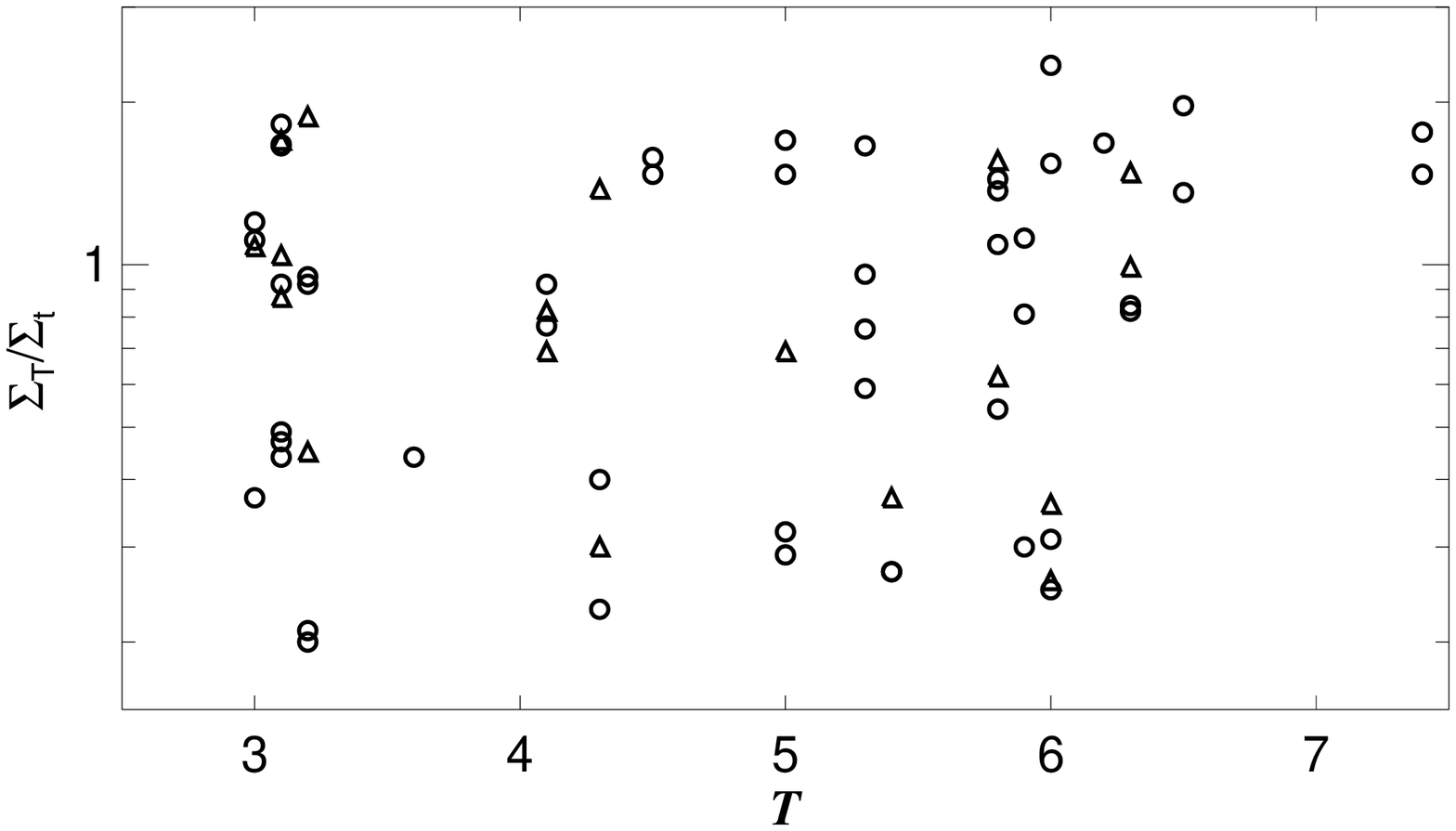}
\end{tabular}
\caption{\label{T_rho} Left panel: central thick to thin disk mass density ratio as a function of galaxy morphological type. Right panel: thick to thin stellar column mass density ratio as a function of galaxy morphological type. Circles denote values fitted at $0.2\,r_{25}<|R|<0.5\,r_{25}$ and triangles values fitted at $0.5\,r_{25}<|R|<0.8\,r_{25}$. Both plots are for fits made using $\Upsilon_{\rm T}/\Upsilon_{\rm t}=1.2$, with-gas. For $\Upsilon_{\rm T}/\Upsilon_{\rm t}=2.4$, points are shifted on average by a factor 1.9 upwards in both plots. Plots for fits in the without-gas case are similar to those with-gas.}
\end{center}
\end{figure*}

There is no variation in the range of possible disk density ratios as a function of galaxy type, $T$, as can be seen in the left panel of Figure~\ref{T_rho}. Moreover, we find that the area populated by points corresponding to $0.2\,r_{25}<|R|<0.5\,r_{25}$ (circles) is very similar to those corresponding to $0.5\,r_{25}<|R|<0.8\,r_{25}$ (triangles) showing that, as previously assumed when not taking into account the line of sight integration, the scalelengths of the thin and the thick disk are similar. 

The shape of the $\Sigma_{T}/\Sigma_{t}$ distribution is very similar to that of the one of $\rho_{T0}/\rho_{t0}$, as can be seen in the right panel of Figure~\ref{T_rho}. We find that $\rm{log}\left(\Sigma_{T}/\Sigma_{t}\right)$ scales linearly with $\rm{log}\left(\rho_{\rm T0}/\rho_{\rm t0}\right)$, with a correlation factor of 0.58. Thus, the ratio of velocity dispersions, $\sigma_{\rm T}/\sigma_{\rm t}$, only accounts for a relatively small scatter.

From the data plotted in Figure~\ref{T_rho} we find that for $\Upsilon_{\rm T}/\Upsilon_{\rm t}=1.2$ only $\sim60\%$ of the data points indicate a higher column mass density in the thin disk than in the thick disk ($\Sigma_{T}/\Sigma_{t}<1$; $\sim70\%$ of the points if an 20\% effect on $\Sigma_{T}/\Sigma_{t}$ is introduced by the halo as described in Section~3.6.2). This fraction goes down to $\sim30\%$ for $\Upsilon_{\rm T}/\Upsilon_{\rm t}=2.4$ ($\sim35\%$ of the points if a 20\% halo bias is considered). In some cases, the column mass density of the thick disk appears to be two or even three times larger than that of the thin disk. As a comparison, the baryonic column mass of the thick disk in the Milky Way has been estimated to be around $\sim20\%$ (Gilmore \& Reid 1983; Chen et al.~2001; Robin et al.~2003; Juri\'c et al.~2008). If we consider the gas fraction of the disk to be one order of magnitude less than that of the stars (Banerjee \& Jog 2007), the Milky Way would appear as a very thin-disk dominated galaxy if $\Upsilon_{\rm T}/\Upsilon_{\rm t}=1.2$, and as an outlier if $\Upsilon_{\rm T}/\Upsilon_{\rm t}=2.4$. On the other hand, one other Milky Way study, by Fuhrmann (2008), considers the mass of Milky Way's thick disk to be comparable with that of the thin disk, which would make it a `typical' galaxy in our sample. If we compare our results with estimates of thick disk masses in external galaxies we see that Yoachim \& Dalcanton (2006) found that for several galaxies (seven out of 34 or $\sim20\%$) the stellar mass of the thick disk is greater than that of the thin disk, but for these galaxies, they assume a large gas fraction in the thin disk, including the far outer parts beyond the optical disk, making its total baryonic mass at least two times greater than that of the thick disk for all the galaxies in their sample.

\begin{figure*}[!t]
\begin{center}
\begin{tabular}{c c}
\includegraphics[width=0.45\textwidth]{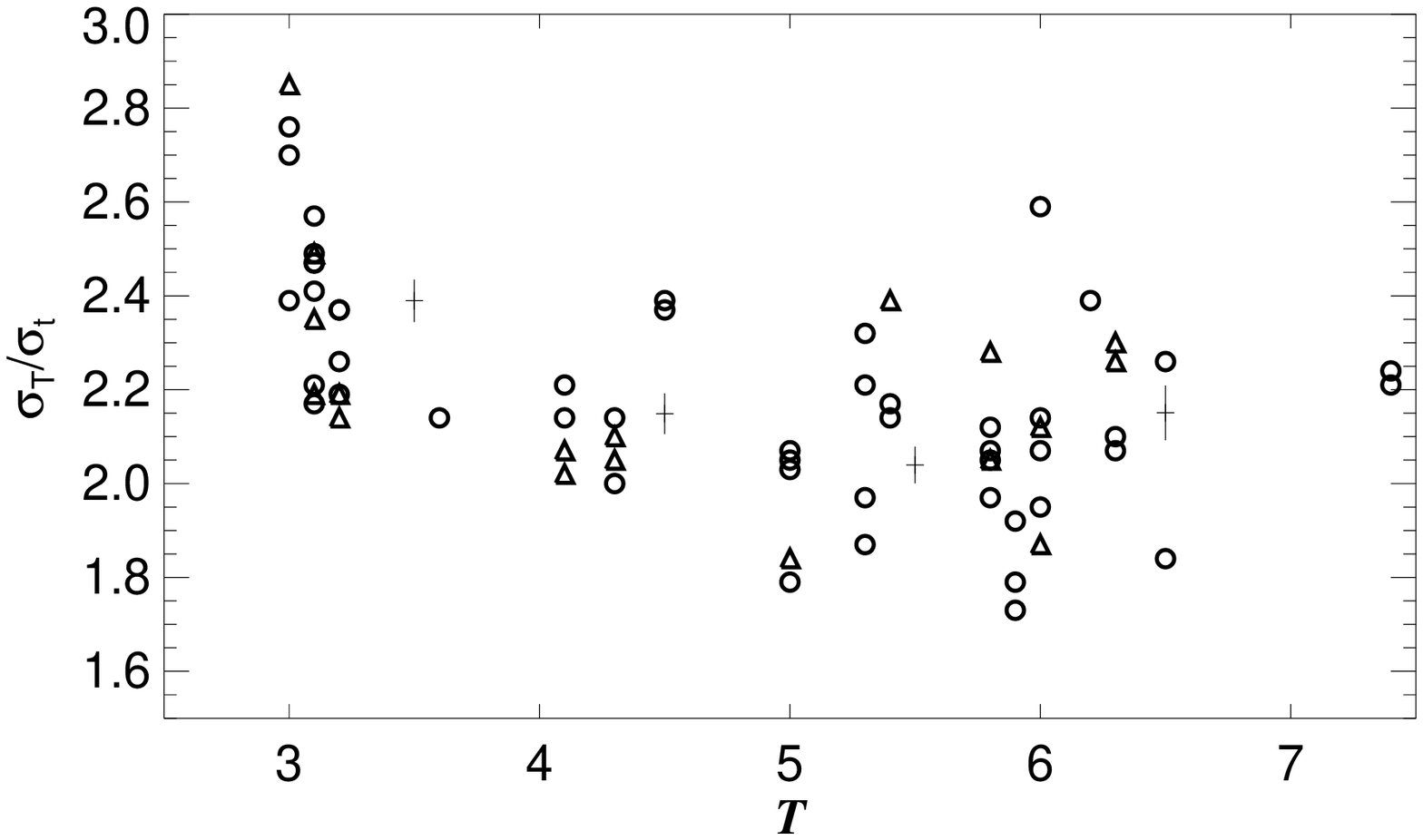}&
\includegraphics[width=0.45\textwidth]{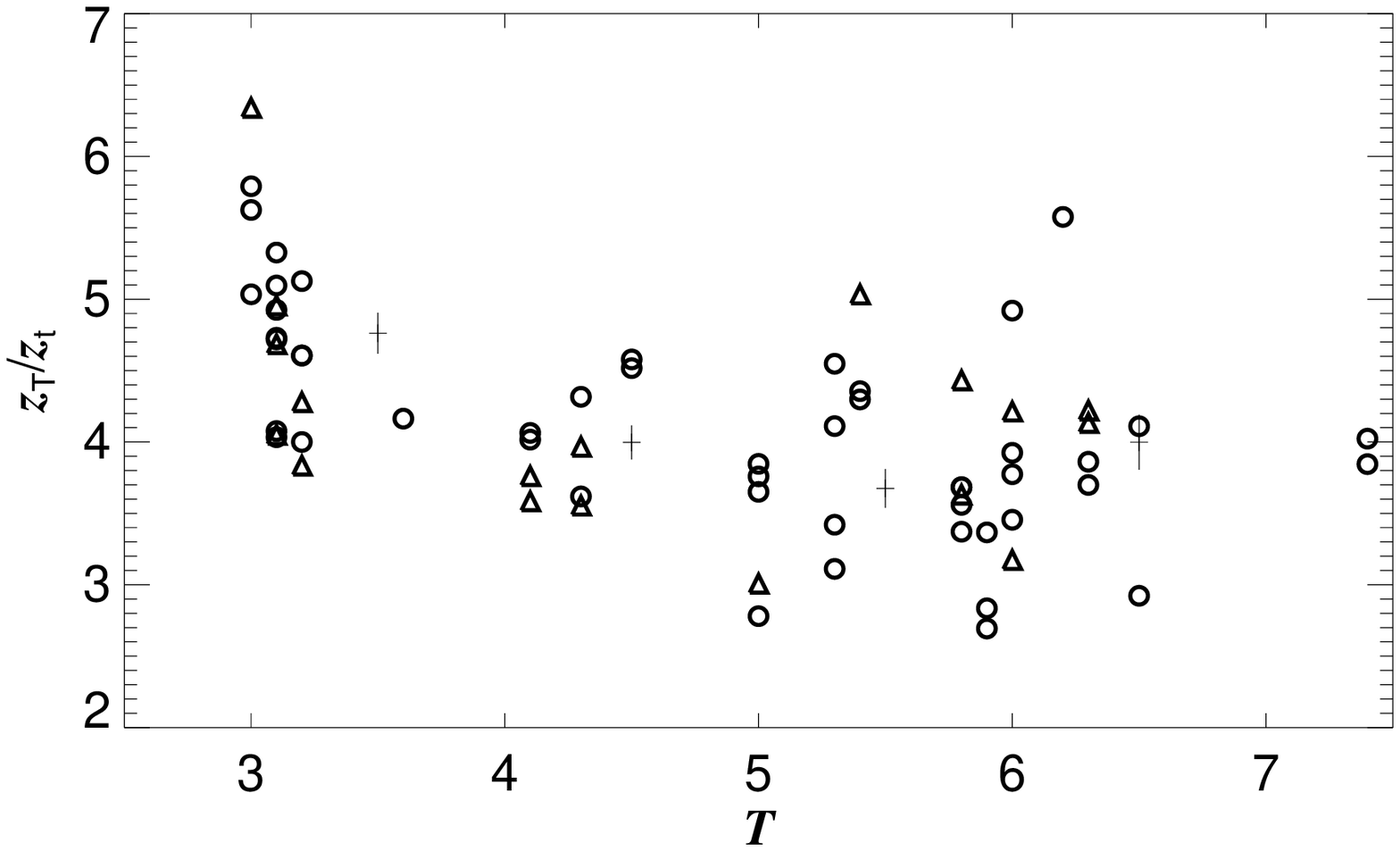}
\end{tabular}
\caption{\label{T_d} Left panel: central thick to thin velocity dispersion ratio in the $z$ direction as a function of galaxy type. Right panel: ratio of thin and thick disk scaleheight as a function of the galaxy type. Circles stand for values fitted at $0.2\,r_{25}<|R|<0.5\,r_{25}$ and triangles stand for values fitted at $0.5\,r_{25}<|R|<0.8\,r_{25}$. The cross symbols denote averages and their errors in bins $T=1$ wide. Both plots are for fits made using $\Upsilon_{\rm T}/\Upsilon_{\rm t}=1.2$, with-gas. For $\Upsilon_{\rm T}/\Upsilon_{\rm t}=2.4$, points are shifted on average by a factor 1.05 up in the velocity dispersion plot. Plots for fits in the without-gas case are similar to those with-gas.}
\end{center}
\end{figure*}

The ratio of vertical velocity dispersions, $\sigma_T/\sigma_t$, shows some correlation with morphological type, with the disks in earlier-type galaxies more kinematically differentiated (left panel in Figure~\ref{T_d}). We find that the degree of differentiation is not linked to the galaxy brightness or to $v_{\rm c}$.

\begin{figure}[!t]
\begin{center}
\begin{tabular}{c}
\includegraphics[width=0.45\textwidth]{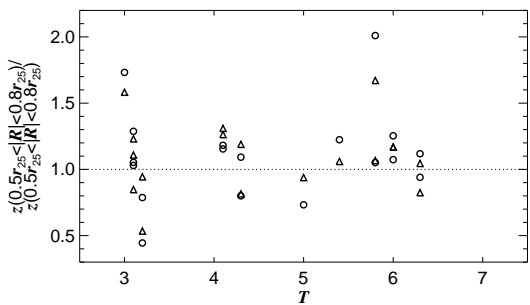}\\
\end{tabular}
\caption{\label{T_zratlat} Scaleheight for the bin $0.5\,r_{25}<|R|<0.8\,r_{25}$, divided by that for the bin $0.2\,r_{25}<|R|<0.5\,r_{25}$. Circles stand for the thin disk, and triangles for the thick disk. Plots for fits using $\Upsilon_{\rm T}/\Upsilon_{\rm t}=2.4$ and without-gas are similar to that presented here. }
\end{center}
\end{figure}

The right panel of Figure~\ref{T_d} shows the ratio of the scaleheights between the thin and the thick disk, $z_{\rm T}/z_{\rm t}$. The scaleheight of each disk was measured by finding the distance between those points where the density of the disk is $\rho_{\rm i}(z)={\rm e}^{-4}\rho_{\rm i0}$ and $\rho_{\rm i}(z)={\rm e}^{-5}\rho_{\rm i0}$, where the subindex $i$ stands for the thin or the thick disk. In the selected regime, the slope of each component is close to exponential, thus our scalelengths can be compared to those of exponential fits and to half of the scaleheight which is found in ${\rm sech}^2(z/z_0)$ fits. The distribution of disk scaleheights is naturally very similar to that of vertical velocity dispersions, as, at first approximation, the velocity dispersion scales with the scaleheight of the disk. As $z_{\rm t}$ is very sensitive to the inclination angle, $i$, part of the scatter in the plot may be caused by the not exactly edge-on orientation of the galaxies.

Disks, both thin and thick, do not systematically flare in a significant way, as seen in Figure~\ref{T_zratlat}. The only points at which the ratio $z_{\rm i}(0.5\,r_{25}<|R|<0.8\,r_{25})/z_{\rm i}(0.2\,r_{25}<|R|<0.5\,r_{25})$ is significantly higher than unity correspond to IC~2135 and NGC~4013. In the case of IC~2135, the flare appears linked to a warp. What we see as a flare in NGC~4013 could be a consequence of a bad fit, as NGC~4013 is one of the two galaxies in our sample which is likely to need more than two disks to be well fitted (see discussion at the start of this Section and Comer\'on et al.~2011b).

\section{Discussion}

\subsection{Comparison of thick to thin disk mass ratios with Yoachim \& Dalcanton}

\begin{figure*}[!t]
\begin{center}
\begin{tabular}{c c}
\includegraphics[width=0.45\textwidth]{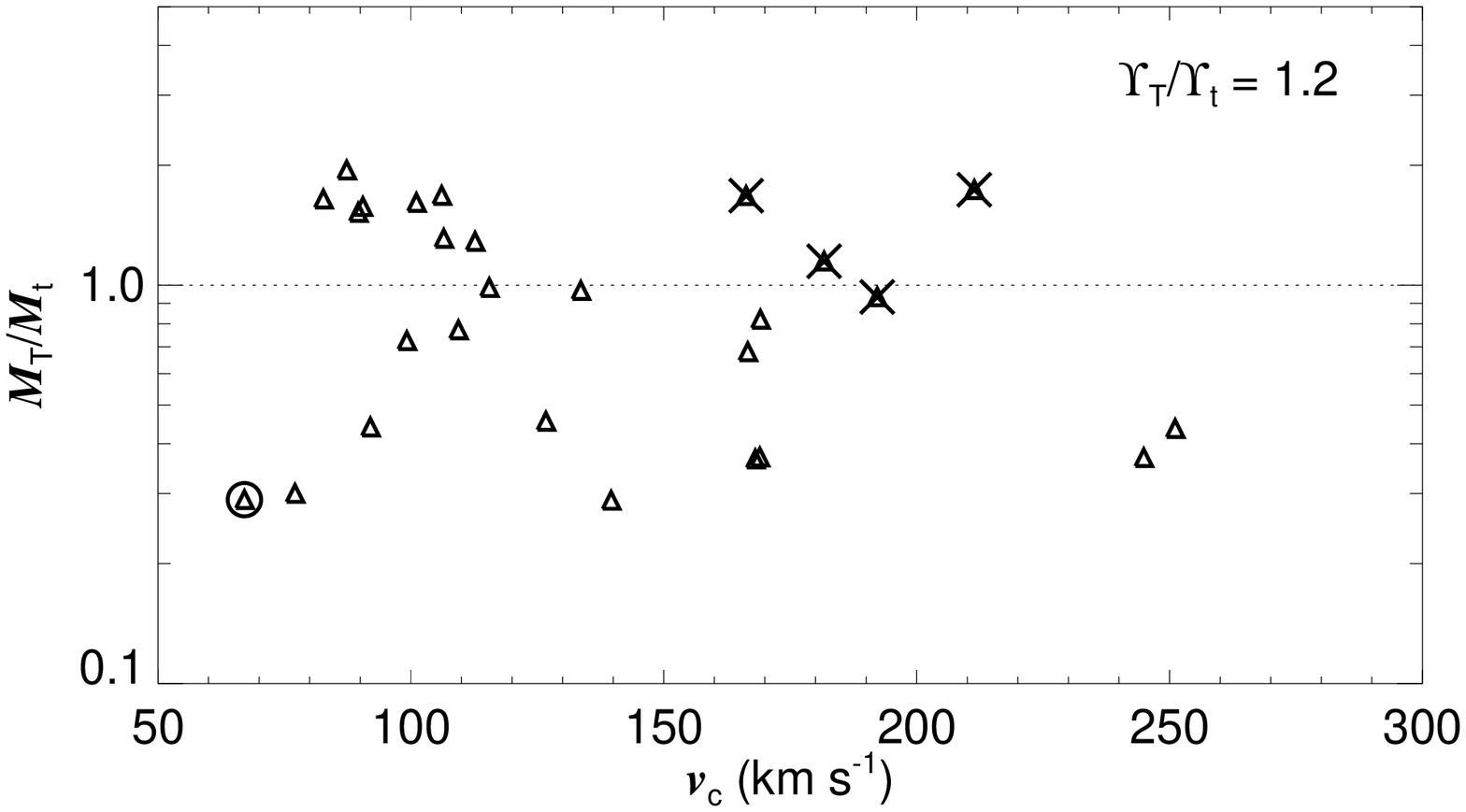}&
\includegraphics[width=0.45\textwidth]{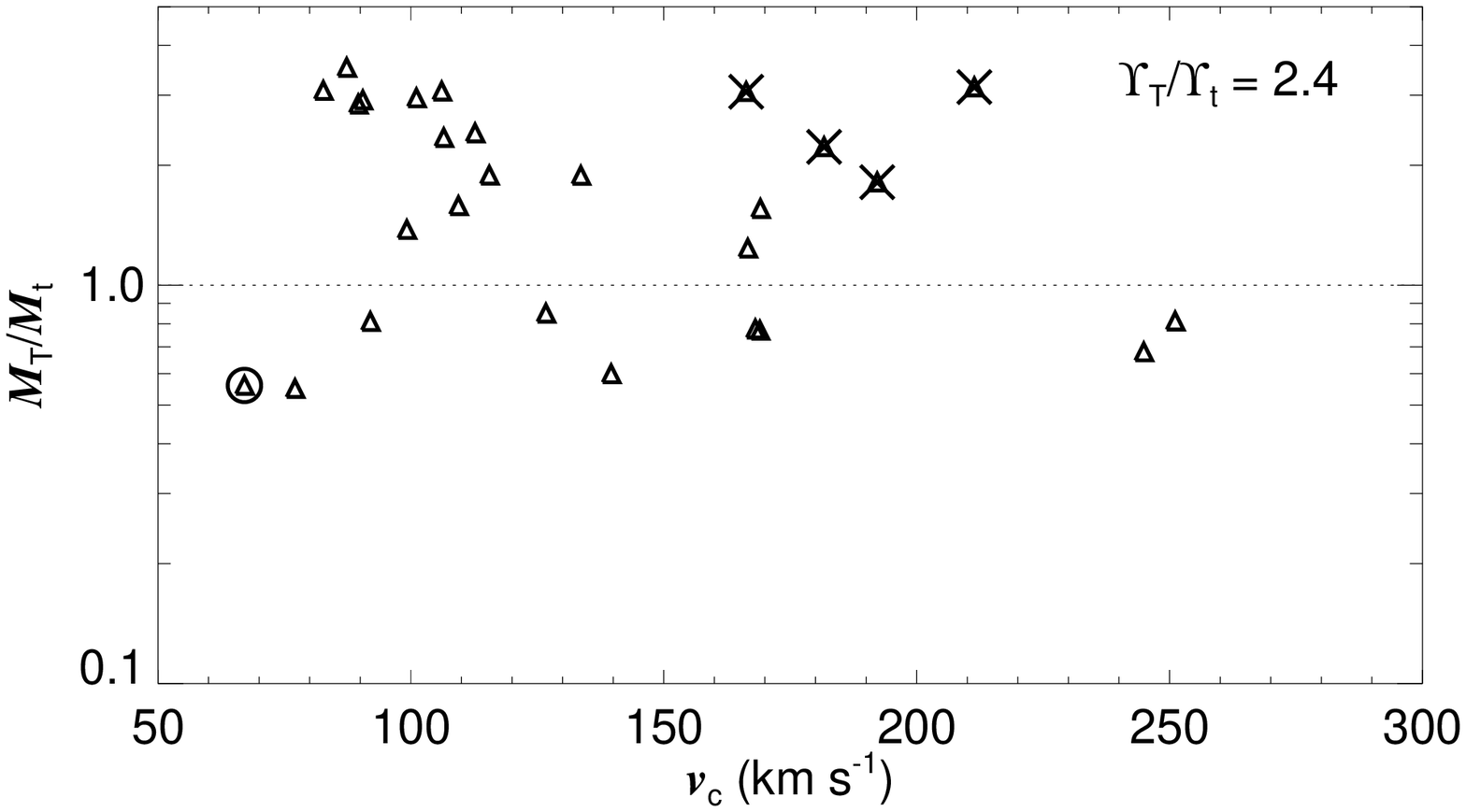}
\end{tabular}
\caption{\label{vmax} Ratio of the thick to thin disk stellar mass as a function of the circular velocity. The plot in the left panel is produced for $\Upsilon_{\rm T}/\Upsilon_{\rm t}=1.2$, and that in the right panel with $\Upsilon_{\rm T}/\Upsilon_{\rm t}=2.4$. The circle indicates a very asymmetrical galaxy, and crosses indicate galaxies with an X-shape bulge (see text). Both plots are in the with-gas case and should be compared to Fig.~22 of Yoachim \& Dalcanton (2006).}
\end{center}
\end{figure*}

A recent study of a statistically significant number of thin/thick disk decompositions was presented by Yoachim \& Dalcanton (2006). Here, we will compare our relative thick disk masses with their results. As mentioned in Section~4, they found that $\sim20\%$ of galaxies have a thick disk stellar mass larger than that of the thin disk. We have not directly measured the disk masses, so in order to obtain an estimate of this quantity which could be compared with their results we have estimated the disk masses by using the following formula:
\begin{equation}
M_{\rm T}/M_{\rm t}=\frac{\sum_{b}\left(10^{-0.4\mu_{\rm c\,b}}\right)\left(\frac{\Sigma_{T}}{\Sigma_{t}}\right)_{\rm b}}{\sum_{b}10^{-0.4\mu_{\rm c\,b}}}
\end{equation}
\noindent where the subindex $b=1-4$ refers to the different bins in galactocentric distance for which the fits have been done and $\mu_{\rm c}=\mu_{\rm l}-\Delta m$ is the mid-plane surface brightness for a given bin. The results are presented in Figure~\ref{vmax}, where we plot $M_{\rm T}/M_{\rm t}$ against the maximum gas circular velocity as obtained from HyperLEDA (Paturel et al.~2003). We see that for $\Upsilon_{\rm T}/\Upsilon_{\rm t}=1.2$, 11 out of 27 galaxies ($\sim40\%$) have a thick to thin disk stellar mass ratio larger than unity and that this number goes up to 18 ($\sim70\%$) when $\Upsilon_{\rm T}/\Upsilon_{\rm t}=2.4$ (the sample of galaxies for which fits were fitted is 30, but three were dropped from this analysis as the fits in all the bins are compatible with a single disk mass distribution).

The two plots in Figure~\ref{vmax} present a hint of a trend with rapidly rotating galaxies having lower relative thick disk mass (lower $M_{\rm T}/M_{\rm t}$) than slowly rotating ones. This trend appears in a much sharper way in the Fig.~22 of Yoachim \& Dalcanton (2006). In order to infer whether the large spread in our Figure~\ref{vmax} may be influenced by sample selection biases, we looked for particularities in our sample galaxies. We found one galaxy -- IC~1553 -- which is significantly distorted and asymmetrical. We also found four galaxies with X-shaped bulges, ESO~079-003, ESO~443-042, NGC~3628 and NGC~4013. Once these galaxies are removed from Fig.~\ref{vmax}, the trend discovered by Yoachim \& Dalcanton (2006) appears clearly. The outlier behavior of the galaxies with X-shaped bulges remains even after considering only the bins at galactocentric radius $0.5\,r_{25}<|R|<0.8\,r_{25}$, implying that their position in the plot is not due to their bulge being extended and affecting some of our fits. In addition, the four galaxies with an X-shaped bulges are known to be warped (S\'anchez-Saavedra et al.~1990; S\'anchez-Saavedra et al.~2002; also seen in S$^4$G images). However, the warp alone does not cause these galaxies to be outliers, as other galaxies in the sample, such as NGC~0522 and NGC~4565, are also warped. 

Why do thick disks in galaxies which host an X-shaped bulge not fit the `normal' behavior of $M_{\rm T}/M_{\rm t}$? Yoachim \& Dalcanton (2006) selected their sample to be made of bulgeless galaxies, so galaxies with an X-shaped bulge were not included. However, our sample has not been selected to be bulgeless and contains several galaxies with a significant bulge which fit in the general trend. For example NGC~0522 and NGC~4565 have boxy bulges (but not X-shaped) and fall into the main relationship. Two of the X-shaped bulges reside in the two galaxies for which they are bins with luminosity profiles that we could not fit down to $\mu_{\rm l}>24.0\,{\rm mag\,arcsec^{-1}}$, so their position in the plot could be a consequence of a bad fit, and in the case of NGC~4013 it is a galaxy which is accreting external material. NGC~3628 has been fitted down to a low $\mu_{\rm l}$, but has obviously been recently perturbed as indicated by its tidal tail (Kormendy \& Bahcall 1974). ESO~443-042 has no noticeable features except for its bulge and warp. So, in at least two out of four cases, the X-shaped bulge may have been caused by bar creation and buckling triggered by a minor merger event, as suggested by Mihos et al.~(1995). Alternatively, the X-shape could be the manifestation of a spiral-like perturbation going down to the center of the galaxy, triggered by a perpendicular interaction (Figs.~6 and 7 in Elmegreen et al.~1995). Other authors, such as Binney \& Petrou (1985) have also suggested mergers as a cause for boxy bulges. An alternative, and more widely followed, possibility is that the X-shaped bulges are edge-on views of strong bars (Athanassoula \& Misiriotis 2002; Mart\'nez-Valpuesta et al.~2006) and do not involve any merging. They would then fall in the same category as the boxy/peanut bulges, the difference being either due to the viewing angle of the bar (e.g. Bureau et al.~2006), or to the strength of the bar (see Athanassoula 2008, for a review). Thus weaker bars seen edge-on would have a boxy shape and stronger ones an X shape, while stronger bars could look boxy-like if viewed sufficiently close to end-on. However, finding a link between the mechanism responsible for the X-shaped bulge and the atypical behavior of the thick disk is not straightforward, although it may be argued that the gravitational influence of a recent minor merger or a strong bar may cause us to observe transient non-equilibrium states for which our equilibrium fitting functions may not hold.

\begin{figure*}[!t]
\begin{center}
\begin{tabular}{c c}
\includegraphics[width=0.45\textwidth]{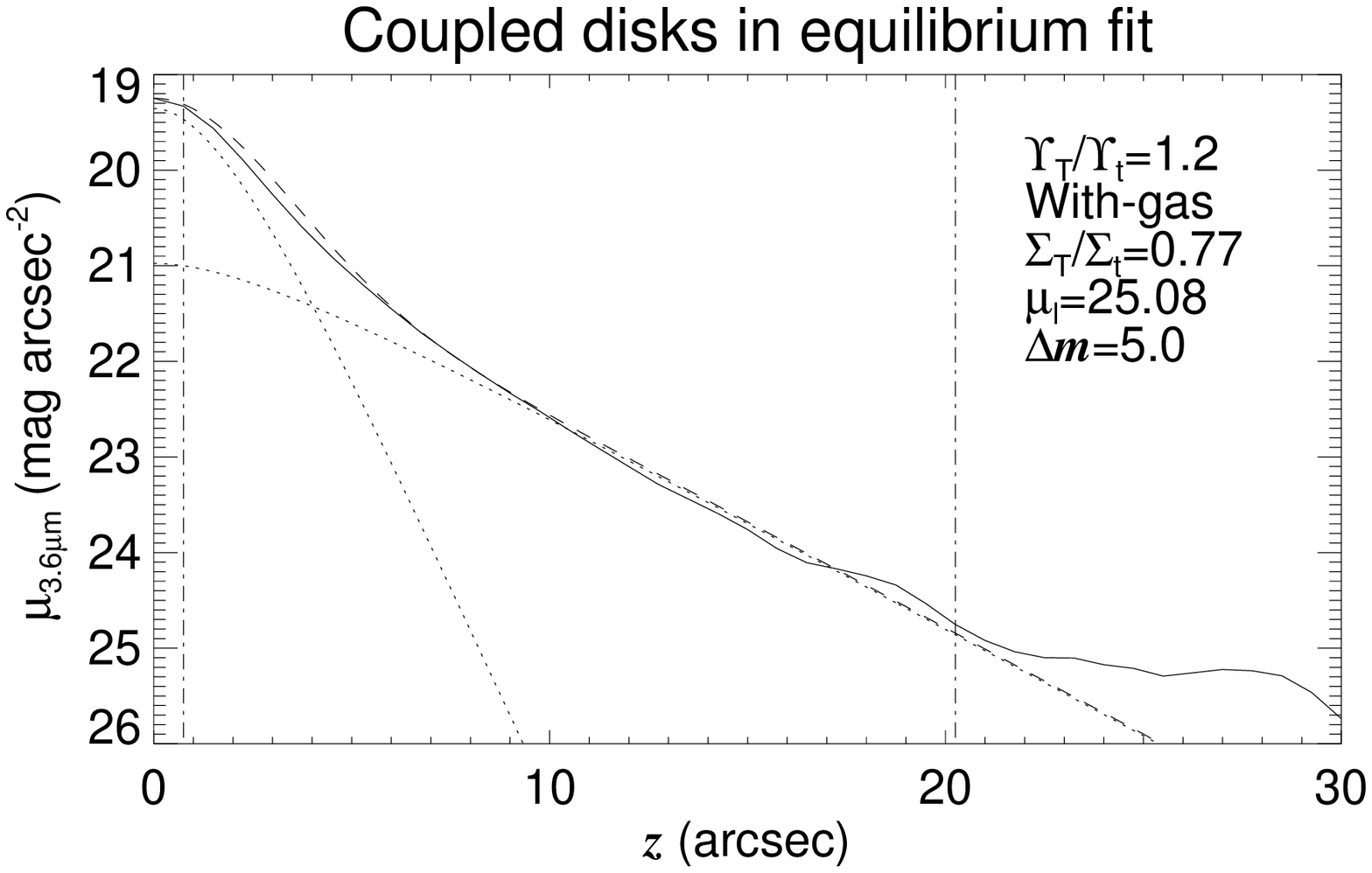}&
\includegraphics[width=0.45\textwidth]{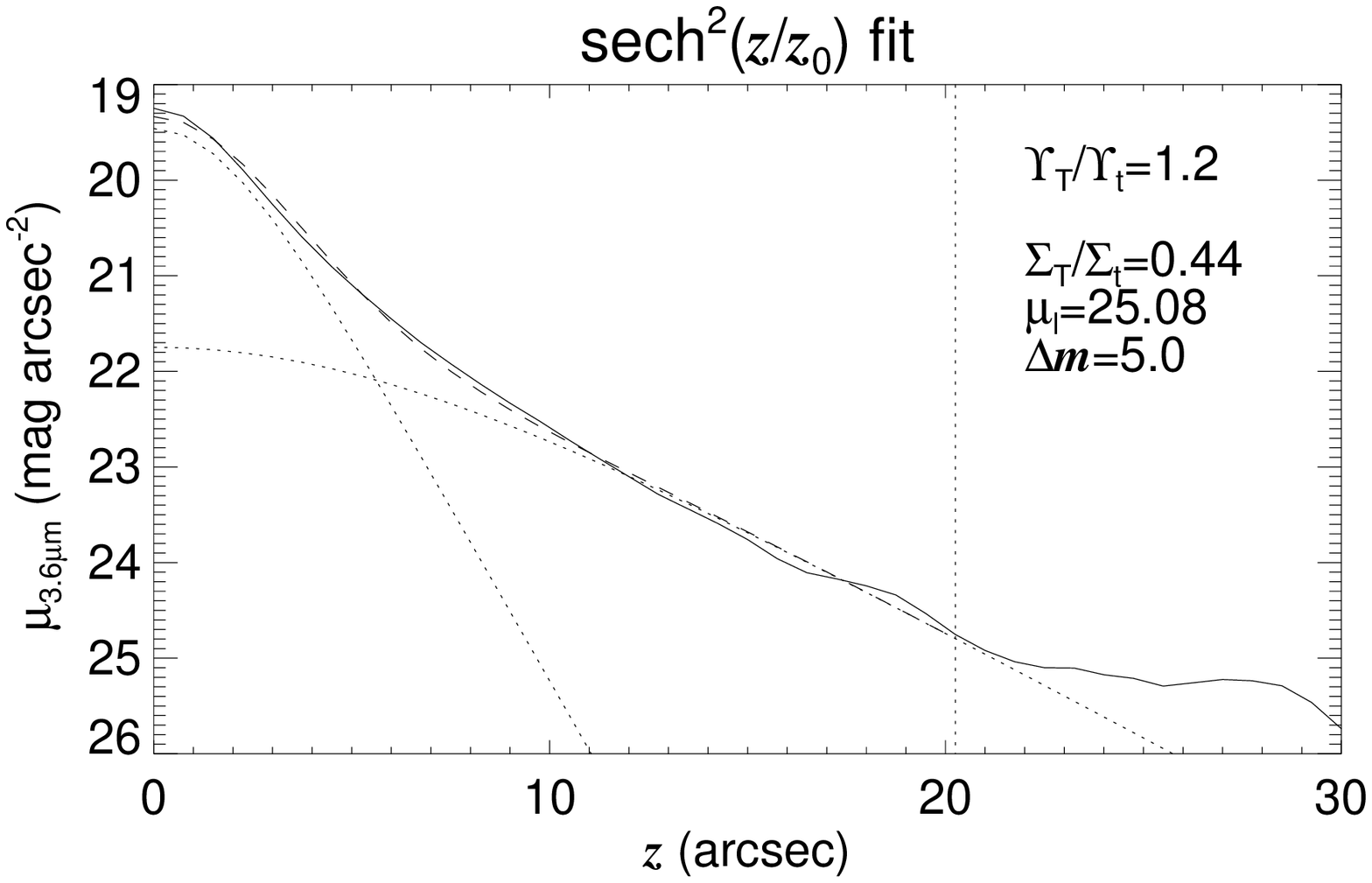}
\end{tabular}
\caption{\label{comparison} Comparison of the fit in the bin $-0.5r_{25}<R<-0.2r_{25}$ for NGC~0522, for the equations of two stellar and one gaseous isothermal coupled disks in equilibrium (left) and the addition of two ${\rm sech}^2(z/z_0)$ functions (right). Solid curves represent the observed luminosity profile, and the dashed curves the best fit. The dotted lines indicate the contributions of the thin and the thick disk. The dash-dotted vertical lines indicate the limits of the range in vertical distance used for the fit.}
\end{center}
\end{figure*}

\begin{figure}[!t]
\begin{center}
\begin{tabular}{c}
\includegraphics[width=0.45\textwidth]{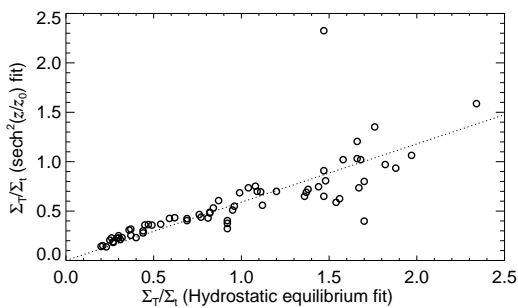}
\end{tabular}
\caption{\label{sech2} Comparison of the column densities obtained from fitting the luminosity profiles by integrating the equations of two stellar and one gaseous isothermal coupled disks in equilibrium with those from fitting the addition of two ${\rm sech}^2(z/z_0)$ functions. The dotted line indicates the best linear fit which crosses the origin. Circles stand for values fitted for $0.2\,r_{25}<|R|<0.5\,r_{25}$, and triangles stand for values fitted for $0.5\,r_{25}<|R|<0.8\,r_{25}$.}
\end{center}
\end{figure}

If we ignore galaxies hosting an X-shaped bulge, our fraction of galaxies in which the thick disk contains more mass than the thin disk is 8 out of 23 ($\sim35\%$) for $\Upsilon_{\rm T}/\Upsilon_{\rm t}=1.2$, and 15 out of 23 ($\sim 65\%$) for $\Upsilon_{\rm T}/\Upsilon_{\rm t}=2.4$. This remains in sharp contrast with the 20\% found by Yoachim \& Dalcanton (2006). One possibility is that this difference is due to the choice of the fitting function. Yoachim \& Dalcanton's (2006) results for each galaxy come from doing the median of the results of six fits with different mid-plane maskings, different smoothings to account for the seeing and different combinations of ${\rm sech}(z/z_0)$ and ${\rm sech}^2(z/z_0)$ accounting for the thin and the thick disk. Of these six fitting models, three make use of a sum of two ${\rm sech}^2(z/z_0)$ functions and are thus likely to dominate the median of the six models.

We tested whether fitting the sum of two ${\rm sech}^2(z/z_0)$ functions gives more weight to the thin disk than the functions we used. We produced ${\rm sech}^2(z/z_0)$ fits over our luminosity profiles using the same procedure than when fitting the equilibrium solutions, except for the fact that, for simplicity, we have not used a PSF convolution and we have not taken into account the effect of dust extinction which we show to be very small in most cases (see Table~\ref{table2}). We found that the shape of the thick disk close to the mid-plane appears rounded in the case of a ${\rm sech}^2(z/z_0)$ fit, but it is significantly more peaked when making a fit using the equations of equilibrium (see Figure~\ref{comparison} for an example).

From a linear regression crossing the origin (Figure~\ref{sech2}), we found that $\left(\Sigma_{\rm T}/\Sigma_{\rm t}\right)_{\rm s}=0.59\pm0.03\left(\Sigma_{\rm T}/\Sigma_{\rm t}\right)_{\rm h}$, where $s$ denotes the fits using two ${\rm sech}^2(z/z_0)$ functions, and $h$ stands for the fits using the equations of equilibrium. This implies that, if we assume our approach to be accurate, and Yoachim \& Dalcanton (2006) mass measurements to be dominated by the models with a sum of two ${\rm sech}^2(z/z_0)$ functions, they might have underestimated $M_{\rm T}/M_{\rm t}$ by a factor of $\sim1.7$. When multiplying by 1.7 the $M_{\rm T}/M_{\rm t}$ values in Fig.~22 of Yoachim \& Dalcanton (2006), we find that 12 out 34 galaxies ($\sim35\%$) would have $M_{\rm T}/M_{\rm t}>1$, a value close to that we obtain with $\Upsilon_{\rm T}/\Upsilon_{\rm t}=1.2$.

The gradient in $\Upsilon_{\rm T}/\Upsilon_{\rm t}$ with $v_{\rm c}$ discussed in Section~3.3.4 is likely to have some effect in the trend seen in Figure~\ref{vmax}. As $\Upsilon_{\rm T}/\Upsilon_{\rm t}$ is larger for galaxies with a small $v_{\rm c}$, it would cause the trend to be strengthened.  

For further comparison with the results of Yoachim \& Dalcanton (2006) we should decompose galaxies from their sample with our methodology using S$^4$G imaging. There is a four galaxy overlap between the S$^4$G and their sample but, unfortunately, at the time of the sample selection of this paper (December 2010) those galaxies have not gone through the automated mosaicking/cleaning process yet. We plan to study this overlap in detail in a follow-up paper.

In summary, we find that our thick to thin disk mass ratios ($M_{\rm T}/M_{\rm t}$) are higher than those found in literature due to the large fraction of mass attributed to the thick disk at low $z$ when fitting luminosity profiles with the equations of coupled isothermal disks in equilibrium. These results have been obtained from data which can reproduce Yoachim \& Dalcanton's (2006) results when fitting a sum of two ${\rm sech}^2(z/z_0)$ functions. As our model is based in physically motivated functions and we are using images less affected by dust than in previous studies, our results are likely to be the most accurate so far for such a large sample. We will now explore the implications of our results, in particular higher thick disk masses.

\subsection{Thick disk formation mechanisms}

The main result of this Paper is that the total stellar mass of thick disks is similar to that of thin disks and that in some cases, the thick disk mass can be up to 2 -- 3 times larger than that in the thin disk. These surprisingly massive thick disks, together with the fact that we have found no significant flares, put some constraints on their formation mechanisms.

The first formation mechanism to be discussed is that of a dynamical heating of an originally thin disk by disk overdensities. Secular heating due to giant molecular clouds (GMCs) has sometimes been suggested as a reason of disk heating. However, this mechanism has been recognized as insufficient to explain the highest velocity dispersion stars (Villumsen 1985; even a fourfold increase in the number of GMCs would not be enough to explain the presence of the hottest populations of stars according, to H\"anninen \& Flynn 2002). This mechanism is thus not likely to create very massive thick disks. Alternatively, the disk may have been mostly heated in a short interval of time at the beginning of the history of the galaxy, an epoch in which there is a large population of clumpy disks whose clumps formed through disk gravitational instabilities and had a mass of a few $10^8M_{\bigodot}$ (Elmegreen \& Elmegreen 2005). These massive clumps have a stellar scattering power much larger than present-day GMCs. Bournaud et al.~(2009) found that this mechanism plus further secular evolution could cause a thick to thin disk mass ratio $M_{\rm T}/M_{\rm t}\sim0.3$ in simulations designed to produce galaxies with Milky Way mass. Thus, Bournaud's et al.~(2009) Milky Way-like simulated galaxies, although a little bit short in thick disk mass, would be compatible with the $v_{\rm c}-M_{\rm T}/M_{\rm t}$ relationship for $\Upsilon_{\rm T}/\Upsilon_{\rm t}=1.2$. However, as these simulations have only been run for Milky Way-like galaxies, the effect caused in a less massive galaxy in unknown. Thick disks created through star scattering and radial mixing by spiral arms has been discussed and modeled by Sch\"onrich \& Binney (2009) who also succeed to reproduce a reasonable thick disk (although again light when put in the $v_{\rm c}-M_{\rm T}/M_{\rm t}$ relationship due to the fact that they fitted the Milky Way which we have shown to have often been measured to be very thin disk-dominated).

Another possibility is that the thick disk has been created by the heating of a thin disk by satellite galaxies or dark matter subhaloes. This mechanism is more efficient for galaxies with orbits far from coplanar with the disk (Read et al.~2008). According to simulations, these interactions may heat the disk to form a thick disk, but they are thought to also cause a significant flare whose exact properties are dependent on the encounter parameters (Quinn et al.~1993; Walker et al.~1996; Kazantzidis et al.~2008; Villalobos \& Helmi 2008; Bournaud et al.~2009). Most of these authors suggest the flare to be at galactocentric radii small enough to be detectable in some cases. The exception is the work by Villalobos \& Helmi (2008), where the flare would only affect the thick disk at very large galactocentric radii, which would need deeper photometry than ours to be detected. Kazantzidis et al.~(2008) suggested that the flares in edge-on galaxies are what we interpret as truncations, as flares would be so substantial that the stars would be scattered over a large vertical range, causing the disk in those regions to have a very low surface brightness when seen edge-on. This would also imply that truncations are more often detected in edge-on galaxies, which is likely to be the case (see discussion in Pohlen et al.~2004b). We have found that strong flares are not common among our galaxies, implying that thick disks are unlikely to form through the heating of a disk by satellites. Even if the flares were so strong to appear as truncations, the heating through encounters would fail to explain the existence of thick disks in some galaxies, since thick disks are ubiquitous and truncations, although very common, do not appear in all galaxies (Kregel et al.~2002; van der Kruit 2007). Minor mergers indeed occur in disk galaxies, but they are likely to have a small effect at high redshift, where they are more frequent, due to the presence of a high gas fraction in the disk (Moster et al.~2010; previous simulations usually do not consider gas effects). The effect of gas for avoiding heating by interactions with satellite galaxies is still significant for galaxies with a gas fraction similar to that of present-day Milky Way, although it allows part of the thin disk to be heated into a flared thick disk (Qu et al.~2011).

Some authors theorize that a thick disk may be a consequence of an {\it in situ} star formation. According to the simulations of Brook et al.~(2004), the thick disk star formation is triggered by the accretion of gas-rich protogalactic fragments during hierarchical clustering at high redshift. The gas disk would still be dynamically hot at the moment of the thick disk creation, which would explain the difference in scaleheight between the thick disk  which formed early and the thin disk formed from a cooled gas disk and from additional gas coming from cold flows. Brook et al.~(2004) found that thick disk scalelengths are larger than those of thin disks, which is in contradiction with observations (Yoachim \& Dalcanton 2006). Later models using disk-disk gas-rich mergers (Robertson et al.~2006; Brook et al.~2007; Richard et al.~2010) are able to produce reasonable thick disks from stars from the progenitor galaxies plus stars created during the merger, with the stars which are created later settling in a thin disk. However, these simulations do not include posterior accretion of cold gas, which will settle in the thin disk, increase its mass, and reduce the scaleheight of the thick disk.   

Another {\it in situ} formation model is that from Elmegreen \& Elmegreen (2006), who study chain and spiral galaxies in the {\it Hubble Space Telescope} Ultra Deep Field (UDF) and find that the scaleheight of those galaxies is similar to that of present-day thick disks, with a mass column density of 4 to 40 $M_{\bigodot}\,{\rm pc}^{-2}$. Recent studies (e.g., Moni Bidin et al.~2010) found the Milky Way column mass density in the solar neighborhood to be $\sim60M_{\bigodot}\,{\rm pc}^{-2}$, thus, if we assume UDF thick disks to be progenitors of present-day thick disks, thick-to-thin disk column mass density fractions up to $\Sigma_{\rm T}/\Sigma_{\rm t}\sim2$ would be compatible with the UDF thick disk mass estimates for Milky Way-like galaxies. However, interpreting these galaxies as direct progenitors of thick disks for which a thin disk has been added after accreting gas through cold flows is problematic, as discussed in Elmegreen \& Elmegreen (2006) as the UDF thick disks are likely to shrink to lower scaleheights when further material is accreted, making them much thinner than observed present-day thick disks. Elmegreen \& Elmegreen (2006) calculated that in order to create a Milky Way-like thick disk with $z_{\rm T}=875$\,pc, the equivalent UDF thick disk should have $z_{\rm T}\sim3$\,kpc, which is far larger than the observed average UDF thick disk. However, as we found that thick disks are much more massive than previously thought, the fraction of added gas needed to generate a thin disk is lower than that estimated by Elmegreen \& Elmegreen (2006), making the decrease in $z_{\rm T}$ of the UDF thick disks lower than what they estimated.

In a $\Lambda$CDM scenario, at least a fraction of the thick disk stars is expected to be accreted from satellite galaxies dragged into the disk plane by dynamical friction. This process would be effective for galaxies with an impact angle $\theta<20\deg$, and larger angle encounters produce spherical-like distributions as well as a thick disk by thin disk heating, as discussed before (Read et al.~2008; but again these simulations do not include gas and thus the effect of disk heating may be overestimated). However, according to these authors, this mechanism applied to the Milky Way would produce a disk $\sim2$ -- 10 times less massive than observed, and thus needs to be combined with other thick disk formation mechanisms. Scannapieco et al.~(2011) found that their simulated thick disks, which have been defined to be made of stars older than 9\,Gyr, have ``relatively high fractions'' of non-accreted stars although the exact value is not specified. A strong indication that this mechanism is likely to be at the origin of a significant fraction of material in at least some galaxies is the discovery of a thick disk with a significant fraction of counterrotating material by Yoachim \& Dalcanton (2008b).

We now summarize our comments on the various proposed thick disk formations scenarios:
\begin{itemize}
 \item Heating of a thin disk by its own overdensities, in the form of, e.g., spiral arms, GMCs or giant clusters: This mechanism may have difficulties to produce massive enough thick disks.
 \item Heating of a thin disk by satellite galaxies or dark matter subhaloes crossing it: Not likely to cause a significant effect as it would cause noticeable flares. Its effect is highly reduced when the disk contains a significant fraction of gas.
 \item {\it In situ} thick disk formation: This formation mechanism had the problem of the shrinking of the thick disk scaleheight due to the accretion of external material after the thick disk formation, which would make thick disks indistinguishable from their thin counterparts. However, our results indicate that thick disks of many galaxies are much more massive than previously thought, lessening the impact of the newer material in the thick disk scaleheight.
 \item Accretion of stars of satellite galaxies (tidal stripping): This mechanism occurs necessarily for some galaxies, as indicated by their kinematical or morphological signatures. However, simulations have shown that this effect can not explain all the mass found in thick disks.
\end{itemize}

Thus, our results plus data from previous studies tend to favor an {\it in situ} origin for most of the stars in the thick disk, where the thin disk created afterwards is not massive enough to reduce $z_{\rm T}$ by too much. In addition the thick disk may contain a significant amount of stars coming from satellites accreted after the initial build-up of the galaxy and an extra fraction of stars coming from the secular heating of the thin disk.

\subsection{Thick disks: lair of missing baryons?} 

Thick disk-dominated galaxies may play some role in what has been called the ``missing baryon problem'' (Persic \& Salucci 1992). WMAP cosmic microwave background observations have constrained the fraction of baryons in the Universe (Spergel et al.~2003), but most of this baryonic matter is apparently not found in the ``obvious'' reservoirs which are stars and cold gas in galaxies (Bell et al.~2003). At a galactic scale, several studies indicate that observations are missing a large fraction of the baryons relative to the dark matter content in the Milky Way halo, and this problem appears to be even more severe for galaxies with smaller masses (Hoekstra et al.~2005; McGaugh 2007; McGaugh 2008; Bregman 2009). Rough counts by McGaugh (2008) for the Milky Way show that we are missing at least $10^{11}\,M_{\bigodot}$, which is comparable or larger than the mass of all the baryonic mass thought to be contained in the disk.

Two mechanisms have been proposed to explain the ``missing baryons'' in galaxy dark matter haloes. The first is that there is a quantity hot gas in the halo which has either failed to cool down, or was blown away by supernova feedback. This hot gas in the halo has been measured to have a mass of $\sim4\times10^8\,M_{\bigodot}$ (Bregman \& Lloyd-Davies 2007), which falls way short of solving the ``local missing baryon'' problem. The second possibility is that supernova feedback has blown away halo gas so efficiently that this material has been lost into the intergalactic medium. The amount of baryons lost in this way is difficult to estimate but it is also apparently not enough to explain the low fraction of baryons in galaxies (McGaugh 2010). Even though they remain insufficient, both explanations have the merit of explaining satisfactorily why lower-mass galaxies have lower baryon fractions, as supernova feedback is more effective in shallower potential wells.

For external galaxies, the stellar masses inside the disks are calculated by assuming a given mass-to-light ratio. However, ignoring the presence of an unexpectedly high fraction of stars in the thick disk may have led to an underestimate of the value of $\Upsilon$. Some numbers may be easily obtained from the data plotted in Figure~\ref{vmax}. Let us consider the more thin disk-dominated galaxies ($M_{\rm T}/M_{\rm t}=0.3$) for the case $\Upsilon_{\rm T}/\Upsilon_{\rm t}=1.2$. If we were ignoring a thick disk, and assign a thin disk $\Upsilon$ for all the light emitted by the galaxy, we would be underestimating the stellar mass of the galaxy by $\sim10\%$. On the other hand, if we consider the most thick disk-dominated galaxy ($M_{\rm T}/M_{\rm t}=3.4$), and $\Upsilon_{\rm T}/\Upsilon_{\rm t}=2.4$, we would underestimate the stellar disk mass by a factor of two. These numbers are calculated for the 3.6$\mu$m-band, which implies that the underestimation of the disk stellar mass would be different when measured at shorter wavelengths. We stress that these numbers only account for the baryons in stellar disks, and that if extended gas disks beyond the optical radius were included in the thin disk mass as done by Yoachim \& Dalcanton (2006) $M_{\rm T}/M_{\rm t}$ would go down.

Thus, our results suggest that a least a fraction of the ``local missing baryons'' are located in the thick disk component of galaxies and that they have been ``lost'' until now due to the underestimation of $M_{\rm T}/M_{\rm t}$. This explanation is complementary, and not an alternative, to those which have already been suggested to solve the problem.

\section{Summary and conclusions}

The nature of the mechanisms creating the thick disks has been a topic of discussion for a long time. Basically four pictures have been suggested, each leaving some distinct observational signature. The first one is that the thick disk is the consequence of the thin disk have been dynamically heated by its own overdensities. The second possibility is that thin disks being dynamically heated by the disk crossing of satellite galaxies or dark matter subhaloes. The third possibility is {\it in situ} star formation during or shortly after the build-up of the galaxy. The fourth is that the thick disk could be made of stars which have been accreted from infalling satellites after the formation of the thin disk.

In order to study the problem of thick disk formation we have used 3.6$\mu$m images of edge-on galaxies from the S$^4$G. The selection criterion used, maximum isophote ellipticity $\epsilon>0.8$, has led to a sample of 46 late-type galaxies with types ranging from $T=3$ to $T=8$. We have fitted the luminosity profiles of these galaxies in radial bins with $0.2\,r_{25}<|R|<0.5\,r_{25}$ and $0.5\,r_{25}<|R|<0.8\,r_{25}$, using solutions of the equations of equilibrium for two stellar and one gaseous coupled isothermal disks. These functions have the advantage of being physically motivated. The largest uncertainty of the fit comes from the assumed ratio of mass-to-light ratios of the thick and the thin disk, $\Upsilon_{\rm T}/\Upsilon_{\rm t}$. We studied cases with $\Upsilon_{\rm T}/\Upsilon_{\rm t}=1.2$ and $\Upsilon_{\rm T}/\Upsilon_{\rm t}=2.4$ in order to account for the range of fitted star formation histories for the Solar neighborhood. The low value of $\Upsilon_{\rm T}/\Upsilon_{\rm t}$ arises when the stars in thick disks form over a long period of time and the high value arises when the thick disk stars formed entirely in a burst in the early Universe. The studied interval in $\Upsilon_{\rm T}/\Upsilon_{\rm t}$ is in broad agreement with that derived from the study by Yoachim \& Dalcanton (2006).

We found that the stellar mass of the thick disk component ranges from one third to two times that of the thin disk if $\Upsilon_{\rm T}/\Upsilon_{\rm t}=1.2$ and from two thirds to more than three times if $\Upsilon_{\rm T}/\Upsilon_{\rm t}=2.4$. We thus found that thick disks are much more massive than generally assumed so far. In order to check the reasons of this difference, we compared our results in detail with those from Yoachim \& Dalcanton (2006). We found that the choice of the fitting function is responsible for the high fraction of stellar light we detect in the thick disk. We also found that disks do not flare significantly within the range of galactocentric radii we have studied, and that the ratio of thick to thin disk scaleheights is higher for earlier-type galaxies in our sample. The lack of flares discards the kinematical heating of a thin disk by satellite galaxies or dark matter subhaloes as the main thick disk formation mechanism.

As a result of these findings we suggest the following tentative formation mechanism for thick disks. First, a thick stellar disk is created during and soon after the build-up of the galaxy by the stirring action from massive disk clumps and the instabilities that form those clumps. A thin disk subsequently forms from gas which has not initially been spent in stars, and from gas that arrives later at a lower rate so that, when combined with the hot disk and bulge that are already present, it is relatively stable and cannot heat itself much. The newly accreted gas makes the thick disk scaleheight, $z_{\rm T}$, shrink, and if this shrinkage is a relatively small amount, then the final thick disk remains a distinct component of the galaxy, with a younger thin disk inside of it. In the case of thick-disk dominated galaxies, the thick disk forms partly at high redshift by internal disk stirring, as above, and then increases its mass over a Hubble time by the heating of the thin disk by internal disk substructure, such as large star clusters and giant molecular clouds, and the accretion of satellite galaxies. 

As our results suggest that thick disks have masses comparable to those of thin disks and in addition they have a large mass-to-light ratio due to the absence of young stars, we suggest that ignoring them has led to an underestimate of the baryonic mass located in disk galaxies. We thus suggest that thick disks contain part of what has been called the ``local missing baryons''. We have calculated this underestimate to be between 10\% and 50\% of the stellar mass of a disk, depending on the ratio of the thick versus thin disk masses and mass-to-light ratios, $M_{\rm T}/M_{\rm t}$ and $\Upsilon_{\rm T}/\Upsilon_{\rm t}$. 

\section*{Acknowledgments}

The authors wish to thank the entire S$^4$G team for their efforts in this project. This work is based on observations and archival data made with the Spitzer Space Telescope, which is operated by the Jet Propulsion Laboratory, California Institute of Technology under a contract with NASA. We are grateful to the dedicated staff at the Spitzer Science Center for their help and support in planning and execution of this Exploration Science program. We also gratefully acknowledge support from NASA JPL/Spitzer grant RSA 1374189 provided for the S$^4$G project.

We thank our anonymous referee for comments that helped to improve the paper.

KS, J-CMM, TKim and TMizusawa acknowledge support from the National Radio Astronomy Observatory, which is a facility of the National Science Foundation operated under cooperative agreement by Associated Universities, Inc. EA and AB thank the Centre National d'\'Etudes Spatiales for financial support.

We thank Peter Yoachim for kindly sharing information he used for his Yoachim \& Dalcanton (2006) paper.

This research has made use of observations made with the NASA/ESA Hubble Space Telescope and obtained from the Hubble Legacy Archive, which is a collaboration between the Space Telescope Science Institute (STScI/NASA), the Space Telescope European Coordinating Facility (ST-ECF/ESA) and the Canadian Astronomy Data Centre (CADC/NRC/CSA).

Funding for the SDSS and SDSS-II has been provided by the Alfred P. Sloan Foundation, the Participating Institutions, the National Science Foundation, the U.S. Department of Energy, the National Aeronautics and Space Administration, the Japanese Monbukagakusho, the Max Planck Society, and the Higher Education Funding Council for England. The SDSS Web Site is http://www.sdss.org/. This research has made use of the NASA/IPAC Extragalactic Database (NED) which is operated by the Jet Propulsion Laboratory, California Institute of Technology, under contract with the National Aeronautics and Space Administration.

This research has made use of the NASA/IPAC Extragalactic Database (NED) which is operated by the Jet Propulsion Laboratory, California Institute of Technology, under contract with the National Aeronautics and Space Administration.

\clearpage

\begin{deluxetable}{lcccccc|cccc|cccc|cccc|cccc}
\tablecolumns{21}
\rotate
\tablewidth{0pt}
\tabletypesize{\scriptsize}
\tablecaption{Galaxy and fit data (part one)}
\tablehead{ID & $T$ & $D$ & PA & $D_{25}$ & $B$ & $v_{\rm c}$ &\multicolumn{4}{c|}{Status}&\multicolumn{4}{c|}{$\rho_{\rm T0}/$} & \multicolumn{4}{c|}{$\sigma_{\rm T}/\sigma_{\rm t}$}&\multicolumn{4}{c}{$f_0$}\\
 & & (Mpc) & (\deg) & (\arcsec) & & (km\,s$^{-1}$) & & & & &\multicolumn{4}{c|}{$\rho_{\rm t0}$}& & & &  &\\
(1) & (2) &(3) &(4)&(5)&(6)&(7)&(8)&(9)&(10)&(11)&(12)&(13)&(14)&(15)&(16)&(17)&(18)&(19)&(20)&(21)&(22)&(23)}
\startdata
\multirow{4}{*}{ESO~079-003} & \multirow{4}{*}{3.2} & \multirow{4}{*}{41.1} & \multirow{4}{*}{130.2} & \multirow{4}{*}{177} &\multirow{4}{*}{-19.97}&\multirow{4}{*}{192.2} & $\star$ & V & V & $\bigtriangledown$ & \nodata &0.33&0.30& \nodata& \nodata&2.28 & 2.51 & \nodata & \nodata & 1.00 & 1.00 & \nodata \\
 &&& & & & & $\star$ & V & V & $\bigtriangledown$ & \nodata & 0.33 & 0.29 & \nodata & \nodata & 2.19 & 2.37 & \nodata & \nodata & 1.00 & 1.00 & \nodata \\
 &&& & & & & $\star$ & V & V & $\bigtriangledown$ & \nodata & 0.61 & 0.56 & \nodata & \nodata & 2.41 & 2.65 & \nodata & \nodata & 1.00 & 1.00 & \nodata \\
 &&& & & & & $\star$ & V & V & $\bigtriangledown$& \nodata & 0.61 & 0.54 & \nodata & \nodata &2.32 & 2.51& \nodata & \nodata & 1.00 & 1.00 & \nodata \\
\hline
\multirow{4}{*}{ESO~443-042} & \multirow{4}{*}{3.1} & \multirow{4}{*}{38.3} & \multirow{4}{*}{127.7} & \multirow{4}{*}{169} & \multirow{4}{*}{-20.46} & \multirow{4}{*}{166.3}& $\perp$ & $\star$ & V & V & \nodata & \nodata & 0.50 & 0.60 & \nodata &\nodata & 2.57 & 2.30&\nodata&\nodata&1.00&1.00\\
&&& & & & & $\perp$ & $\star$ & V & V&\nodata&\nodata&0.51&0.61&\nodata&\nodata&2.47&2.19&\nodata&\nodata&1.00&1.00\\
&&& & & & & $\perp$ & $\star$ & V & V&\nodata&\nodata&0.96& 1.08 &\nodata&\nodata &2.76 &2.41 &\nodata&\nodata& 1.00 & 1.00\\
&&& & & & & $\perp$ & $\star$ & V & V&\nodata&\nodata&0.91&1.11&\nodata&\nodata&2.59&2.32&\nodata&\nodata&1.00&1.00\\
\hline
\multirow{4}{*}{ESO~544-027} & \multirow{4}{*}{3.6} & \multirow{4}{*}{40.3} & \multirow{4}{*}{153.3} & \multirow{4}{*}{95} & \multirow{4}{*}{-18.46}& \multirow{4}{*}{92.0} & $\bigtriangledown$ & V & * & $\bigtriangledown$ &\nodata&0.16&\nodata&\nodata&\nodata&2.19&\nodata&\nodata&\nodata&1.00&\nodata&\nodata\\
&& & & & & & $\bigtriangledown$ & V & * & $\bigtriangledown$ &\nodata & 0.15 &\nodata&\nodata&\nodata&2.14&\nodata&\nodata&\nodata&1.00&\nodata&\nodata \\
&& & & & & & $\bigtriangledown$ & V & * & $\bigtriangledown$ &\nodata & 0.30 &\nodata&\nodata&\nodata& 2.28 &\nodata&\nodata&\nodata& 1.00 &\nodata&\nodata\\
&& & & & & & $\bigtriangledown$ & V & * & $\bigtriangledown$ &\nodata&0.27 &\nodata &\nodata &\nodata&2.24 &\nodata &\nodata &\nodata&1.00 &\nodata &\nodata\\
\hline
\multirow{4}{*}{ESO~548-063} & \multirow{4}{*}{4.2} & \multirow{4}{*}{25.1} & \multirow{4}{*}{37.6} & \multirow{4}{*}{85} &\multirow{4}{*}{-18.14} &\multirow{4}{*}{74.0} & $\bigtriangledown$ & $\perp$ & V & $\bigtriangledown$ &\nodata &\nodata &{\it 1.29} &\nodata&\nodata&\nodata&{\it 1.73} &\nodata&\nodata&\nodata&{\it 1.00} &\nodata\\
&& & & & & &$\bigtriangledown$ & $\perp$ & V & $\bigtriangledown$&\nodata&\nodata&{\it 1.29}&\nodata&\nodata&\nodata&{\it 1.64}&\nodata&\nodata&\nodata&{\it 1.00}&\nodata\\
&& & & & & &$\bigtriangledown$ & $\perp$ & V & $\bigtriangledown$&\nodata&\nodata&{\it 2.25}&\nodata&\nodata&\nodata&{\it 1.76}&\nodata&\nodata&\nodata&{\it 1.00}&\nodata\\
&& & & & & &$\bigtriangledown$ & $\perp$ & V & $\bigtriangledown$&\nodata&\nodata&{\it 2.25}&\nodata&\nodata&\nodata&{\it 1.70}&\nodata&\nodata&\nodata&{\it 1.00}&\nodata\\
\hline
\multirow{4}{*}{IC~0217} & \multirow{4}{*}{5.8} & \multirow{4}{*}{24.2} & \multirow{4}{*}{35.7} & \multirow{4}{*}{120}&\multirow{4}{*}{-18.16}&\multirow{4}{*}{99.2} &$\bigtriangledown$ & V & V & $\perp$&\nodata&0.42&0.37&\nodata&\nodata&2.14&2.12&\nodata&\nodata&1.00&1.00&\nodata\\
&& & & & & &$\bigtriangledown$ & V & V & $\perp$&\nodata&0.42&0.21&\nodata&\nodata&2.05&1.97&\nodata&\nodata&1.00&1.00&\nodata\\
&& & & & & &$\bigtriangledown$ & V & V & $\perp$&\nodata&0.78&0.69&\nodata&\nodata&2.26&2.24&\nodata&\nodata&1.00&1.00&\nodata\\
&& & & & & &$\bigtriangledown$ & V & V & $\perp$&\nodata&0.78&0.39&\nodata&\nodata&2.17&2.05&\nodata&\nodata&1.00&1.00&\nodata\\
\hline
\multirow{4}{*}{IC~1197} & \multirow{4}{*}{6.0} & \multirow{4}{*}{25.7} & \multirow{4}{*}{56.4} & \multirow{4}{*}{154} &\multirow{4}{*}{-18.50}&\multirow{4}{*}{87.3} & $\bigtriangledown$ & V & V & $\bigtriangledown$ &\nodata&0.57&0.66&\nodata&\nodata&2.24&2.66&\nodata&\nodata&0.98&1.00&\nodata\\
&& & & & & &$\bigtriangledown$ & V & V & $\bigtriangledown$&\nodata&0.57&0.69&\nodata&\nodata&2.14&2.59&\nodata&\nodata&0.97&1.00&\nodata\\
&& & & & & &$\bigtriangledown$ & V & V & $\bigtriangledown$&\nodata&1.03&1.17&\nodata&\nodata&2.35&2.79&\nodata&\nodata&0.99&1.00&\nodata\\
&& & & & & &$\bigtriangledown$ & V & V & $\bigtriangledown$&\nodata&1.05&1.22&\nodata&\nodata&2.26&2.70&\nodata&\nodata&0.98&1.00&\nodata\\
\hline
\multirow{4}{*}{IC~1553} & \multirow{4}{*}{5.4} & \multirow{4}{*}{41.9} & \multirow{4}{*}{15.0} & \multirow{4}{*}{81} &\multirow{4}{*}{-19.62} &\multirow{4}{*}{67.1} & $\bigtriangledown$ & V & V & V &\nodata & 0.44 & 0.12 & 0.10 &\nodata & 1.97 & 2.14 & 2.47&\nodata & 1.00 & 1.00 & 1.00\\
&& & & & & &$\bigtriangledown$ & V & V & V&\nodata & 0.09 &0.09&0.10&\nodata&2.14&2.17&2.39&\nodata&1.00&1.00&1.00\\
&& & & & & &$\bigtriangledown$ & V & V & V&\nodata & 0.80 & 0.24 & 0.19 &\nodata&2.05 &2.21 & 2.57 &\nodata&1.00 & 1.00 & 1.00\\
&& & & & & &$\bigtriangledown$ & V & V & V&\nodata & 0.15 & 0.18 & 0.19 &\nodata&2.28 &2.24 & 2.49 &\nodata&1.00 & 1.00 & 1.00\\
\hline
\multirow{4}{*}{IC~1970} & \multirow{4}{*}{3.1} & \multirow{4}{*}{19.8} & \multirow{4}{*}{75.1} & \multirow{4}{*}{181}& \multirow{4}{*}{-18.61}& \multirow{4}{*}{126.7}& $\bigtriangledown$ & V & V &$\bigtriangledown$&\nodata&0.15&0.12&\nodata&\nodata&2.45&2.55&\nodata&\nodata&1.00&1.00&\nodata\\
&& & & & & & $\bigtriangledown$ & V & V &$\bigtriangledown$ &\nodata&0.13& 0.12&\nodata&\nodata&2.41 & 2.47&\nodata&\nodata&1.00&1.00&\nodata\\
&& & & & & & $\bigtriangledown$ & V & V &$\bigtriangledown$ &\nodata&0.27& 0.21 &\nodata&\nodata&2.57&2.68&\nodata&\nodata&1.00&1.00&\nodata\\
&& & & & & & $\bigtriangledown$ & V & V &$\bigtriangledown$&\nodata&0.25&0.21&\nodata&\nodata&2.51&2.59&\nodata&\nodata&1.00&1.00&\nodata\\
\hline
\multirow{4}{*}{IC~2058} & \multirow{4}{*}{6.5} & \multirow{4}{*}{19.4} & \multirow{4}{*}{17.4} & \multirow{4}{*}{203}& \multirow{4}{*}{-18.97}& \multirow{4}{*}{82.7}& $\bigtriangledown$ & V & V &$\bigtriangledown$&\nodata& 0.61 & 0.66&\nodata&\nodata&1.92 & 2.32&\nodata&\nodata&1.00&0.99&\nodata\\
&& & & & & &$\bigtriangledown$ & V & V &$\bigtriangledown$  &\nodata & 0.61&0.69 &\nodata & \nodata& 1.84& 2.26& \nodata &\nodata & 1.00&0.98&\nodata\\
&& & & && & $\bigtriangledown$ & V & V &$\bigtriangledown$&\nodata&1.14&1.18&\nodata&\nodata& 2.02 & 2.43&\nodata&\nodata&1.00&0.99&\nodata\\
&& & & & & & $\bigtriangledown$ & V & V &$\bigtriangledown$&\nodata&1.15&1.23&\nodata&\nodata&1.95&2.37&\nodata&\nodata&1.00&0.99&\nodata\\
\hline
\tablebreak
\multirow{4}{*}{IC~2135} & \multirow{4}{*}{5.8} & \multirow{4}{*}{29.2} & \multirow{4}{*}{108.6} & \multirow{4}{*}{194}&  \multirow{4}{*}{-19.01}&  \multirow{4}{*}{106.5}&$\perp$ & V & V & V&\nodata&0.53&0.51&0.19&\nodata&2.21&2.17&2.35&\nodata&1.00&1.00&1.00\\
&& & & & & & V & V & V & V& 0.61 & 0.54 & 0.53 & 0.19 & 2.05 & 2.12 & 2.07 & 2.28 & 1.00 &1.00&1.00&0.99\\
&& & & & & & $\perp$ & V & V & V&\nodata&0.96&0.93&0.37&\nodata&2.32&2.28&2.47&\nodata&1.00 &1.00 &1.00\\
&& & & & & & $\perp$ & V & V & V&\nodata&0.98&0.95&0.36&\nodata&2.24&2.19&2.39&\nodata&1.00&1.00&1.00\\
\hline
\multirow{4}{*}{IC~5052} & \multirow{4}{*}{7.1} & \multirow{4}{*}{8.1} & \multirow{4}{*}{141.4} & \multirow{4}{*}{425}& \multirow{4}{*}{\nodata}& \multirow{4}{*}{79.8}&V&V&$\bigtriangledown$&$\bigtriangledown$&{\it 0.93}&{\it 1.02}&\nodata&\nodata&{\it 1.95}&{\it 2.07}&\nodata&\nodata&{\it 1.00}&{\it 0.96}&\nodata&\nodata\\
&& &  &  &  & &V&V&$\bigtriangledown$&$\bigtriangledown$&{\it 0.95}&{\it 1.08}&\nodata&\nodata&{\it 1.84}&{\it 2.00}&\nodata&\nodata&{\it 1.00}&{\it 0.96}&\nodata&\nodata\\
&& &  &  &  & &V&V&$\bigtriangledown$&$\bigtriangledown$&{\it 1.65}&{\it 1.80}&\nodata&\nodata&{\it 2.00}&{\it 2.14}&\nodata&\nodata&{\it 1.00}&{\it 0.96}&\nodata&\nodata\\
&& &  &  &  & &V&V&$\bigtriangledown$&$\bigtriangledown$&{\it 1.71}&{\it 1.88}&\nodata&\nodata&{\it 1.95}&{\it 2.10}&\nodata&\nodata&{\it 1.00}&{\it 0.96}&\nodata&\nodata\\
\hline
\multirow{4}{*}{NGC~0522} & \multirow{4}{*}{4.1} & \multirow{4}{*}{35.0} & \multirow{4}{*}{32.9} & \multirow{4}{*}{144} & \multirow{4}{*}{-20.69} & \multirow{4}{*}{169.1} &V & V & V & V & 0.29 & 0.29 & 0.33 & 0.33 & 2.17 & 2.26 & 2.37 & 2.10 & 1.00 & 1.00 & 1.00 & 1.00\\
&& &  &  &  & &V&V&V&V&0.25&0.27&0.32&0.32&2.07&2.14&2.21&2.02&1.00&1.00&1.00&1.00\\
&& &  &  &  & &V&V&V&V&0.69&0.60&0.60&0.61&2.35&2.45&2.47&2.21&1.00&1.00&1.00&1.00\\
&& &  &  &  & &V&V&V&V&0.50&0.48&0.59&0.60&2.17&2.24&2.35&2.12&1.00&1.00&1.00&1.00\\
\hline
\multirow{4}{*}{NGC~0678} & \multirow{4}{*}{3.0} & \multirow{4}{*}{27.1} & \multirow{4}{*}{77.4} & \multirow{4}{*}{185}&\multirow{4}{*}{-20.95}&\multirow{4}{*}{169.0}&$\bigtriangledown$&V& $\approx$ & $\approx$ &\nodata&0.25&\nodata&\nodata&\nodata&2.14&\nodata&\nodata&\nodata&1.00&\nodata&\nodata\\
&& &  &  &  & &$\bigtriangledown$&V& $\approx$ & $\approx$ &\nodata& 0.10  &\nodata &\nodata &\nodata&2.39 &\nodata &\nodata &\nodata&1.00 &\nodata &\nodata\\
&& &  &  &  & &$\bigtriangledown$&V& $\approx$ & $\approx$&\nodata&0.48&\nodata&\nodata&\nodata&2.24&\nodata&\nodata&\nodata&1.00&\nodata&\nodata\\
&& &  &  &  & &$\bigtriangledown$&V& $\approx$ & $\approx$&\nodata&0.23&\nodata&\nodata&\nodata&2.43&\nodata&\nodata&\nodata&1.00&\nodata&\nodata\\
\hline
\multirow{4}{*}{NGC~1351A} & \multirow{4}{*}{4.5} & \multirow{4}{*}{20.9} & \multirow{4}{*}{132.1} & \multirow{4}{*}{147}&\multirow{4}{*}{-18.17}&\multirow{4}{*}{89.6}&$\perp$&V&V&$\bigtriangledown$&\nodata&0.46&0.50&\nodata&\nodata&2.51&2.47&\nodata&\nodata&1.00&1.00&\nodata\\
&& &  &  &  & &$\perp$&V&V&$\bigtriangledown$&\nodata&0.46&0.51&\nodata&\nodata&2.39&2.37&\nodata&\nodata&1.00&1.00&\nodata\\
&& &  &  &  & &$\perp$&V&V&$\bigtriangledown$&\nodata&0.83&0.89&\nodata&\nodata&2.65&2.61&\nodata&\nodata&1.00&1.00&\nodata\\
&& &  &  &  & &$\perp$&V&V&$\bigtriangledown$&\nodata&0.86&0.91&\nodata&\nodata&2.55&2.51&\nodata&\nodata&1.00&1.00&\nodata\\
\hline
\multirow{4}{*}{NGC~1495} & \multirow{4}{*}{5.0} & \multirow{4}{*}{17.3} & \multirow{4}{*}{104.5} & \multirow{4}{*}{131}&\multirow{4}{*}{-18.62}&\multirow{4}{*}{90.5}&$\perp$&V&V&$\perp$&\nodata&0.56&0.80&\nodata&\nodata&2.17&1.87&\nodata&\nodata&1.00&1.00&\nodata\\
&& &  &  &  & &$\perp$&V&V&$\perp$&\nodata&0.57&0.81&\nodata&\nodata&2.07&1.79&\nodata&\nodata&1.00&1.00&\nodata\\
&& &  &  &  & &$\perp$&V&V&$\perp$&\nodata&1.00&0.44&\nodata&\nodata&2.28&1.95&\nodata&\nodata&1.00&1.00&\nodata\\
&& &  &  &  & &$\perp$&V&V&$\perp$&\nodata&1.02&0.46&\nodata&\nodata&2.19&1.87&\nodata&\nodata&1.00&1.00&\nodata\\
\hline
\multirow{4}{*}{NGC~1827} & \multirow{4}{*}{5.9} & \multirow{4}{*}{11.6} & \multirow{4}{*}{119.6} & \multirow{4}{*}{194}& \multirow{4}{*}{-18.68}& \multirow{4}{*}{77.1}& $\bigtriangledown$&V&V&$\bigtriangledown$&\nodata&0.16&{\it 1.23}&\nodata&\nodata&1.90&{\it 2.07}&\nodata&\nodata&1.00&{\it 1.00}&\nodata\\
&& &  &  &  & &$\bigtriangledown$&V&V&$\bigtriangledown$&\nodata&0.12&{\it 1.29}&\nodata&\nodata&1.92&{\it 2.00}&\nodata&\nodata&1.00&{\it 1.00}&\nodata\\
&& &  &  &  & &$\bigtriangledown$&V&V&$\bigtriangledown$&\nodata&1.18&{\it 2.16}&\nodata&\nodata&1.97&{\it 2.12}&\nodata&\nodata&0.97& {\it 1.00}&\nodata\\
&& &  &  &  & &$\bigtriangledown$&V&V&$\bigtriangledown$&\nodata&0.21&{\it 2.25}&\nodata&\nodata&2.00&{\it 2.07}&\nodata&\nodata&1.00&{\it 1.00}&\nodata\\
\hline
\multirow{4}{*}{NGC~3501} & \multirow{4}{*}{5.9} & \multirow{4}{*}{23.3} & \multirow{4}{*}{28.0} & \multirow{4}{*}{262}& \multirow{4}{*}{-19.11}& \multirow{4}{*}{133.6}& $\star$&V&V&$\bigtriangledown$&\nodata&0.45&0.53&\nodata&\nodata&1.82&1.87&\nodata&\nodata&1.00&1.00&\nodata\\
&& &  &  &  & &$\star$&V&V&$\bigtriangledown$&\nodata&0.39&0.53&\nodata&\nodata&1.73&1.79&\nodata&\nodata&1.00&1.00&\nodata\\
&& &  &  &  & &$\star$&V&V&$\bigtriangledown$&\nodata&0.84&0.96&\nodata&\nodata&1.90&1.95&\nodata&\nodata&1.00&1.00&\nodata\\
&& &  &  &  & &$\star$&V&V&$\bigtriangledown$&\nodata&0.78&0.95&\nodata&\nodata&1.82&1.87&\nodata&\nodata&1.00&1.00&\nodata\\
\hline
\multirow{4}{*}{NGC~3628} & \multirow{4}{*}{3.1} & \multirow{4}{*}{12.2} & \multirow{4}{*}{102.6} & \multirow{4}{*}{658}&\multirow{4}{*}{-21.44}&\multirow{4}{*}{211.4}& $\perp$&V&V&$\perp$&\nodata&0.48&0.53&\nodata&\nodata&2.68&2.59&\nodata&\nodata&1.00&1.00&\nodata\\
&& &  &  &  & &$\perp$&V&V&$\perp$&\nodata&0.48&0.56&\nodata&\nodata&2.57&2.49&\nodata&\nodata&0.98&1.00&\nodata\\
&& &  &  &  & &$\perp$&V&V&$\perp$&\nodata&0.86&0.93&\nodata&\nodata&2.85&2.72&\nodata&\nodata&1.00&1.00&\nodata\\
&& &  &  &  & &$\perp$&V&V&$\perp$&\nodata&0.86&0.98&\nodata&\nodata&2.74&2.63&\nodata&\nodata&0.98&1.00&\nodata\\
\hline
\tablebreak
\multirow{4}{*}{NGC~4013} & \multirow{4}{*}{3.0} & \multirow{4}{*}{18.6} & \multirow{4}{*}{65.1} & \multirow{4}{*}{294}&\multirow{4}{*}{-19.38}&\multirow{4}{*}{181.7}& $\perp$&V&V&V&\nodata&0.29&0.30&0.25&\nodata&2.85&2.90&3.00&\nodata&1.00&1.00&1.00\\
&& &  &  &  & &$\perp$&V&V&V&\nodata&0.29&0.29&0.30&\nodata&2.70&2.76&2.85&\nodata&1.00&1.00&1.00\\
&& &  &  &  & &$\perp$&$\perp$&$\perp$&$\perp$&\nodata&\nodata&\nodata&\nodata&\nodata&\nodata&\nodata&\nodata&\nodata&\nodata&\nodata&\nodata\\
&& &  &  &  & &$\perp$&V&V&V&\nodata&0.54&0.56&0.50&\nodata&2.93&2.95&3.15&\nodata&1.00&1.00&1.00\\
\hline
\multirow{4}{*}{NGC~4330} & \multirow{4}{*}{6.3} & \multirow{4}{*}{19.5} & \multirow{4}{*}{59.2} & \multirow{4}{*}{137}&\multirow{4}{*}{-19.87}&\multirow{4}{*}{115.5}&V&V&V&V&0.48&0.33&0.32&0.33&2.43&2.17&2.19&2.32&1.00&1.00&1.00&1.00\\
&& &  &  &  & &V&V&V&V&0.50&0.32&0.30&0.33&2.30&2.07&2.10&2.26&1.00&0.99&1.00&0.97\\
&& &  &  &  & &V&V&V&V&0.87&0.61&0.74&0.61&2.57&2.28&2.43&2.47&1.00&1.00&1.00&1.00\\
&& &  &  &  & &V&V&V&V&0.90&0.60&0.56&0.61&2.47&2.19&2.19&2.37&1.00&1.00&1.00&0.99\\
\hline
\multirow{4}{*}{NGC~4437} & \multirow{4}{*}{6.0} & \multirow{4}{*}{9.8} & \multirow{4}{*}{82.4} & \multirow{4}{*}{547}&\multirow{4}{*}{-21.47}&\multirow{4}{*}{139.6}&V&V&V&V&0.21&0.12&0.10&0.12&1.87&2.00&2.10&2.07&0.94&0.99&1.00&1.00\\
&& &  &  &  & &V&V&V&V&0.15&0.12&0.09&0.09&1.87&1.95&2.07&2.12&0.94&0.94&1.00&1.00\\
&& &  &  &  & &V&V&V&V&0.44&0.25&0.19&0.24&1.92&2.05&2.17&2.12&0.94&1.00&1.00&1.00\\
&& &  &  &  & &V&V&V&V&0.35&0.24&0.18&0.19&1.90&2.00&2.12&2.14&0.94&0.95&1.00&1.00\\
\hline
\multirow{4}{*}{NGC~4565} & \multirow{4}{*}{3.2} & \multirow{4}{*}{13.3} & \multirow{4}{*}{135.1} & \multirow{4}{*}{1019}& \multirow{4}{*}{-22.51}& \multirow{4}{*}{244.9}&V&V&V&V&0.68&0.06&0.06&0.19&2.26&2.32&2.43&2.19&1.00&1.00&1.00&1.00\\
&& &  &  &  & &V&V&V&V&0.70&0.06&0.06&0.15&2.14&2.26&2.37&2.19&1.00&1.00&1.00&1.00\\
&& &  &  &  & &V&V&V&V&1.20&0.12&0.10&0.36&2.35&2.37&2.55&2.28&1.00&1.00&1.00&1.00\\
&& &  &  &  & &V&V&V&V&1.24&0.10&0.10&0.30&2.26&2.37&2.47&2.26&1.00&1.00&1.00&1.00\\
\hline
\multirow{4}{*}{NGC~5470} & \multirow{4}{*}{3.1} & \multirow{4}{*}{14.7} & \multirow{4}{*}{62.2} & \multirow{4}{*}{158}&\multirow{4}{*}{-18.01}&\multirow{4}{*}{109.4}&V&V&V&V&0.29&0.33&0.25&0.30&2.45&2.32&2.24&2.59&1.00&1.00&1.00&1.00\\
&& &  &  &  & &V&V&V&V&0.27&0.32&0.16&0.30&2.35&2.21&2.17&2.49&0.97&1.00&1.00&1.00\\
&& &  &  &  & &V&V&V&V&0.53&0.60&0.64&0.56&2.57&2.45&2.43&2.76&1.00&1.00&1.00&1.00\\
&& &  &  &  & &V&V&V&V&0.53&0.59&0.42&0.56&2.49&2.32&2.24&2.63&1.00&1.00&1.00&1.00\\
\hline
\multirow{4}{*}{NGC~5981} & \multirow{4}{*}{4.3} & \multirow{4}{*}{29.2} & \multirow{4}{*}{139.8} & \multirow{4}{*}{165}& \multirow{4}{*}{-20.54}& \multirow{4}{*}{251.1}&V&V&V&V&0.12&0.08&0.15&0.53&2.12&2.21&2.05&2.12&1.00&1.00&0.99&1.00\\
&& &  &  &  & &V&V&V&V&0.10&0.08&0.15&0.54&2.10&2.14&2.00&2.05&0.99&1.00&0.96&1.00\\
&& &  &  &  & &V&V&V&V&0.27&0.21&0.27&0.96&2.14&2.14&2.12&2.26&1.00&0.99&1.00&1.00\\
&& &  &  &  & &V&V&V&V&0.23&0.13&0.27&0.98&2.12&2.24&2.07&2.17&1.00&0.99&0.96&1.00\\
\hline
\multirow{4}{*}{PGC~012439} & \multirow{4}{*}{6.2} & \multirow{4}{*}{39.4} & \multirow{4}{*}{0.3} & \multirow{4}{*}{102}&\multirow{4}{*}{-19.04}&\multirow{4}{*}{106.1}&$\bigtriangledown$&$\perp$&V&$\bigtriangledown$&\nodata&\nodata&0.53&\nodata&\nodata&\nodata&2.49&\nodata&\nodata&\nodata&1.00&\nodata\\
&& &  &  &  & &$\bigtriangledown$&$\perp$&V&$\bigtriangledown$&\nodata&\nodata&0.54&\nodata&\nodata&\nodata&2.39&\nodata&\nodata&\nodata&1.00&\nodata\\
&& &  &  &  & &$\bigtriangledown$&$\perp$&V&$\bigtriangledown$&\nodata&\nodata&0.93&\nodata&\nodata&\nodata&2.61&\nodata&\nodata&\nodata&1.00&\nodata\\
&& &  &  &  & &$\bigtriangledown$&$\perp$&V&$\bigtriangledown$&\nodata&\nodata&0.96&\nodata&\nodata&\nodata&2.53&\nodata&\nodata&\nodata&1.00&\nodata\\
\hline
\multirow{4}{*}{PGC~012798} & \multirow{4}{*}{7.4} & \multirow{4}{*}{22.6} & \multirow{4}{*}{9.7} & \multirow{4}{*}{161}&\multirow{4}{*}{-18.97}&\multirow{4}{*}{101.1}&$\bigtriangledown$&V&V&$\bigtriangledown$&\nodata&0.61&0.50&\nodata&\nodata&2.30&2.32&\nodata&\nodata&1.00&1.00&\nodata\\
&& &  &  &  & &$\bigtriangledown$&V&V&$\bigtriangledown$&\nodata&0.63&0.51&\nodata&\nodata&2.21&2.24&\nodata&\nodata&1.00&1.00&\nodata\\
&& &  &  &  & &$\bigtriangledown$&V&V&$\bigtriangledown$&\nodata&1.10&0.90&\nodata&\nodata&2.39&2.45&\nodata&\nodata&1.00&1.00&\nodata\\
&& &  &  &  & &$\bigtriangledown$&V&V&$\bigtriangledown$&\nodata&1.12&0.91&\nodata&\nodata&2.32&2.37&\nodata&\nodata&1.00&1.00&\nodata\\
\hline
\multirow{4}{*}{PGC~013646} & \multirow{4}{*}{5.0} & \multirow{4}{*}{32.6} & \multirow{4}{*}{34.4} & \multirow{4}{*}{203}& \multirow{4}{*}{-19.97}& \multirow{4}{*}{168.1}&V&V&V&V&{\it 0.83}&0.13&0.13&0.33&{\it 1.92}&2.07&2.05&1.92&{\it 1.00}&1.00&1.00&1.00\\
&& &  &  &  & &V&V&V&V&{\it 0.84}&0.12&0.10&0.30&{\it 1.84}&2.03&2.05&1.84&{\it 1.00}&1.00&1.00&1.00\\
&& &  &  &  & &V&V&V&V&{\it 1.47}&0.24&0.24&0.63&{\it 2.00}&2.14&2.14&2.00&{\it 1.00}&1.00&1.00&1.00\\
&& &  &  &  & &V&V&V&V&{\it 1.50}&0.25&0.24&0.57&{\it 1.92}&2.07&2.07&1.92&{\it 1.00}&1.00&1.00&1.00\\
\hline
\tablebreak
\multirow{4}{*}{UGC~09977} & \multirow{4}{*}{5.3} & \multirow{4}{*}{31.0} & \multirow{4}{*}{77.6} & \multirow{4}{*}{233}& \multirow{4}{*}{-19.79}& \multirow{4}{*}{112.7}&$\bigtriangledown$&V&V&$\bigtriangledown$&\nodata&0.42&0.68&\nodata&\nodata&1.92&2.05&\nodata&\nodata&0.99&1.00&\nodata\\
&& &  &  &  & &$\bigtriangledown$&V&V&$\bigtriangledown$&\nodata&0.42&0.69&\nodata&\nodata&1.87&1.97&\nodata&\nodata&0.96&1.00&\nodata\\
&& &  &  &  & &$\bigtriangledown$&V&V&$\bigtriangledown$&\nodata&0.78&1.22&\nodata&\nodata&2.00&2.14&\nodata&\nodata&1.00&1.00&\nodata\\
&& &  &  &  & &$\bigtriangledown$&V&V&$\bigtriangledown$&\nodata&0.78&1.24&\nodata&\nodata&1.95&2.07&\nodata&\nodata&0.97&1.00&\nodata\\
\hline
\multirow{4}{*}{UGC~10288} & \multirow{4}{*}{5.3} & \multirow{4}{*}{32.4} & \multirow{4}{*}{90.4} & \multirow{4}{*}{287}&\multirow{4}{*}{-20.32}&\multirow{4}{*}{166.6}&$\bigtriangledown$&V&V&$\bigtriangledown$&\nodata&0.25&0.18&\nodata&\nodata&2.28&2.39&\nodata&\nodata&1.00&1.00&\nodata\\
&& &  &  &  & &$\bigtriangledown$&V&V&$\bigtriangledown$&\nodata&0.25&0.18&\nodata&\nodata&2.21&2.32&\nodata&\nodata&1.00&1.00&\nodata\\
&& &  &  &  & &$\bigtriangledown$&V&$\perp$&$\bigtriangledown$&\nodata&0.46&\nodata&\nodata&\nodata&2.39&\nodata&\nodata&\nodata&1.00&\nodata&\nodata\\
&& &  &  &  & &$\bigtriangledown$&V&V&$\bigtriangledown$&\nodata&0.45&0.33&\nodata&\nodata&2.30&2.43&\nodata&\nodata&1.00&1.00&\nodata\\
\hline
\multirow{4}{*}{UGC~10297} & \multirow{4}{*}{5.1} & \multirow{4}{*}{39.2} & \multirow{4}{*}{2.9} & \multirow{4}{*}{128}& \multirow{4}{*}{-19.32}& \multirow{4}{*}{102.8}&$\bigtriangledown$&$\bigtriangledown$&V&$\bigtriangledown$&\nodata&\nodata&{\it 0.59}&\nodata&\nodata&\nodata&{\it 1.73}&\nodata&\nodata&\nodata&{\it 1.00}&\nodata\\
&& &  &  &  & &$\bigtriangledown$&$\bigtriangledown$&V&$\bigtriangledown$&\nodata&\nodata&{\it 0.46}&\nodata&\nodata&\nodata&{\it 1.64}&\nodata&\nodata&\nodata&{\it 1.00}&\nodata\\
&& &  &  &  & &$\bigtriangledown$&$\bigtriangledown$&V&$\bigtriangledown$&\nodata&\nodata&{\it 1.07}&\nodata&\nodata&\nodata&{\it 1.79}&\nodata&\nodata&\nodata&{\it 1.00}&\nodata\\
&& &  &  &  & &$\bigtriangledown$&$\bigtriangledown$&V&$\bigtriangledown$&\nodata&\nodata&{\it 0.89}&\nodata&\nodata&\nodata&{\it 1.70}&\nodata&\nodata&\nodata&{\it 1.00}&\nodata\\
\enddata
\tablecomments{\label{table1} Galaxy ID (col.~1), morphological type (HyperLEDA; col.~2), distance (using NASA/IPAC Extragalactic Database mean value of redshift-independent distances except for IC~1553, NGC~5470 and PGC~012349 for which we estimated the distance using HyperLEDA's heliocentric radial velocity from radio measurement and using a Hubble constant $H_{0}=70\,{\rm km\,s^{-1}}$; col.~3), PA (col.~4), $D_{25}$ (HyperLEDA; col.~5), absolute blue magnitude (HyperLEDA; col.~6), apparent maximum rotation velocity of gas (HyperLEDA; col.~7), luminosity profile fit status for projected galactocentric distances $-0.8r_{25}<R<-0.5r_{25}$, $-0.5r_{25}<R<-0.2r_{25}$, $0.2r_{25}<R<0.5r_{25}$ and $0.5r_{25}<R<0.8r_{25}$ (V meaning that the fit has been successful, $\star$ meaning that a bright star made the fit impossible, $\bigtriangledown$ meaning that the mid-plane surface brightness was $\mu>21.5\,{\rm mag\,arcsec}^{-2}$ thus preventing $\Delta m\ge4.5$, $\perp$ meaning that although the mid-plane brightness was $\mu\le21.5\,{\rm mag\,arcsec}^{-2}$, $\Delta m<4.5$ and $\approx$ meaning that dust lanes not located in the mid-plane are affecting the fit; cols.~8 to 11), fitted thick to thin disk mid-plane stellar density (cols.~12 to 15), fitted thick to thin disk velocity dispersion ratio in the direction of the $z$ axis (cols.~16 to 19) and fitted fraction of the mid-plane light not absorbed by dust (cols.~20 to 23). For each galaxy, the first row corresponds to a fit using $\Upsilon_{\rm T}/\Upsilon_{\rm t}=1.2$ and without-gas, the second one to $\Upsilon_{\rm T}/\Upsilon_{\rm t}=1.2$ and with-gas, the third one to $\Upsilon_{\rm T}/\Upsilon_{\rm t}=2.4$ and without-gas and the fourth one to $\Upsilon_{\rm T}/\Upsilon_{\rm t}=2.4$ and with-gas. Data in italics indicate fitted values which are compatible with those of a single disk structure.}
\end{deluxetable}

\begin{deluxetable}{l|cccc|cccc|cccc|cccc|cccc}
\tablecolumns{21}
\rotate
\tablewidth{0pt}
\tabletypesize{\scriptsize}
\tablecaption{Galaxy and fit data (part two)}
\tablehead{ID &\multicolumn{4}{c|}{$z_t$}&\multicolumn{4}{c|}{$z_T$}& \multicolumn{4}{c|}{$\Sigma_{\rm T}/\Sigma_{\rm t}$}&\multicolumn{4}{c|}{$\mu_{\rm l}$}&\multicolumn{4}{c}{$\Delta m$}\\
   &\multicolumn{4}{c|}{(pc)}&\multicolumn{4}{c|}{(pc)}& & & & &\multicolumn{4}{c|}{${\rm (mag\,arcsec}^{-2})$}&\multicolumn{4}{c}{${\rm (mag\,arcsec}^{-2})$}\\
(1) & (2) &(3) &(4)&(5)&(6)&(7)&(8)&(9)&(10)&(11)&(12)&(13)&(14)&(15)&(16)&(17)&(18)&(19)&(20)&(21)}
\startdata
\multirow{4}{*}{ESO~079-003}& \nodata & 200 & 180 & \nodata & \nodata & 850 & 860 & \nodata & \nodata & 0.95 & 0.99 & \nodata & \nodata & 24.27 & 23.19 & \nodata & \nodata & 5.5 & 4.5 &\nodata\\
& \nodata&210&200& \nodata& \nodata&850&910& \nodata& \nodata&0.95&0.92& \nodata& \nodata&24.27&23.19& \nodata& \nodata&5.5&4.5& \nodata\\
& \nodata & 180 & 170 & \nodata & \nodata & 820 & 840 & \nodata & \nodata & 1.83 & 1.87 & \nodata& \nodata & 24.27 & 23.19 & \nodata& \nodata& 5.5 & 4.5 & \nodata\\
& \nodata&200&180& \nodata& \nodata&830&870& \nodata& \nodata&1.83&1.80& \nodata& \nodata&24.27&23.19& \nodata& \nodata&5.5&4.5& \nodata\\
\hline
\multirow{4}{*}{ESO~443-042}& \nodata & \nodata & 180 & 220 & \nodata & \nodata & 870 & 920 & \nodata & \nodata & 1.63 & 1.69 & \nodata & \nodata & 25.21 & 25.28 & \nodata & \nodata & 5.5 & 4.5 \\
& \nodata& \nodata&190&230& \nodata& \nodata&880&930& \nodata& \nodata&1.67&1.70& \nodata& \nodata&25.21&25.28& \nodata& \nodata&5.5&4.5\\
& \nodata&\nodata&160 & 220 & \nodata&\nodata&820& 900 &\nodata&\nodata&3.26 & 3.11 &\nodata&\nodata&24.71 &25.28 & \nodata& \nodata& 5.0 & 4.5 \\
&\nodata&\nodata&180&220&\nodata&\nodata&860&900&\nodata&\nodata&3.02&3.17&\nodata&\nodata&25.21&25.28&\nodata&\nodata&5.5&4.5\\
\hline
\multirow{4}{*}{ESO~544-027} & \nodata& 190 & \nodata& \nodata& \nodata& 850 & \nodata& \nodata& \nodata& 0.47 & \nodata& \nodata&\nodata&25.07 & \nodata&\nodata&\nodata&5.0& \nodata&\nodata\\
&\nodata&210&\nodata&\nodata&\nodata&900&\nodata&\nodata&\nodata&0.44&\nodata&\nodata&\nodata&25.07&\nodata&\nodata&\nodata&5.0&\nodata&\nodata\\
 & \nodata & 190 & \nodata& \nodata& \nodata&840 & \nodata& \nodata& \nodata & 0.87 & \nodata& \nodata& \nodata& 25.07 & \nodata& \nodata& \nodata& 5.0 & \nodata& \nodata\\
 &\nodata&210 &\nodata &\nodata &\nodata&890 &\nodata &\nodata &\nodata&0.81 &\nodata &\nodata &\nodata&25.07 &\nodata &\nodata &\nodata&5.0 &\nodata &\nodata\\
\hline
\multirow{4}{*}{ESO~548-063} & \nodata& \nodata& {\it 130}&\nodata&\nodata&\nodata&{\it 340}&\nodata&\nodata&\nodata&{\it 2.50}&\nodata&\nodata&\nodata&{\it 25.67}&\nodata&\nodata&\nodata&{\it 4.5}&\nodata\\
 & \nodata & \nodata&{\it 140} & \nodata & \nodata & \nodata&{\it 340} & \nodata & \nodata & \nodata&{\it 2.40} & \nodata & \nodata & \nodata&{\it 25.67} & \nodata & \nodata & \nodata&{\it 4.5} & \nodata\\
& \nodata& \nodata& {\it 120}& \nodata& \nodata& \nodata& {\it 340}& \nodata& \nodata& \nodata& {\it 4.39}& \nodata& \nodata& \nodata& {\it 25.67}& \nodata& \nodata& \nodata& {\it 4.5}& \nodata\\
&\nodata&\nodata&{\it 130}&\nodata&\nodata&\nodata&{\it 340}&\nodata&\nodata&\nodata&{\it 4.29}&\nodata&\nodata&\nodata&{\it 25.67}&\nodata&\nodata&\nodata&{\it 4.5}&\nodata\\
\hline
\multirow{4}{*}{IC~0217}&\nodata& 180&160&\nodata&\nodata&690&590&\nodata&\nodata&1.11&0.98&\nodata&\nodata&25.65&24.91&\nodata&\nodata&5.0&5.0&\nodata\\
&\nodata&200&210&\nodata&\nodata&700&720&\nodata&\nodata&1.09&0.54&\nodata&\nodata&25.65&25.41&\nodata&\nodata&5.0&5.5&\nodata\\
&\nodata&170&140&\nodata&\nodata&670&580&\nodata&\nodata&2.12&1.86&\nodata&\nodata&25.65&24.91&\nodata&\nodata&5.0& 5.0&\nodata\\
&\nodata&180&200&\nodata&\nodata&680&720&\nodata&\nodata&2.09&1.02&\nodata&\nodata&25.65&25.41&\nodata&\nodata&5.0&5.5&\nodata\\
\hline
\multirow{4}{*}{IC~1197}&\nodata&170&120&\nodata&\nodata&680&610&\nodata&\nodata&1.56&2.22&\nodata&\nodata&25.45&25.42&\nodata&\nodata&4.5&4.5&\nodata\\
&\nodata&180&130&\nodata&\nodata&690&610&\nodata&\nodata&1.54&2.34&\nodata&\nodata&25.45&25.42&\nodata&\nodata&4.5&4.5&\nodata\\
&\nodata&160&110&\nodata&\nodata&670&600&\nodata&\nodata&2.89&3.98&\nodata&\nodata&25.45&25.42&\nodata&\nodata&4.5&4.5&\nodata\\
&\nodata&170&120&\nodata&\nodata&670&600&\nodata&\nodata&2.90&4.12&\nodata&\nodata&25.45&25.42&\nodata&\nodata&4.5&4.5&\nodata\\
\hline
\multirow{4}{*}{IC~1553}&\nodata&200& 220& 240&\nodata&680& 910&1280&\nodata&1.03 & 0.33 & 0.36&\nodata& 24.23& 24.49& 25.59&\nodata&4.5&5.5&5.5\\
&\nodata&310&240&250&\nodata&1320&1040&1280&\nodata&0.27&0.27&0.37&\nodata&25.23&25.49&25.59&\nodata&5.5&6.5&5.5\\
&\nodata&190&210&230&\nodata&660&870&1250&\nodata&1.92&0.68&0.69&\nodata&24.23 &24.49& 25.59&\nodata&4.5&5.5&5.5\\
&\nodata&310&230&250&\nodata&1400&1020&1270&\nodata&0.48&0.55&0.70&\nodata&24.73&25.49&25.59&\nodata&5.0&6.5&5.5\\
\hline
\multirow{4}{*}{IC~1970} &\nodata&190&180 &\nodata &\nodata&990&1030 &\nodata &\nodata&0.50&0.43 &\nodata &\nodata& 24.49&25.00&\nodata&\nodata&5.0&5.5&\nodata\\
&\nodata&220&200&\nodata&\nodata&1070&1050&\nodata&\nodata&0.47&0.44&\nodata&\nodata&24.49&25.00&\nodata&\nodata&5.0&5.5&\nodata\\
&\nodata&190 & 180 &\nodata&\nodata &990 &1040&\nodata&\nodata&0.93&0.78&\nodata&\nodata&24.49&25.00&\nodata&\nodata&5.0&5.5&\nodata\\
&\nodata&200&200&\nodata&\nodata&1030&1050&\nodata&\nodata&0.91&0.79&\nodata&\nodata&24.49&25.00&\nodata&\nodata&5.0&5.5&\nodata\\
\hline
\multirow{4}{*}{IC~2058}&\nodata&160&120&\nodata&\nodata&490&490&\nodata&\nodata&1.39&1.87&\nodata&\nodata&25.59&25.73&\nodata&\nodata&4.5&4.5&\nodata\\
& \nodata&170 &120 &\nodata & \nodata&500 &490 & \nodata&\nodata & 1.36& 1.97&\nodata &\nodata & 25.59& 25.73& \nodata&\nodata&4.5&4.5&\nodata\\
&\nodata&150&110&\nodata&\nodata&490&480&\nodata&\nodata&2.67&3.43&\nodata&\nodata&25.59&25.73&\nodata&\nodata&4.5&4.5&\nodata\\
&\nodata&160&110&\nodata&\nodata&490&480&\nodata&\nodata&2.66&3.56&\nodata&\nodata&25.59&25.73&\nodata&\nodata&4.5&4.5&\nodata\\
\hline
\tablebreak
\multirow{4}{*}{IC~2135}&\nodata&230&220&370&\nodata&930&840&1700&\nodata&1.42&1.35&0.60&\nodata&24.65&24.86&25.92&\nodata&4.5& 5.0 & 5.0\\
&270 &250 &230 &390 &990&940&850&1720 &1.56 & 1.44&1.37&0.62&25.35&24.65&24.86&25.92&4.5&4.5&5.0&5.0\\
&\nodata&220&210&340&\nodata&910&830&1640&\nodata&2.67&2.53&1.20&\nodata&24.65&24.86&25.92&\nodata&4.5&5.0&5.0\\
&\nodata&230&230&370&\nodata&910&840&1680&\nodata&2.68&2.54&1.16&\nodata&24.65&24.86&25.92&\nodata&4.5&5.0&5.0\\
\hline
\multirow{4}{*}{IC~5052}& {\it 170}& {\it 140}& \nodata& \nodata& {\it 530}& {\it 470}& \nodata& \nodata& {\it 2.10}& {\it 2.47}& \nodata& \nodata& {\it 26.00}& {\it 25.53}& \nodata& \nodata& {\it 4.5}& {\it 4.5}& \nodata& \nodata\\
& {\it 190}& {\it 140}& \nodata& \nodata& {\it 530}& {\it 470}& \nodata& \nodata& {\it 2.05}& {\it 2.57}& \nodata& \nodata& {\it 26.00}& {\it 25.53}& \nodata& \nodata& {\it 4.5}& {\it 4.5}& \nodata& \nodata\\
& {\it 160}& {\it 130}& \nodata& \nodata& {\it 530}& {\it 470}& \nodata& \nodata& {\it 3.75}& {\it 4.42}& \nodata& \nodata& {\it 26.00}& {\it 25.53}& \nodata& \nodata& {\it 4.5}& {\it 4.5}& \nodata& \nodata\\
&{\it 170}&{\it 130}&\nodata&\nodata&{\it 530}&{\it 480}&\nodata&\nodata&{\it 3.86}&{\it 4.59}&\nodata&\nodata&{\it 26.00}&{\it 25.53}&\nodata&\nodata&{\it 4.5}&{\it 4.5}&\nodata&\nodata\\
\hline
\multirow{4}{*}{NGC~0522}& 230 & 190 & 160 & 220 & 930 & 810 & 730 & 850 & 0.78 & 0.82 & 1.00 & 0.85 & 25.08 & 24.75 & 24.13 & 25.51 & 5.0 & 5.5 & 5.0 & 5.5\\
&260&210&190&240&980&830&760&880&0.69&0.77&0.92&0.82&25.08&24.75&24.13&25.51&5.0&5.5&5.0&5.5\\
&190&160&150&210&800&740&710&840&1.98&1.83&1.85&1.65&24.58&24.25&24.13&25.51&4.5&5.0&5.0&5.5\\
&240&200&170&220&950&830&740&850&1.37&1.39&1.77&1.59&25.08&24.75&24.13&25.51&5.0&5.5&5.0&5.5\\
\hline
\multirow{4}{*}{NGC~0678}&\nodata&450&\nodata&\nodata&\nodata&1760&\nodata&\nodata&\nodata&0.69&\nodata&\nodata&\nodata&25.18&\nodata&\nodata&\nodata&4.5&\nodata&\nodata\\
&\nodata&580&\nodata&\nodata&\nodata&2890&\nodata&\nodata&\nodata&0.37&\nodata&\nodata&\nodata&25.68&\nodata&\nodata&\nodata&5.0&\nodata&\nodata\\
&\nodata&420&\nodata&\nodata&\nodata&1710&\nodata&\nodata&\nodata&1.32&\nodata&\nodata&\nodata&25.18&\nodata&\nodata&\nodata&4.5&\nodata&\nodata\\
&\nodata&540&\nodata&\nodata&\nodata&2640&\nodata&\nodata&\nodata&0.77&\nodata&\nodata&\nodata&25.68&\nodata&\nodata&\nodata&5.0&\nodata&\nodata\\
\hline
\multirow{4}{*}{NGC~1351A}&\nodata&120&110&\nodata&\nodata&580&530&\nodata&\nodata&1.49&1.55&\nodata&\nodata&24.90&25.48&\nodata&\nodata&4.5&5.0&\nodata\\
&\nodata&130&120&\nodata&\nodata&590&530&\nodata&\nodata&1.47&1.58&\nodata&\nodata&24.90&25.48&\nodata&\nodata&4.5&5.0&\nodata\\
&\nodata&120&110&\nodata&\nodata&570&510&\nodata&\nodata&2.70&2.83&\nodata&\nodata&24.90&25.48&\nodata&\nodata&4.5&5.0&\nodata\\
&\nodata&130&110&\nodata&\nodata&570&520&\nodata&\nodata&2.79&2.91&\nodata&\nodata&24.90&25.49&\nodata&\nodata&4.5&5.0&\nodata\\
\hline
\multirow{4}{*}{NGC~1495}&\nodata&100&120&\nodata&\nodata&370&350&\nodata&\nodata&1.46&1.71&\nodata&\nodata&24.61&24.80&\nodata&\nodata&5.0&5.0&\nodata\\
&\nodata&100&130&\nodata&\nodata&380&350&\nodata&\nodata&1.47&1.70&\nodata&\nodata&24.61&24.80&\nodata&\nodata&5.0&5.0&\nodata\\
&\nodata&90&110&\nodata&\nodata&370&340&\nodata&\nodata&2.72&3.20&\nodata&\nodata&24.61&24.80&\nodata&\nodata&5.0&5.0&\nodata\\
&\nodata&90&120&\nodata&\nodata&370&340&\nodata&\nodata&2.72&3.15&\nodata&\nodata&24.61&24.80&\nodata&\nodata&5.0&5.0&\nodata\\
\hline
\multirow{4}{*}{NGC~1827}&\nodata&180&{\it 90}&\nodata&\nodata&600&{\it 330}&\nodata&\nodata&0.38&{\it 2.95}&\nodata&\nodata&25.75&{\it 25.54}&\nodata&\nodata&5.0&{\it 4.5}&\nodata\\
&\nodata&210&{\it 100}&\nodata&\nodata&700&{\it 330}&\nodata&\nodata&0.30&{\it 3.04}&\nodata&\nodata&25.75&{\it 25.54}&\nodata&\nodata&5.0&{\it 4.5}&\nodata\\
&\nodata&120&{\it 90}&\nodata&\nodata&370&{\it 330}&\nodata&\nodata&2.69&{\it 5.21}&\nodata&\nodata&25.25&{\it 25.54}&\nodata&\nodata&4.5&{\it 4.5}&\nodata\\
&\nodata&200&{\it 90}&\nodata&\nodata&710&{\it 330}&\nodata&\nodata&0.55&{\it 5.39}&\nodata&\nodata&25.75&{\it 25.54}&\nodata&\nodata&5.0&{\it 4.5}&\nodata\\
\hline
\multirow{4}{*}{NGC~3501}&\nodata&210&180&\nodata&\nodata&610&560&\nodata&\nodata&0.95&1.15&\nodata&\nodata&25.70&25.64&\nodata&\nodata&5.5&5.5&\nodata\\
&\nodata&240&200&\nodata&\nodata&650&560&\nodata&\nodata&0.81&1.12&\nodata&\nodata&25.70&25.64&\nodata&\nodata&5.5&5.5&\nodata\\
&\nodata&210&170&\nodata&\nodata&600&550&\nodata&\nodata&1.84&2.16&\nodata&\nodata&25.70&25.64&\nodata&\nodata&5.5&5.5&\nodata\\
&\nodata&220&180&\nodata&\nodata&630&560&\nodata&\nodata&1.67&2.09&\nodata&\nodata&25.70&25.64&\nodata&\nodata&5.5&5.5&\nodata\\
\hline
\multirow{4}{*}{NGC~3628}&\nodata&290&300&\nodata&\nodata&1470&1520&\nodata&\nodata&1.67&1.73&\nodata&\nodata&24.41&25.01&\nodata&\nodata&5.0&5.5&\nodata\\
&\nodata&300&330&\nodata&\nodata&1510&1540&\nodata&\nodata&1.66&1.82&\nodata&\nodata&24.41&25.01&\nodata&\nodata&5.0&5.5&\nodata\\
&\nodata&270&290&\nodata&\nodata&1460&1510&\nodata&\nodata&3.05&3.12&\nodata&\nodata&24.41&25.01&\nodata&\nodata&5.0&5.5&\nodata\\
&\nodata&280&310&\nodata&\nodata&1470&1510&\nodata&\nodata&3.04&3.26&\nodata&\nodata&24.41&25.01&\nodata&\nodata&5.0&5.5&\nodata\\
\hline
\tablebreak
\multirow{4}{*}{NGC~4013}&\nodata&120&110&180&\nodata&720&680&1190&\nodata&1.12&1.20&1.09&\nodata&22.99&22.93&24.67&\nodata&4.5&4.5&4.5\\
&\nodata&130&120&190&\nodata&730&700&1210&\nodata&1.11&1.20&1.08&\nodata&22.99&22.93&24.67&\nodata&4.5&4.5&4.5\\
&\nodata&\nodata&\nodata&\nodata&\nodata&\nodata&\nodata&\nodata&\nodata&\nodata&\nodata&\nodata&\nodata&\nodata&\nodata&\nodata&\nodata&\nodata&\nodata&\nodata\\
&\nodata&120&110&180&\nodata&700&680&1140&\nodata&2.19&2.26&2.22&\nodata&22.99&22.93&24.67&\nodata&4.5&4.5&4.5\\
\hline
\multirow{4}{*}{NGC~4330}&150&190&170&190&700&730&710&820&1.47&0.84&0.87&0.98&25.54&25.65&25.44&25.93&5.0&5.5&5.5&5.5\\
&170&200&190&200&710&760&740&830&1.48&0.84&0.82&0.99&25.54&25.65&25.44&25.93&5.0&5.5&5.5&5.5\\
&150&180&140&170&680&710&620&800&2.74&1.71&2.19&1.89&25.54&25.65&24.94&25.93&5.0&5.5&5.0&5.5\\
&150&190&180&190&690&730&710&820&2.81&1.66&1.55&1.88&25.54&25.65&25.44&25.93&5.0&5.5&5.5&5.5\\
\hline
\multirow{4}{*}{NGC~4437}&330&300&300&350&1040&1110&1200&1400&0.47&0.30&0.28&0.32&25.98&25.97&25.98&25.87&5.0&6.0&6.0&5.0\\
&380&320&330&390&1200&1120&1300&1630&0.36&0.31&0.25&0.26&25.98&25.97&25.98&25.87&5.0&6.0&6.0&5.0\\
&310&280&290&350&990&1050&1190&1330&0.99&0.65&0.54&0.64&25.98&25.97&25.98&25.87&5.0&6.0&6.0&5.0\\
&340&300&320&370&1100&1080&1250&1500&0.81&0.62&0.51&0.56&25.98&25.97&25.98&25.87&5.0&6.0&6.0&5.0\\
\hline
\multirow{4}{*}{NGC~4565}&240&420&450&380&930&2080&2290&1600&1.85&0.19&0.21&0.55&25.94&25.70&25.74&25.84&4.5&6.0&6.0&4.5\\
&250&460&460&430&950&2140&2350&1850&1.88&0.20&0.21&0.45&25.94&25.70&25.74&25.84&4.5&6.0&6.0&4.5\\
&220&410&430&360&930&2010&2360&1560&3.33&0.39&0.38&1.04&25.94&25.70&25.74&25.84&4.5&6.0&6.0&4.5\\
&240&440&450&410&930&2200&2400&1740&3.40&0.36&0.39&0.91&25.94&25.70&25.74&25.84&4.5&6.0&6.0&4.5\\
\hline
\multirow{4}{*}{NGC~5470}&90&70&90&90&430&330&380&460&0.92&0.98&0.73&1.03&25.58&24.64&24.70&25.99&5.0&5.0&5.0&5.0\\
&100&90&110&100&440&350&460&470&0.87&0.92&0.49&1.04&25.58&24.64&25.70&25.99&5.0&5.0&6.0&5.0\\
&80&70&70&80&420&330&310&440&1.72&1.83&1.94&1.97&25.58&24.64&24.20&25.99&5.0&5.0&4.5&5.0\\
&90&80&90&90&420&330&390&450&1.73&1.75&1.23&1.95&25.58&24.64&24.70&25.99&5.0&5.0&5.0&5.0\\
\hline
\multirow{4}{*}{NGC~5981}&310&270&240&190&1230&1190&880&700&0.33&0.22&0.39&1.35&25.80&25.87&25.93&25.70&5.0&6.5&6.5&5.0\\
&330&280&250&200&1310&1200&890&710&0.30&0.23&0.40&1.38&25.80&25.87&25.93&25.70&5.0&6.5&6.5&5.0\\
&290&250&220&170&1110&970&880&690&0.73&0.57&0.72&2.58&25.80&25.37&25.93&25.70&5.0&6.0&6.5&5.0\\
&310&280&240&180&1220&1210&890&700&0.63&0.42&0.73&2.58&25.80&25.87&25.93&25.70&5.0&6.5&6.5&5.0\\
\hline
\multirow{4}{*}{PGC~012439}&\nodata&\nodata&130&\nodata&\nodata&\nodata&620&\nodata&\nodata&\nodata&1.65&\nodata&\nodata&\nodata&25.41&\nodata&\nodata&\nodata&4.5&\nodata\\
&\nodata&\nodata&110&\nodata&\nodata&\nodata&640&\nodata&\nodata&\nodata&1.68&\nodata&\nodata&\nodata&25.41&\nodata&\nodata&\nodata&4.5&\nodata\\
&\nodata&\nodata&120&\nodata&\nodata&\nodata&620&\nodata&\nodata&\nodata&2.97&\nodata&\nodata&\nodata&25.41&\nodata&\nodata&\nodata&4.5&\nodata\\
&\nodata&\nodata&130&\nodata&\nodata&\nodata&620&\nodata&\nodata&\nodata&3.07&\nodata&\nodata&\nodata&25.41&\nodata&\nodata&\nodata&4.5&\nodata\\
\hline
\multirow{4}{*}{PGC~012798}&\nodata&130&130&\nodata&\nodata&540&560&\nodata&\nodata&1.73&1.43&\nodata&\nodata&25.70&25.66&\nodata&\nodata&4.5&4.5&\nodata\\
&\nodata&140&140&\nodata&\nodata&540&570&\nodata&\nodata&1.76&1.47&\nodata&\nodata&25.70&25.66&\nodata&\nodata&4.5&4.5&\nodata\\
&\nodata&120&120&\nodata&\nodata&530&550&\nodata&\nodata&3.12&2.67&\nodata&\nodata&25.70&25.66&\nodata&\nodata&4.5&4.5&\nodata\\
&\nodata&130&130&\nodata&\nodata&540&550&\nodata&\nodata&3.20&2.71&\nodata&\nodata&25.70&25.66&\nodata&\nodata&4.5&4.5&\nodata\\
\hline
\multirow{4}{*}{PGC~013646}&{\it 180}&250&240&230&{\it 570}&960&950&750&{\it 1.84}&0.36&0.35&0.76&{\it 25.40}&24.93&25.44&26.00&{\it 4.5}&5.5&6.0&5.5\\
&{\it 200}&270&280&260&{\it 580}&1010&1070&790&{\it 1.83}&0.32&0.29&0.69&{\it 25.40}&24.93&25.94&26.00&{\it 4.5}&5.5&6.5&5.5\\
&{\it 180}&250&250&220&{\it 570}&970&970&740&{\it 3.36}&0.65&0.65&1.49&{\it 25.40}&24.93&25.44&26.00&{\it 4.5}&5.5&6.0&5.5\\
&{\it 190}&250&260&240&{\it 580}&950&980&770&{\it 3.35}&0.69&0.65&1.34&{\it 25.40}&24.93&25.44&26.00&{\it 4.5}&5.5&6.0&5.5\\
\hline
\tablebreak
\multirow{4}{*}{UGC~09977}&\nodata&280&230&\nodata&\nodata&870&800&\nodata&\nodata&0.96&1.64&\nodata&\nodata&25.79&25.93&\nodata&\nodata&4.5&4.5&\nodata\\
&\nodata&280&240&\nodata&\nodata&880&810&\nodata&\nodata&0.96&1.66&\nodata&\nodata&25.79&25.93&\nodata&\nodata&4.5&4.5&\nodata\\
&\nodata&250&210&\nodata&\nodata&860&770&\nodata&\nodata&1.83&3.03&\nodata&\nodata&25.79&25.93&\nodata&\nodata&4.5&4.5&\nodata\\
&\nodata&280&230&\nodata&\nodata&860&790&\nodata&\nodata&1.83&3.07&\nodata&\nodata&25.79&25.93&\nodata&\nodata&4.5&4.5&\nodata\\
\hline
\multirow{4}{*}{UGC~10288}&\nodata&230&270&\nodata&\nodata&1030&1310&\nodata&\nodata&0.75&0.57&\nodata&\nodata&25.58&25.71&\nodata&\nodata&5.5&5.5&\nodata\\
&\nodata&260&300&\nodata&\nodata&1050&1340&\nodata&\nodata&0.76&0.59&\nodata&\nodata&25.58&25.71&\nodata&\nodata&5.5&5.5&\nodata\\
&\nodata&230&\nodata&\nodata&\nodata&1010&\nodata&\nodata&\nodata&1.40&\nodata&\nodata&\nodata&25.58&\nodata&\nodata&\nodata&5.5&\nodata&\nodata\\
&\nodata&240&280&\nodata&\nodata&1040&1300&\nodata&\nodata&1.36&1.10&\nodata&\nodata&25.58&25.71&\nodata&\nodata&5.5&5.5&\nodata\\
\hline
\multirow{4}{*}{UGC~10297}&\nodata&\nodata&{\it 250}&\nodata&\nodata&\nodata&{\it 680}&\nodata&\nodata&\nodata&{\it 1.16}&\nodata&\nodata&\nodata&{\it 25.89}&\nodata&\nodata&\nodata&{\it 4.5}&\nodata\\
&\nodata&\nodata&{\it 290}&\nodata&\nodata&\nodata&{\it 720}&\nodata&\nodata&\nodata&{\it 0.89}&\nodata&\nodata&\nodata&{\it 25.89}&\nodata&\nodata&\nodata&{\it 4.5}&\nodata\\
&\nodata&\nodata&{\it 240}&\nodata&\nodata&\nodata&{\it 670}&\nodata&\nodata&\nodata&{\it 2.16}&\nodata&\nodata&\nodata&{\it 25.89}&\nodata&\nodata&\nodata&{\it 4.5}&\nodata\\
&\nodata&\nodata&{\it 280}&\nodata&\nodata&\nodata&{\it 720}&\nodata&\nodata&\nodata&{\it 1.74}&\nodata&\nodata&\nodata&{\it 25.89}&\nodata&\nodata&\nodata&{\it 4.5}&\nodata\\
\enddata
\tablecomments{\label{table2} Galaxy ID (col.~1), fitted thin disk scaleheight for projected galactocentric distances $-0.8r_{25}<R<-0.5r_{25}$, $-0.5r_{25}<R<-0.2r_{25}$, $0.2r_{25}<R<0.5r_{25}$ and $0.5r_{25}<R<0.8r_{25}$ (cols.~2 to 5), fitted thick disk scaleheight (cols.~6 to 9), fitted thick to thin stellar column mass density (cols.~10 to 13), limiting magnitude of the fit (cols.~14 to 17) and dynamical range of the fit (cols.~18 to 21). For each galaxy, the first row corresponds to a fit using $\Upsilon_{\rm T}/\Upsilon_{\rm t}=1.2$ and  without-gas, the second one to $\Upsilon_{\rm T}/\Upsilon_{\rm t}=1.2$ and with-gas, the third one to $\Upsilon_{\rm T}/\Upsilon_{\rm t}=2.4$ and without-gas and the fourth one to $\Upsilon_{\rm T}/\Upsilon_{\rm t}=2.4$ and with-gas. Data in italics indicate fitted values which are compatible with those of a single disk structure.}
\end{deluxetable}
 
\end{document}